\documentclass[structabstract]{aa}

\usepackage{graphicx}
\usepackage{txfonts}
\usepackage{natbib}

\begin{document}

\title{Detecting stars, galaxies, and asteroids with Gaia}

\author{
  J.H.J. de Bruijne\inst{\ref{inst1}}
  \and
  M. Allen\inst{\ref{inst1},\ref{inst2}}
  \and
  S. Azaz\inst{\ref{inst1}}
  \and
  A. Krone--Martins\inst{\ref{inst3}}
  \and
  T. Prod'homme\inst{\ref{inst4}}
  \and
  D. Hestroffer\inst{\ref{inst5}}
}

\institute{
  Scientific Support Office, Directorate of Science and Robotic Exploration, European Space Research and Technology Centre (ESA/ESTEC), Keplerlaan 1, 2201AZ, Noordwijk, The Netherlands\ \email{\ jos.de.bruijne@esa.int}\label{inst1}
  \and
  Cardiff School of Physics and Astronomy, Cardiff University, Queens Buildings, The Parade, Cardiff, CF24 3AA, United Kingdom\label{inst2}
  \and
  Universidade de Lisboa, Faculdade de Ci\^encias, CENTRA/SIM, 1749--016 Lisboa, Portugal\label{inst3}
  \and
  Directorate of Technical and Quality Management ESA/ESTEC), Keplerlaan 1, 2201AZ, Noordwijk, The Netherlands\label{inst4}
  \and
  Institut de m\'ecanique c\'eleste et de calcul des \'eph\'em\'erides (IMCCE), Observatoire de Paris, UPMC, Universit\'e Lille 1, CNRS, 77 Avenue Denfert--Rochereau, F-75014 Paris, France\label{inst5}
}


\abstract
{Gaia is Europe's space astrometry mission, aiming to make a three-dimensional map of $1,000$ million stars in our Milky Way to unravel its kinematical, dynamical, and chemical structure and evolution.}
{We present a study of Gaia's detection capability of objects, in particular non-saturated stars, double stars, unresolved external galaxies, and asteroids. Gaia's on-board detection software autonomously discriminates stars from spurious objects like cosmic rays and Solar protons. For this, parametrised point-spread-function-shape criteria are used, which need to be calibrated and tuned. This study aims to provide an optimum set of parameters for these filters.}
{We developed a validated emulation of the on-board detection software, which has $20$ free, so-called rejection parameters which govern the boundaries between stars on the one hand and sharp (high-frequency) or extended (low-frequency) events on the other hand. We evaluate the detection and rejection performance of the algorithm using catalogues of simulated single stars, resolved and unresolved double stars, cosmic rays, Solar protons, unresolved external galaxies, and asteroids.}
{We optimised the rejection parameters, improving -- with respect to the functional baseline -- the detection performance of single stars and of unresolved and resolved double stars, while, at the same time, improving the rejection performance of cosmic rays and of Solar protons. The optimised rejection parameters also remove the artefact of the functional-baseline parameters that the reduction of the detection probability of stars as function of magnitude already sets in before the nominal faint-end threshold at $G = 20$~mag. We find, as a result of the rectangular pixel size, that the minimum separation to resolve a close, equal-brightness double star is $0.23$~arcsec in the along-scan and $0.70$~arcsec in the across-scan direction, independent of the brightness of the primary. To resolve double stars with $\Delta G > 0$~mag, larger separations are required. We find that, whereas the optimised rejection parameters have no significant impact on the detectability of pure de Vaucouleurs profiles, they do significantly improve the detection of pure exponential-disk profiles, and hence also the detection of unresolved external galaxies with intermediate profiles. We also find that the optimised rejection parameters provide detection gains for asteroids fainter than $20$~mag and for fast-moving near-Earth objects fainter than $18$~mag, albeit this gain comes at the expense of a modest detection-probability loss for bright, fast-moving near-Earth objects. The major side effect of the optimised parameters is that spurious ghosts in the wings of bright stars essentially pass unfiltered.}
{}


\maketitle

\section{Introduction}

Gaia \citep[e.g.,][]{2001A&A...369..339P,2008IAUS..248..217L} is the current astrometry mission of the European Space Agency (ESA), following up on the success of the Hipparcos mission \citep{1997ESASP1200.....P,1997A&A...323L..49P,2009aaat.book.....P}. Gaia's objective is to unravel the kinematical, dynamical, and chemical structure and evolution of our Galaxy, the Milky Way \citep[e.g.,][]{2010MNRAS.408..935G}. In addition, Gaia's data will revolutionise many other areas of astronomy, e.g., stellar structure and evolution, stellar variability, double and multiple stars, Solar-system bodies, extra-galactic objects, fundamental physics, and exo-planets \citep[e.g.,][]{2008IAUS..248...59P,2012P&SS...73....1T,2010IAUS..261..306M,2011EAS....45..161E,2011EAS....45..273S,2011PhRvD..84l2001M,Tsalmantza2009,KroneMartins2013}. During its five-year lifetime, Gaia will survey the full sky and repeatedly observe the brightest $1,000$ million objects, down to $20^{\rm th}$ magnitude \citep[e.g.,][]{2010SPIE.7731E..35D}. Gaia's science data comprises absolute astrometry, broad-band photometry, and low-resolution spectro-photometry. Medium-resolution spectroscopic data will be obtained for the brightest $150$ million sources, down to $17^{\rm th}$ magnitude. The final Gaia catalogue, due in $2022$, will contain astrometry (positions, parallaxes, and proper motions) with standard errors less than $10$~micro-arcsecond ($\mu$as, $\mu$as~yr$^{-1}$ for proper motions) for stars brighter than $12^{\rm th}$ magnitude, $25~\mu$as for stars at $15^{\rm th}$ magnitude, and $300~\mu$as at magnitude $20$ \citep{2012Ap&SS.341...31D}. Milli-magnitude-precision photometry \citep{2010A&A...523A..48J} allows to get a handle on effective temperature, surface gravity, metallicity, and reddening of all stars \citep{2010MNRAS.403...96B,2012MNRAS.426.2463L}. The spectroscopic data may allow the determination of radial velocities with errors of $1~{\rm km~s}^{-1}$ at the bright end and $15~{\rm km~s}^{-1}$ at magnitude $17$ \citep{2005MNRAS.359.1306W,2011EAS....45..189K} as well as astrophysical diagnostics such as effective temperature and metallicity for the brightest few million objects \citep{2011A&A...535A.106K}. Clearly, these performances will only be reached with a total of five years of collected data and after careful calibration and extensive data processing.

Gaia is a survey mission and the spacecraft continuously scans the sky. The inertial rotation rate is $60$~arcsec per second -- which means the rotation period is $6$ hours -- and a slow precession of the spin axis at a fixed, $45$-degree angle to the Sun allows to reach full-sky coverage after some $6$ months. On average, stars are seen about $70$ times during the five-year mission. The slow rotation of the spacecraft causes stars to drift through the focal plane. The CCD detectors in the focal plane are hence operated in Time-Delayed Integration (TDI) mode, which means that the charges are clocked in the scanning direction -- also called along-scan (AL) direction, as opposed to the orthogonal direction, which is referred to as the across-scan (AC) direction -- at the same speed as the optical image moves over the CCD surface. The object images thus gradually build up in intensity before reaching the read-out register of each CCD. The precession of the spin axis causes a small, time-variable across-scan motion of the optical image on the CCD, up to $4$~across-scan pixels over a $4.42$-second CCD transit.

The Gaia focal-plane assembly \citep[e.g.,][]{2012SPIE.8442E..1PK}, with $106$ CCD detectors, has five dedicated functions: $4$ CCDs for metrology, i.e., basic-angle monitoring and wave-front sensing \citep{2012SPIE.8442E..1RG,2012SPIE.8442E..1QM}, $14$ Sky Mapper (SM) CCDs for object detection and rejection of prompt-particle events, $62$ Astrometric Field (AF) CCDs, $14$ Blue-Photometer/Red-Photometer (BP/RP) CCDs for low-resolution spectro-photometry, and $12$ Radial-Velocity-Spectrograph (RVS) CCDs for radial velocities and medium-resolution spectra. The AF, BP/RP, and RVS CCDs see the superimposed light coming from the two telescopes, which look at the sky separated by a basic angle of $106.5$~deg along the scan direction. The SM CCDs, in contrast, either see the light from one telescope or the light from the other telescope. The CCDs are distributed over seven independent rows; a star transiting the focal plane sees the following CCDs in time order: either SM$1$ or SM$2$, AF$1$$\ldots$AF$9$, BP, RP, and RVS$1$$\ldots$RVS$3$; RVS is only present for four of the seven rows. Two particular aspects of Gaia's design worth recalling here are its rectangular aperture ratio ($1.45 \times 0.50$~m$^2$, i.e., $3:1$) and its rectangular pixel size ($10 \times 30~\mu$m$^2$, i.e., $1:3$). This configuration allows the along- and across-scan images -- at least of point sources -- to roughly have the same size expressed in units of pixels.

Contrary to the Hipparcos mission, which selected its targets for observation based on a pre-defined input catalogue loaded on board \citep{1992ESASP1136.....T}, Gaia will perform an unbiased survey of the sky. Since an all-sky input catalogue at the Gaia spatial resolution complete down to $20^{\rm th}$ magnitude does not exist, there has essentially been no choice but to implement on-board object detection, with the associated advantage that transient sources (supernovae, near-Earth asteroids, etc.) will not escape Gaia's eyes. The downside of on-board object detection is the associated need for hard- and software, which needs to be fully autonomous and near-perfect for all scientific targets over the magnitude range $6$--$20$~mag (which represents a dynamic range of $400,000$) yet at the same time needs to be robust against real-sky complexities like double stars, extended objects (such as external galaxies, near-Earth asteroids, or planets like Jupiter), nebulosity, crowding, and Galactic cosmic rays and Solar protons, and, in addition, needs to process ``full-frame'' SM data (in TDI mode) in real-time: the continuous spin of the spacecraft causes a new TDI line with information to enter the CCD read-out register every milli-second. And all that, of course, running on space-qualified hardware operated in the hostile environment called ``space'' with severe requirements on and limitations of processing margins, reliability, mass, power, heat dissipation,~etc.

Each CCD row in the focal plane is controlled by a separate Video Processing Unit (VPU). A VPU is a combination of hardware (composed of a pre-processing and a PowerPC board) and associated software which, based on time strobes delivered by the atomic clock, commands and controls the CCDs and associated electronics, extracts and processes the science data, and delivers star packets with science data to the on-board storage area, from where the data is (later) transmitted to ground. The VPU software responsible for the science-data acquisition and processing is called the Video Processing Algorithms (VPAs). The VPA prototypes have been developed by Gaia's industrial prime contractor Airbus Defence \& Space in Toulouse, France, and implemented by Airbus Defence \& Space Ltd in Stevenage, United Kingdom, under ESA~contract.

Among the many functional responsibilities of the VPAs (e.g., supporting attitude-control-loop convergence and maintenance, metrology functions, etc.), the object detection in the SM CCDs is of key importance to the success of the Gaia mission. A critical task of the detection stage is to discriminate stars from prompt-particle events, like Galactic cosmic rays and Solar protons, which provide a continuous background of spurious events on the CCDs. These events need to be filtered out as much as possible at the detection stage since they could otherwise unnecessarily consume telemetry bandwidth and could even prevent stars from being observed. The problem essentially boils down to a trade-off between catalogue completeness and false-detection rates, and this trade-off is at the core of this work. The detection algorithms, described in detail in Section~\ref{sec:vpa}, contain a large number of configurable parameters. In this paper, we focus on $20$ of the most important parameters and describe a method to optimise these in Section~\ref{sec:optimisation} based on simulated data sets of single stars, double stars, Galactic cosmic rays, Solar protons, unresolved external galaxies, and asteroids which are described in Section~\ref{sec:simulations}. Our results are presented in Section~\ref{sec:results} and further discussed in Section~\ref{sec:discussion}. Scientific implications and conclusions of our work can be found in Sections~\ref{sec:implications} and \ref{sec:conclusion}, respectively. Readers primarily interested in the main results of this work are advised to read Sections~\ref{sec:vpa}, \ref{sec:results}, \ref{sec:implications}, and \ref{sec:conclusion}.

\section{Video Processing Algorithms (VPAs)}\label{sec:vpa}

The Video Processing Algorithms (VPAs; e.g., \citealt{2007ESASP.638E..39P}) are responsible for the science-data acquisition and processing, including object detection in the SM CCDs. Object detection has two branches: one for saturated and one for non-saturated objects. For Gaia, saturation of stellar images in the SM CCDs sets in for objects brighter than $G \sim 12$~mag\footnote{Gaia's $G$ magnitude refers to the unfiltered, white-light response of the astrometric CCDs combined with the telescope. Its relation to canonical filter systems is addressed in \citet{2010A&A...523A..48J}.}. The saturated-object-detection branch, based on ``extremity matching'' in Airbus Defence \& Space terminology, has limited freedom for user configuration and is outside the scope of this work. The non-saturated-object-detection branch, on the other hand, has a significant number of user-configurable parameters leaving ample room for scientific optimisation. As a result of real-time constraints in high-density fields which cannot be met with a software implementation, this branch is primarily implemented in hardware -- through field-programmable gate arrays -- and the processing can roughly be decomposed into two modules: pre-processing of raw SM data (Section~\ref{subsec:pre-processing}), followed by the actual non-saturated-object detection (Section~\ref{subsec:detection}).

\subsection{Pre-processing of raw SM data}\label{subsec:pre-processing}

Raw SM samples, composed of $2 \times 2$ hardware-binned pixels, are continuously read and temporarily stored in a moving buffer inside the VPU covering several hundred TDI lines. The pre-processing step identifies, through a user-defined mask, dead columns and interpolates SM flux values in such cases from neighbouring samples. The pre-processing also checks the raw SM data for saturated samples, allowing the VPAs to enter either the saturated-object-detection or the non-saturated-object-detection branch. Finally, any sample which is not saturated has a linear flux correction performed on it to account for dark-signal non-uniformity and column-response non-uniformity (pixel-response non-uniformity integrated over a CCD column). Effectively, the next step in the process, detection of non-saturated objects, only applies to non-saturated samples which have not been dead-column corrected.

\subsection{Detection of non-saturated objects}\label{subsec:detection}

\begin{figure}[t]
  \centering
  \includegraphics[trim=50 100 0 0,clip,width = 0.75\columnwidth]{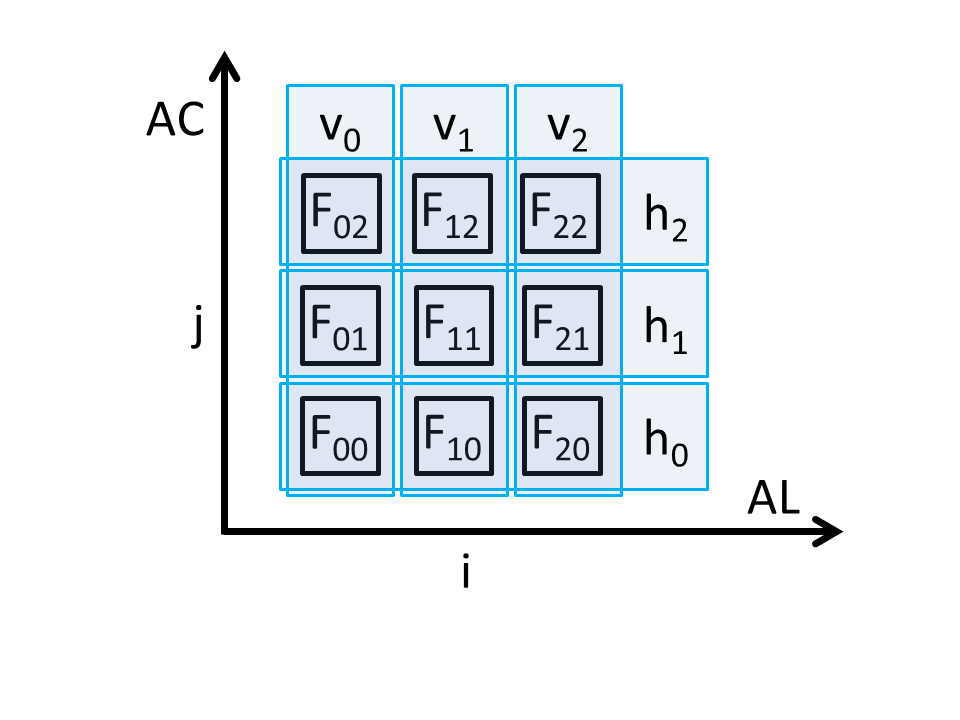}
  \caption{Each SM sample under scrutiny, itself composed of $2 \times 2$ pixels, has a so-called working window, centred on it, associated with it. Object detection uses the $5 \times 5$-samples working window (not shown) for background subtraction and the $3 \times 3$-samples working window (depicted here for sample $i,j = 1,1$) for shape assessment of detections. The three-dimensional summed-flux / shape vectors $\mathbf{h}$ and $\mathbf{v}$ contain, respectively, the along-scan-integrated (AL) and across-scan-integrated (AC) sum of the working-window background-subtracted flux values $F_{ij}$, in LSB units (Equation~\ref{eq:h_and_v}). The total, background-subtracted flux $F$ in the working window equals $F = v_{0} + v_{1} + v_{2} = h_{0} + h_{1} + h_{2}$.}\label{fig:working_window}
\end{figure}

The detection part of the algorithms essentially searches for local maxima of flux, then analyses the shape of these local maxima, subsequently interprets from this shape what type of object it is -- faint star, prompt-particle event (PPE), or ripple -- and finally applies a flux thresholding on the local maxima (see also Section~\ref{subsec:nomenclature}). This logic may seem simple and sub-optimal -- compared to more sophisticated, commonly-used packages such as SExtractor \citep{1996A&AS..117..393B} -- but this is an unavoidable result of the (forced) choice of a hardware implementation.

To detect and analyse local maxima, the VPAs sequentially process all samples in the moving VPU buffer containing the pre-processed SM samples (``continuous, full-frame SM data stream''). Each sample under scrutiny has a so-called working window, a square, finite grid of SM samples centred on the sample of interest, associated with it (Figure~\ref{fig:working_window}).

The first step in the processing of each sample of interest is background determination. The sky background is estimated by default as the $5^{\rm th}$-lowest flux value from the $16$ samples composing the outer ring of the $5 \times 5$-samples working window. This background flux value is subtracted from the sample to give a background-corrected flux. On-board Gaia, fluxes are recorded on a 16-bit analogue-to-digital scale, referred to as LSB (Least Significant Bit) units; the nominal conversion gain equals $0.2566$~LSB per electron.

The second part of the detection uses a smaller, $3 \times 3$-samples, working window (Figure~\ref{fig:working_window}). The VPAs check for a local maximum of flux in this window, centred in our notation on $(i,j) = (1,1)$, by first calculating two three-dimensional summed-flux / shape vectors $\mathbf{h}$ and $\mathbf{v}$ (for horizontal and vertical, respectively):
\begin{eqnarray}
h_j = \sum_{i=0}^{2} F_{ij} && {\rm \ \ \ \ for\ } j = 0, 1, 2;\nonumber\\
&&\label{eq:h_and_v}\\
v_i = \sum_{j=0}^{2} F_{ij} && {\rm \ \ \ \ for\ } i = 0, 1, 2,\nonumber
\end{eqnarray} 
where $F_{ij}$ denotes the background-subtracted flux of sample $(i,j)$ in LSBs; the TDI-coordinate associated with index $i$ is often referred to as along-scan direction ($\rightarrow$), whereas the CCD-column coordinate associated with index $j$ is often referred to as across-scan direction ($\uparrow$). The total, background-subtracted flux $F$ in the $3 \times 3$-samples working window is calculated as $F = v_{0} + v_{1} + v_{2}\ (= h_{0} + h_{1} + h_{2})$. A local maximum is defined~as:
\begin{eqnarray}
v_1 \geq v_0 &\wedge& v_1 > v_2;\nonumber\\
             &\wedge& \label{eq:local_maximum}\\
h_1 \geq h_0 &\wedge& h_1 > h_2,\nonumber
\end{eqnarray} 
where $\wedge$ denotes the ''logical AND'' operator. The vectors $\mathbf{v}$ and $\mathbf{h}$ describe the overall shape of the local maximum in the along- and across-scan directions, respectively: if $h_1$ is much larger than $h_0$ and $h_2$, then the detection has a narrow peak in intensity in the across-scan direction, whereas if $h_1$ is approximately equal to $h_0$ and $h_2$, then the object's Point-Spread Function (PSF) is rather flat (broad) in the across-scan direction. Similar arguments hold for $\mathbf{v}$ and the along-scan direction. The shape vectors $\mathbf{h}$ and $\mathbf{v}$ are hence used on board to distinguish between three different object types. Since the implementation in the VPA detection hardware is primarily based on signed $64$-bit integer operations, we need to define the operators:
\begin{equation}
[x]_n = \left\{
\begin{array}{rcl}
0     &{\rm \ \ \ if\ \ \ }& x < 0;\\
x     &{\rm \ \ \ if\ \ \ }& 0 \leq x \leq 2^n-1;\\
2^n-1 &{\rm \ \ \ if\ \ \ }& 2^n -1 < x,\\
\end{array}
\right.
\label{eq:saturation}
\end{equation}
denoting saturation of $x$ to $n$ bits, and
\begin{equation}
(x)_{n} = x / 2^n,
\label{eq:truncation}
\end{equation}
denoting truncation of $x$ to $n$ bits (truncation refers to elimination of the $n$ least significant bits, which is equivalent to integer division by $2^n$). In general, the truncation and saturation operators are used on board to control under- and overflow situations and to allow casting variables into several integer types, for instance unsigned $32$-bit integers and signed $64$-bit integers. The actual shape discrimination applied on board is user-configurable through $2 \times 5 = 10$ so-called rejection parameters, denoted $(a,b,c,d,e)_{\rm HF}$ and $(a,b,c,d,e)_{\rm LF}$, which are signed integers in the range $[-32768, +32767]$. Objects which satisfy:
\begin{eqnarray}
&&\left[\left(([h_{0} + a_{\rm HF}]_{18} \cdot [h_{2} + b_{\rm HF}]_{18})_{4} \cdot c_{\rm HF}\right)_{8}\right]_{32}\nonumber \\
&&<  \left[(\left[(F)_{2}+d_{\rm HF}\right]_{18}^2 + e_{\rm HF})_{4}\right]_{32}
\label{eq:RejEqPPE}
\end{eqnarray}
are labeled as (sharply-peaked, i.e., with a high spatial frequency, or HF) ``prompt-particle event'' in the across-scan direction, while objects which satisfy:
\begin{eqnarray}
&&\left[\left(([h_{0} + a_{\rm LF}]_{18} \cdot [h_{2} + b_{\rm LF}]_{18})_{4} \cdot c_{\rm LF}\right)_{8}\right]_{32}\nonumber \\
&&>  \left[(\left[(F)_{2}+d_{\rm LF}\right]_{18}^2 + e_{\rm LF})_{4}\right]_{32}
\label{eq:RejEqRipple}
\end{eqnarray}
are labeled as (broadly-peaked, i.e., with a low spatial frequency, or LF) ``ripple'' in the across-scan direction (roughly reminiscent of a higher-order diffraction maximum in a PSF). Objects which violate both conditions, which means with a PSF which is neither too peaked nor too broad in the across-scan direction, are labeled as ``faint star'' in the across-scan direction, where ``faint'' refers to non-saturated.

\begin{figure*}[t]
  \centering
  \includegraphics[width = 0.45\textwidth]{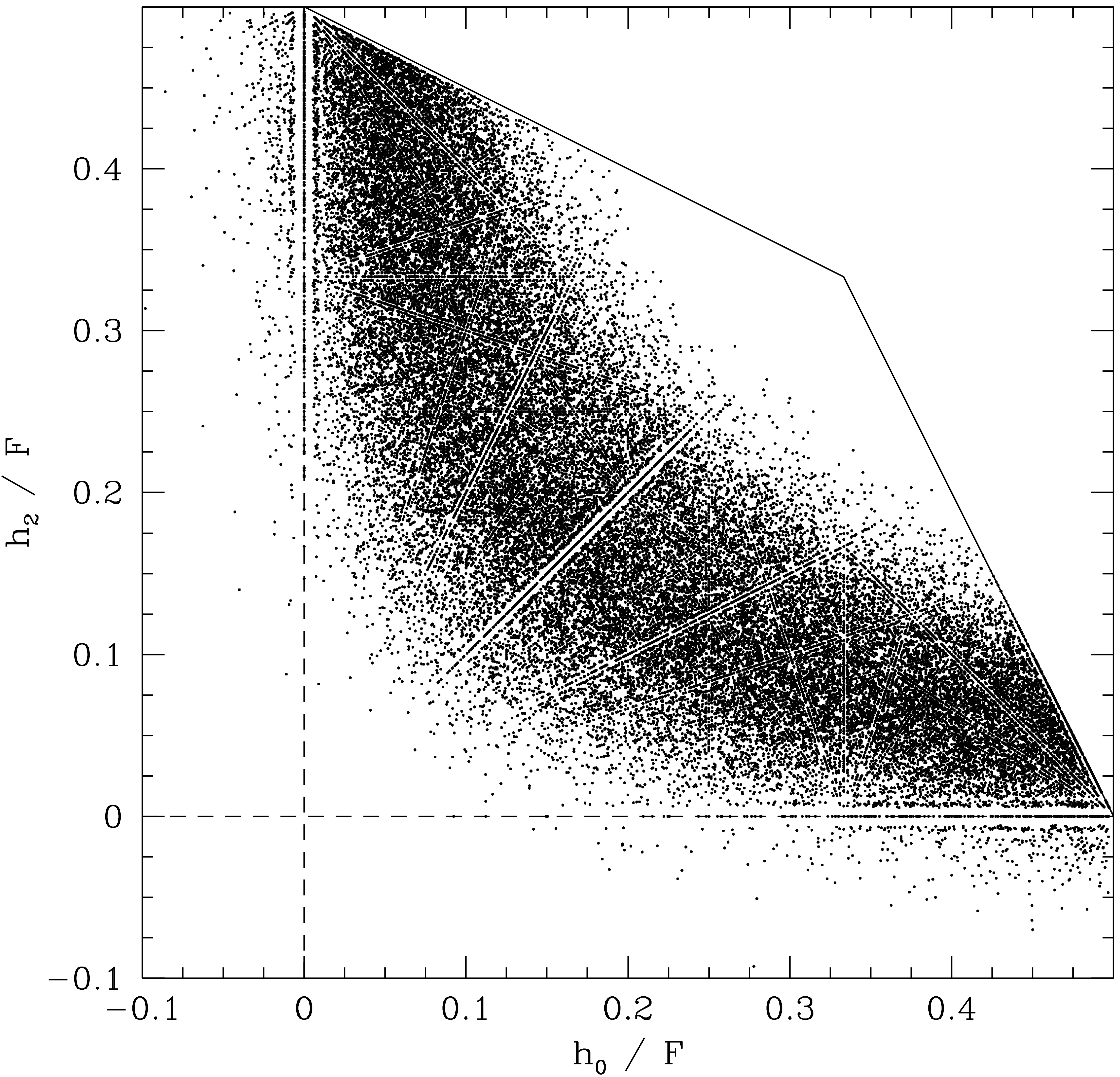}
  \includegraphics[width = 0.45\textwidth]{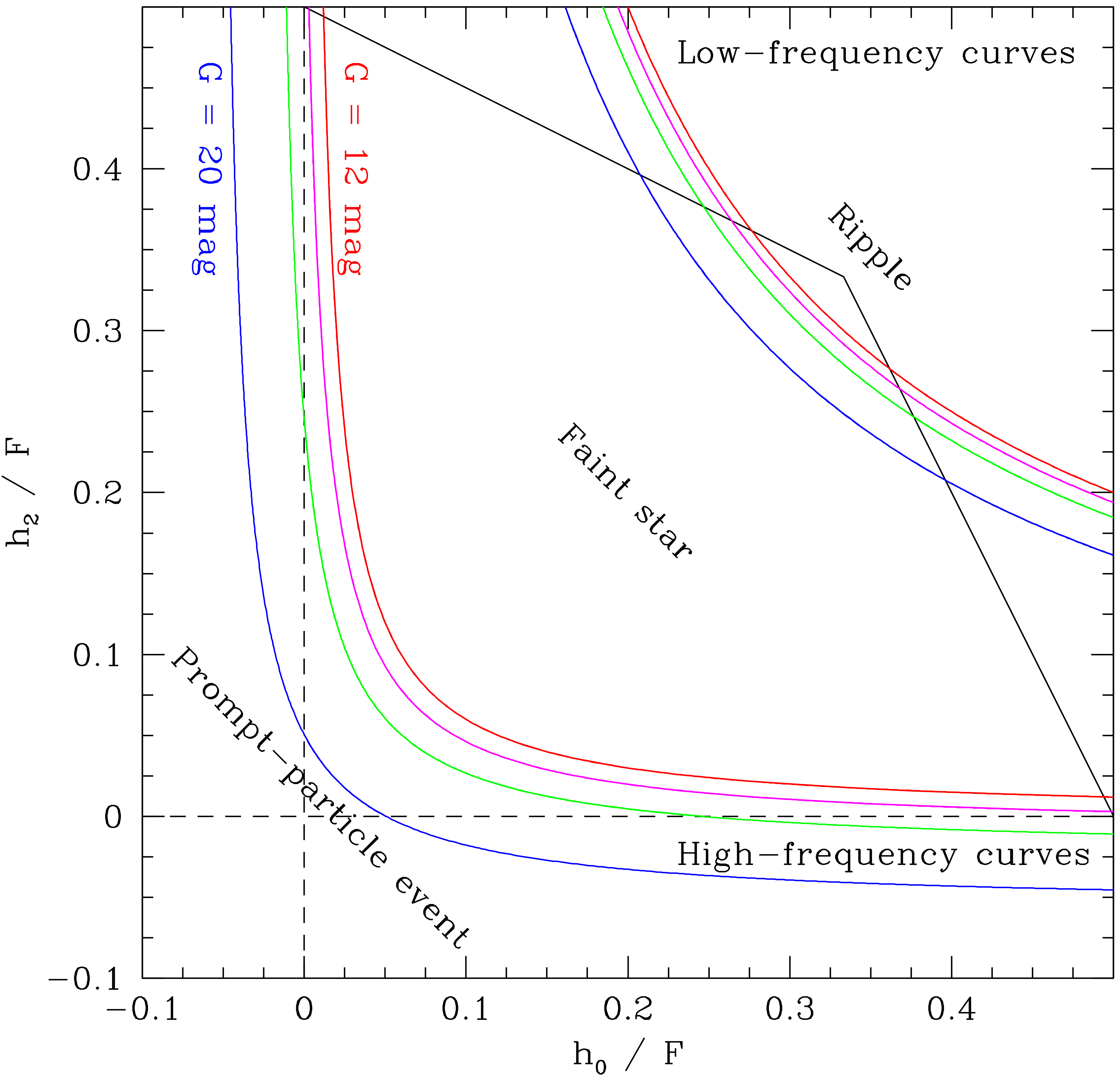}
  \caption{{\it Left panel:} example across-scan (AC) rejection plot, based on the along-scan-integrated flux vector $\mathbf{h}$, for 50,000 single stars (Section~\ref{subsec:single_star_dataset}) with magnitudes between $G = 19.5$ and $20$~mag (so that typical flux values are $F \sim 140$~LSB). Stars with a symmetric PSF which are centred in an SM sample fall on the diagonal 1:1 relation. Stars with sharp PSFs fall close to the origin whereas stars with broad PSFs move diagonally up towards the vertex $h_{0}/F = h_{2}/F = 1/3$. When, for a given PSF size, the PSF centring inside the sample is varied, objects move on a hyperbolic curve either towards the top left or towards the bottom right. In the absence of Poisson noise, stars with different brightnesses occupy the same hyperbolic curves. The effect of Poisson noise is to broaden this curve into a hyperbolically-shaped cloud; the spread is larger for faint stars since Poisson noise is relatively more important for faint than for bright stars. This effect, combined with background-subtraction errors, can lead to negative $h_0$ and/or $h_2$ values and hence negative data points, in particular for faint stars. Due to the finite number of LSB units in the working window ($F \sim 140$~LSB), discretisation effects in $h_{0}/F$ and $h_{2}/F$ can be seen in the data. {\it Right panel:} example across-scan rejection plot with high- and low-frequency curves associated with, respectively, Equations~(\ref{eq:RejEqPPE}) and (\ref{eq:RejEqRipple}) for fluxes $F$ associated with magnitudes $G = 12$ (red), $18$ (magenta), $19$ (green), and $20$ (blue) mag. The saturation and truncation operators from, respectively, Equations~(\ref{eq:saturation}) and (\ref{eq:truncation}) have not been included in the curves; they therefore merely serve illustration purposes. The curves, defined through the user-defined VPA parameters $(a,b,c,d,e)_{{\rm HF},\uparrow}$ and $(a,b,c,d,e)_{{\rm LF},\uparrow}$ which have here -- for illustration -- been set to the functional-baseline values, are flux dependent although the effect of flux on the curves is minimal for bright stars (the curves essentially superimpose for stars brighter than $G \sim 15$~mag). The upper set of curves is referred to as low frequency (LF) whereas the lower set of curves is referred to as high frequency (HF). Objects above the upper curve are labeled ``ripple'' while objects below the lower curve are labeled ``prompt-particle event''; objects in between the lower and upper curves -- for the applicable flux level -- are labeled ``faint star''. Gaia's on-board object detection is based on an along-scan rejection plot using shape vector $\mathbf{v}$ (not shown) {\bf and} an across-scan rejection plot using shape vector $\mathbf{h}$ (shown here). The domain of possible $h_{0}/F$ and $h_{2}/F$ values is limited by the definition of a local maximum in the VPAs: since a local maximum is defined as $h_{1} \geq h_{0}$ {\bf and} $h_{1} > h_{2}$ (Equation~\ref{eq:local_maximum}), the maximum values that $h_{0}/F$ and $h_{2}/F$ can (asymptotically) take are $1/3$ each. Similarly, the maximum value that each of them can (asymptotically) take is $1/2$, with the other then (asymptotically) taking the value $0$. More generally, Equation~(\ref{eq:local_maximum}) induces boundaries on the rejection plot (solid lines), below which a data point must fall to obey the VPA local-maximum definition.}\label{fig:example_rejection_plot}
\end{figure*}

In a plot of $h_{0}/F$ versus $h_{2}/F$ (Figure~\ref{fig:example_rejection_plot}, also referred to as rejection plot), the above inequalities define two hyperbolic curves for a fixed value of flux $F$. The $10$ rejection parameters determine the shape and position of these hyperbolic curves for a fixed value of $F$; more generally, when considering the three-dimensional space of $h_{0}/F$ versus $h_{2}/F$ versus $F$, the above inequalities define two hyperbolic surfaces.

The above discussion, and in particular Equations (\ref{eq:RejEqPPE}) and (\ref{eq:RejEqRipple}), is focused on the horizontal shape vector $\mathbf{h}$ applicable to the across-scan direction. There are similar criteria to Equations~(\ref{eq:RejEqPPE})--(\ref{eq:RejEqRipple}) for prompt-particle-event and ripple definitions in the along-scan direction based on the vertical $\mathbf{v}$ vector. A genuine faint-star detection then requires a faint-star classification along scan (based on $\mathbf{v}$ and $(a,b,c,d,e)_{{\rm HF}, \rightarrow}$ and $(a,b,c,d,e)_{{\rm LF}, \rightarrow}$) {\bf and} a faint-star classification across scan (based on $\mathbf{h}$ and $(a,b,c,d,e)_{{\rm HF}, \uparrow}$ and $(a,b,c,d,e)_{{\rm LF}, \uparrow}$).

The last step in the object detection is a flux-thresholding stage. This step essentially defines Gaia's faint limit (nominally $G = 20$~mag). Since the thresholding works on on-board (background-subtracted) fluxes collected in the SM CCD, its functional default value is a (non-intuitive) $110$~LSB.

All in all, there are $2\ ({\rm \rightarrow,\uparrow})\ \times 2\ ({\rm HF,LF})\ \times 5\ (a,b,c,d,e) = 20$ free parameters which govern the classification of local maxima into faint stars, ripples, prompt-particle events. The functional-baseline values for these rejection parameters are not the outcome of a detailed scientific optimisation but are based on limited simulations and laboratory data and essentially ensure that ``normal, single stars'' are detected while extremely sharp, elongated, and broad cosmic rays and Solar protons are rejected. In reality, however, prompt-particle events, and also stars with their various multiplicity configurations, take a wide variety of (PSF) shapes and wanted objects and unwanted objects are really mixed populations in ($h_{0}/F$, $h_{2}/F$, $F$)- and ($v_{0}/F$, $v_{2}/F$, $F$)-space. This study aims to establish scientifically-optimum separation surfaces in these spaces.

\subsection{Our VPA emulation}\label{subsec:validation}

We have emulated the VPA object detection of non-saturated objects described in Section~\ref{subsec:detection} in a standalone piece of software. It covers background subtraction, application of the rejection equations (\ref{eq:RejEqPPE}--\ref{eq:RejEqRipple}) (both along and across scan), and flux thresholding, but, since it is irrelevant in the scope of this investigation, not the pre-processing stage described in Section~\ref{subsec:pre-processing}. We have successfully tested our emulation against the Airbus Defence \& Space VPA prototype which has been integrated into the Gaia Instrument and Basic Image Simulator (GIBIS; \citealt{2005ESASP.576..417B,2011ascl.soft07002B}) and against a stand-alone version of this prototype running, in a controlled environment with validation test cases, in Gaia's science operations centre in Spain.

\subsection{From detection to catalogue completeness}\label{subsec:nomenclature}

\begin{figure}[t]
  \centering
  \includegraphics[width = \columnwidth]{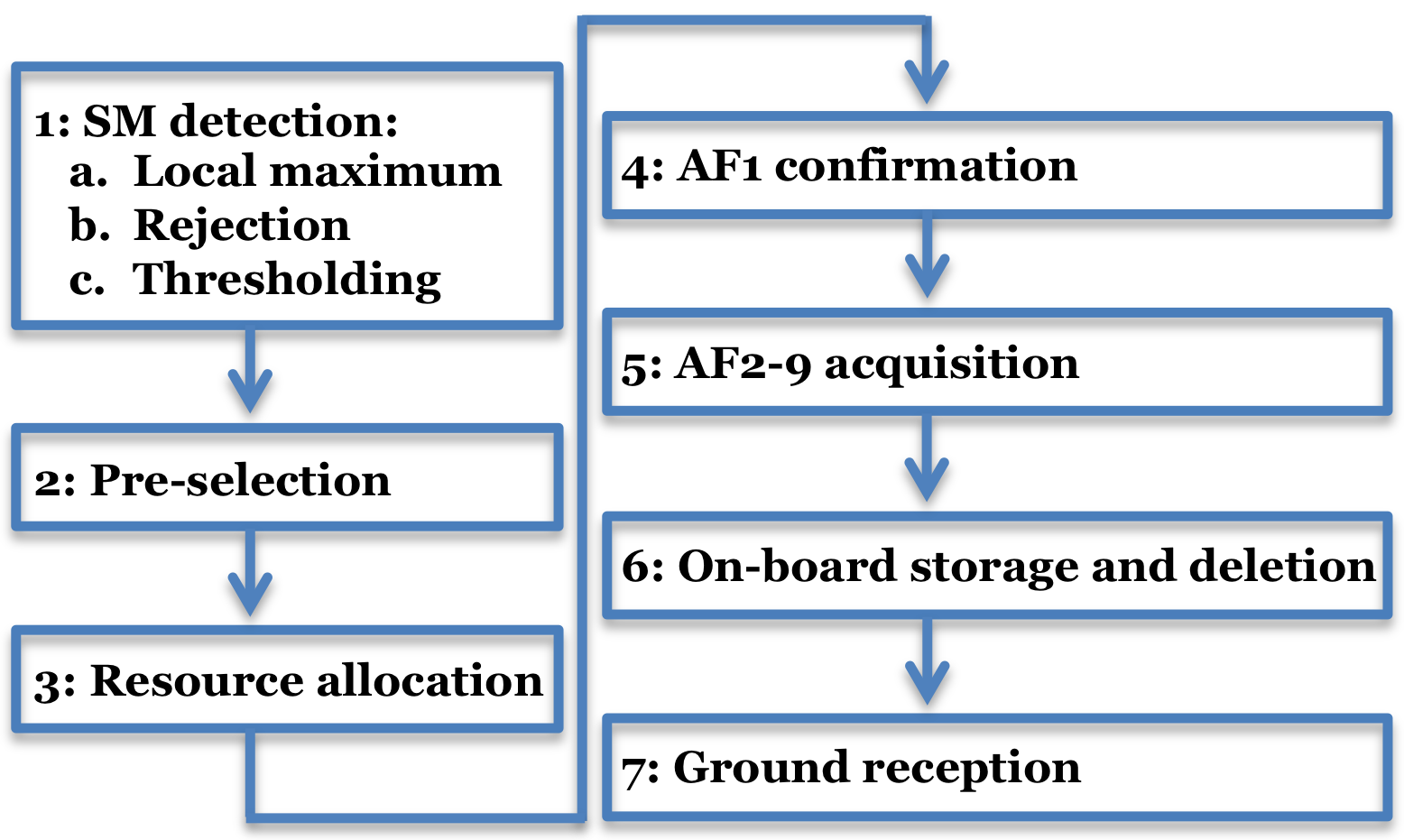}
  \caption{Schematic summary of steps involved in the observation process, i.e., detection, selection, confirmation, acquisition, and survival of objects. Steps 1 and 2 have been implemented in hardware.}\label{fig:nomenclature}
\end{figure}

Although the derivation of Gaia's selection function and catalogue completeness is outside the scope of this paper, we provide a short summary of the observation process of objects with the aim to warn the reader that detection and observation probability are distinct quantities. Schematically speaking, an object (transit) has to survive all of the following steps to contribute to the final Gaia catalogue (see Figure~\ref{fig:nomenclature}):
\begin{enumerate}
\item SM detection: the three-step process described in Section~\ref{subsec:detection}, consisting of (i) the search for local maxima of flux, (ii) the assessment of the shape of these local maxima allowing object classification through application of the rejection equations, and (iii) application of a flux threshold. An object that survives these three steps is denoted as ``detected'';
\item Pre-selection: every TDI line, all detections are first merged with the user-defined ``virtual objects'' required for calibration and then sorted in priority (flux). This list is then subject to an object-flow-limitation condition allowing only the five highest-priority objects to pass to the next step. The associated limiting density is $\sim$$3$ million objects per square degree;
\item Resource allocation: after merging the lists of pre-selected objects from both telescopes (SM1 and SM2), a final selection of objects to be followed throughout the Astrometric Field (AF) is made. The AF CCDs are not read out full frame but only small areas (``windows'') around objects of interest are read out. The window size is $12$~pixels in the across-scan direction and varies from $18$~pixels in the along-scan direction for $G \leq 16$~mag to $12$~pixels for $G > 16$~mag. For stars fainter than $G = 13$~mag, the $12$~pixels in the across-scan direction are normally binned into one sample during read-out leading to effectively one-dimensional data. At each TDI line, the VPAs can simultaneously handle $W = 20$ samples (``resources'' in Airbus Defence \& Space terminology) in the read-out register. Depending on the particular, instantaneous configuration of detected-object magnitudes, this corresponds to a limiting object of at most $\sim$$1$ million objects per square degree.
The VPA uses a prioritised allocation of resources to bright detections, meaning that, when there is a shortage of windows, faint stars will be sacrificed to allow assigning a window to a bright(er), i.e., high(er)-priority, object. In short, in dense areas, not all detected objects will receive a resource (window);
\item AF1 confirmation: the VPAs implement, following the detection stage in the SM CCDs, a confirmation stage in the first AF strip (AF1). This stage has two purposes, namely (i) to confirm, by re-detection of the object using the AF1 samples, the presence of the object detected in SM, and (ii) to estimate the velocities of a subset of the stars to produce measurements for the closed-loop spacecraft attitude and control sub-system. The confirmation essentially involves a pre-processing of raw AF1 samples similar to the SM pre-processing, then constructs a working window around the expected position of the object obtained from forward propagation from the SM detection, then performs background estimation similar to the SM process, and finally runs a local-maximum detection similar to the SM concept. If a local maximum is found {\bf and} if the background-subtracted AF1 flux is consistent with the background-subtracted SM flux, where ``consistent'' is defined through user-configurable criteria, then the object is confirmed and considered for further observation throughout the focal plane. The confirmation criterion is hence purely flux-based: the ``PSF shape'' of the confirmed object is not tested. Clearly, since the confirmation stage is not $100$\% perfect, there is a risk of a detected object to be adversely killed by the confirmation step;
\item AF2--9 acquisition: the acquisition of the bulk astrometric window data in CCD strips AF2--AF9 is not guaranteed to be successful. The scanning-law-induced across-scan motion of objects, for instance, may cause them to drift out of the CCD in the across-scan direction. There is also a finite probability that the window of a star is polluted, for instance by straylight caused by very bright stars or planets or by an injected line of charge used for radiation-damage mitigation. Similarly, windows can be affected, for instance, by a reduced CCD integration time (activated TDI gate) induced by a simultaneously-transiting bright star or by a dead~column;
\item On-board storage and deletion: after the focal-plane transit, the window data are collected into star packets which are temporarily stored into the on-board solid-state mass memory before being transmitted to ground. The downlink to ground uses a prioritised scheme. Since the mass memory has a finite size, it occasionally fills up necessitating activation of an on-board deletion scheme. This scheme is also prioritised. So, even if a detected star manages to get all its window data properly collected into a star packet, there is a finite probability that the data gets deleted on board;
\item Ground reception: finally, even when a star packet is transmitted to ground there is a small but finite probability that it is lost as a result of unplanned ground-station outages or unrecoverable transmission(-frame) anomalies.
\end{enumerate}
This summary clearly demonstrates that near-perfect object detection, being the first element in the chain, is a pre-requisite but not a guarantee for a high observation probability.

\section{Simulated data sets}\label{sec:simulations}

In order to investigate the performance of the on-board detection algorithms on various object categories, we need representative image libraries of various types of objects. As explained in Section~\ref{sec:vpa}, they should cover the non-saturated-object regime in the SM CCDs. Since saturation in SM starts at $G \sim 12$~mag, we decided to use the range $G = 12.5$--$20$~mag, keeping $0.5$~mag as margin. We should stress at this stage that the precise bright-star limit adopted in this study is not an important parameter: the flux dependence of the rejection equations (\ref{eq:RejEqPPE}--\ref{eq:RejEqRipple}) at the bright end ($G \la 15$~mag) is very weak (see also the curves in Figure~\ref{fig:example_rejection_plot}) which means that if we optimise the detection including stars at $G = 12.5$~mag, this solution also applies to any brighter stars (provided they do not saturate).

\subsection{Single stars}\label{subsec:single_star_dataset}

For single stars, we need a library of two-dimensional images covering the magnitude range $G = 12.5$--$20$~mag and, in view of the VPA background subtraction, covering at least $7 \times 7$ SM samples (i.e., $14 \times 14$ CCD pixels).

\begin{figure}[t]
  \centering
  \includegraphics[width = \columnwidth]{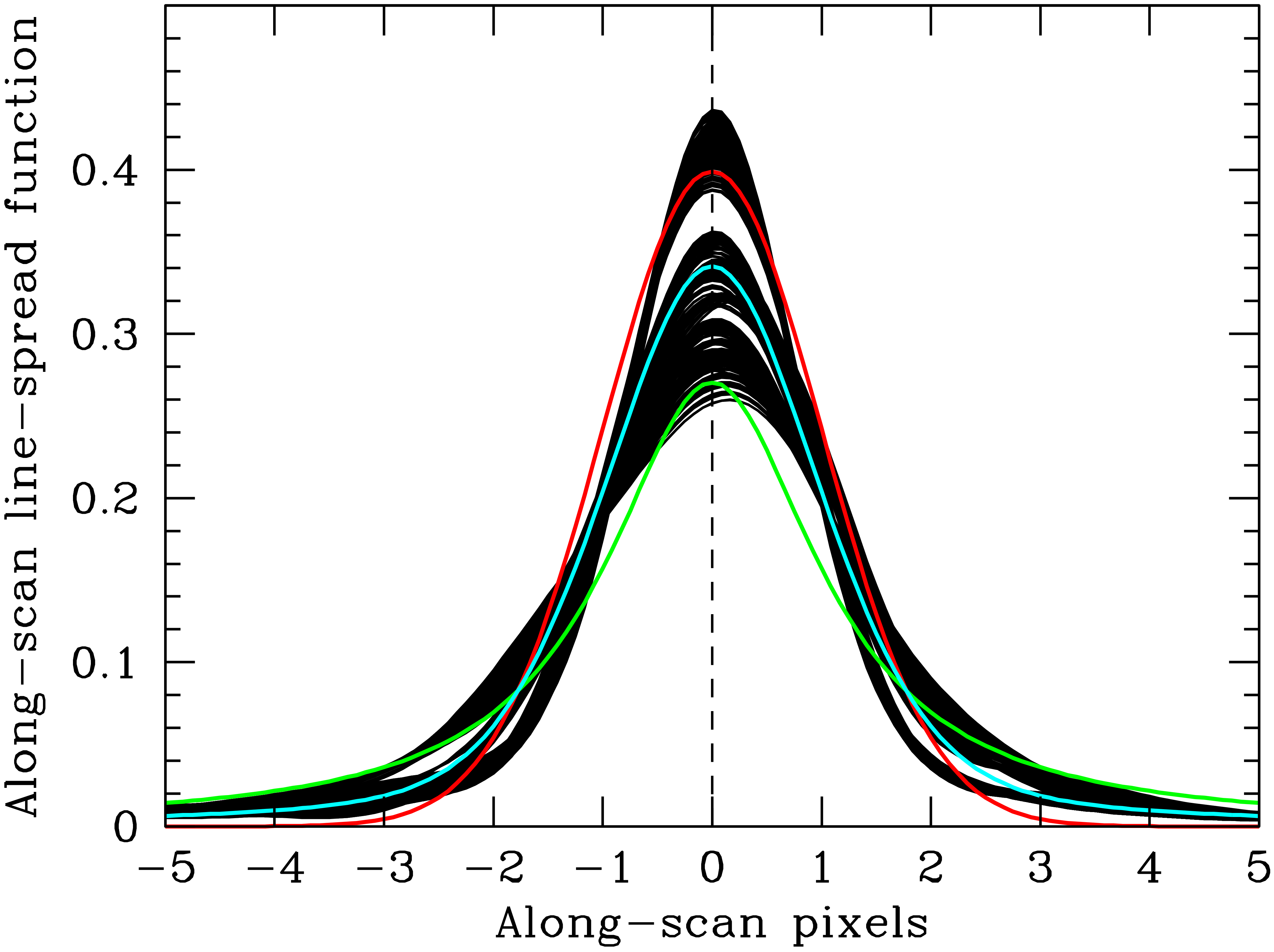}
  \caption{Star-Mapper (SM) along-scan (AL) line-spread functions for all $448$ combinations of 14 CCDs, 16 stellar spectral-energy distributions, and two values of interstellar extinction. Since the LSF size is primarily determined by the (local) optical quality of the telescope, which varies over the SM field of view (the wave-front error varies between $40$ and $105$~nm), the curves cluster in various ``families'' (see also Figure~\ref{fig:parametrisation_histograms}). The curves do not include the effect of the on-chip binning of the SM pixels in $2 \times 2$ samples. Overplotted, for reference, are a Lorentzian in green (Equation~\ref{eq:Lorentzian_def}), a Gaussian in red (Equation~\ref{eq:Gaussian_def}), and a sum of a Gaussian with weight 55\% and a Lorentzian with weight 45\% in cyan (Equation~\ref{eq:Gaussian_Lorentzian_def}).}\label{fig:Stancik_and_Brauns_library_LSF}
\end{figure}

Gaia's optical design allows near-diffraction-limited imaging: the system wave-front error in the astrometric field equals $\sim$$50$~nm RMS so the Strehl ratio exceeds $80$\% for $\lambda > 665$~nm (i.e., unreddened mid-K and later spectral types), applicable to the majority of Gaia targets. Gaia's PSF is hence symmetric to first order and PSF asymmetries, caused by optical aberrations, are modest and mainly visible in the (far) wings of the PSF. In the SM fields of view, at the edges of the telescope's fields of view, the average wave-front error is $\sim$$63$~nm RMS, which means that diffraction-limited imaging is only achieved for the reddest objects ($\lambda > 838$~nm, i.e., reddened M-type stars). Figure~\ref{fig:Stancik_and_Brauns_library_LSF} shows $448$ predicted SM along-scan LSFs. They have been obtained through full-fledged, realistic, time-consuming simulations combining 14 SM wavefront-error maps (delivered by Airbus Defence \& Space) with 16 stellar spectral-energy distributions from \citeauthor{1998PASP..110..863P}'s library \citep[][spectral types B1V, A0V, A3V, A5V, F2V, F6V, F8V, G2V, K3V, M0V, M6V, G8III, K3III, M0III, M7III, and B0I]{1998PASP..110..863P} with two values of interstellar extinction (unreddened and $A_{550~\rm{nm}} = 5$~mag). Overplotted, for reference, are a Gaussian (red) and a Lorentzian (green); both have the same FWHM, corresponding to $\sigma = 1.0$~AL pixel for the Gaussian. Also overplotted for reference (in cyan) is the sum of the Gaussian (weight 55\%) and the Lorentzian (weight 45\%), which is often used as approximation to a Voigt function, i.e., the convolution of a Lorentzian with a Gaussian. Such a sum, after parameter tuning, actually provides a remarkably\footnote{
For {\it spectral} LSFs, a Voigt profile can be physically understood realising that the Gaussian refers to Doppler broadening while the Lorentzian refers to radiation damping and collisional (pressure) broadening. Voigt profiles also feature frequently in crystallography because X-ray diffraction profiles are well represented by (pseudo-)Voigt profiles \citep[e.g.,][]{1947ApJ...106..121V,Wertheim:1974,Langford:a16721,deKeijser:a21783} since particle-size broadening corresponds to a Lorentzian and instrumental contributions and lattice-strain broadening can be represented by a Gaussian. It is therefore not surprising that also for {\it optical} LSFs, where the physical expectation is a convolved Fraunhofer-diffraction profile, Voigt profiles provide a convenient representation. Gaia has a rectangular aperture and an associated monochromatic Fraunhofer diffraction pattern described by the square of a sinc function: $I_\lambda \propto \sin^2(\alpha)/\alpha^2$, with $\alpha = [\pi D v] / [F \lambda]$, with $D$ the aperture dimension ($1.45$~m along and $0.50$~m across scan), $F = 35$~m the focal length, and $v$ the spatial coordinate in the focal plane / on the CCD. This profile, after spectral superposition, is convolved with Gaussian and boxcar functions representing various smearing contributors caused by spacecraft attitude jitter during the CCD integration, the scanning-law induced differences between the optical and electronic speed of the image, the detector modulation transfer function which includes charge diffusion of electrons inside the CCD, optical distortions, the electrodes / phases corresponding to the TDI integration stages in a pixel, and pixel binning.} good approximation to the individual LSFs. Since the SM LSFs do show small asymmetries, a more suitable, empirically-motivated, parametrisation of the LSF in SM is a summation of a Gaussian and a Lorentzian LSF including LSF asymmetry \citep[e.g., ][]{2008VibSpec47...66S}:
\begin{equation}
{\rm LSF}(v) = f \cdot L(v) + (1 - f) \cdot G(v),\label{eq:Gaussian_Lorentzian_def}
\end{equation}
where:
\begin{equation}
L(v) = {{[2 A]/[\pi \gamma(v)]}\over{1 + 4 [(v-v_0)/\gamma(v)]^2}} {\rm ~is~a~Lorentzian},\label{eq:Lorentzian_def}
\end{equation}
\begin{equation}
G(v) = {{A}\over{\gamma(v)}} \sqrt{{{4 {\rm ln}2}\over{\pi}}} {\rm exp}\left[-4 {\rm ln}2\left({{v-v_0}\over{\gamma(v)}}\right)^2\right]{\rm ~is~a~Gaussian},\label{eq:Gaussian_def}
\end{equation}
and
\begin{equation}
\gamma(v) = {{2 \gamma_0}\over{1+{\rm exp}[\alpha \cdot (v-v_0)]}}{\rm ~is~a~sigmoid},\label{eq:sigmoid_def}
\end{equation}
where $v$ is the along-scan pixel coordinate, $0 \leq f \leq 1$ is the fraction of the Lorentzian character contributing to the LSF ($f = 0 = {\rm Gaussian}$ and $f = 1 = {\rm Lorentzian}$), $A$ is the area (intensity) of the LSF, and $v_0$ is the mean (centre) position of the LSF. The parameter $\alpha$ describes LSF asymmetry: negative values skew the LSF towards higher values of $v$, while positive values skew the LSF towards lower values of $v$. When $\alpha = 0$, $\gamma(v)$ in Equation~(\ref{eq:sigmoid_def}) reduces to $\gamma_0$ and the LSF is a standard, symmetric Gaussian or Lorentzian with a constant width. The parameter $\gamma_0$ denotes the FWHM of the Gaussian or Lorentzian for $\alpha = 0$ (for the Gaussian, we have $\gamma_0 = 2\sqrt{2{\rm ln}2}\ \sigma$ when $\alpha = 0$). The particular sigmoidal functional form of $\gamma(v)$ in Equation~(\ref{eq:sigmoid_def}) is advantageous since the width asymptotically approaches upper and lower bounds.

\begin{figure*}[t]
  \centering
  \includegraphics[width = \textwidth]{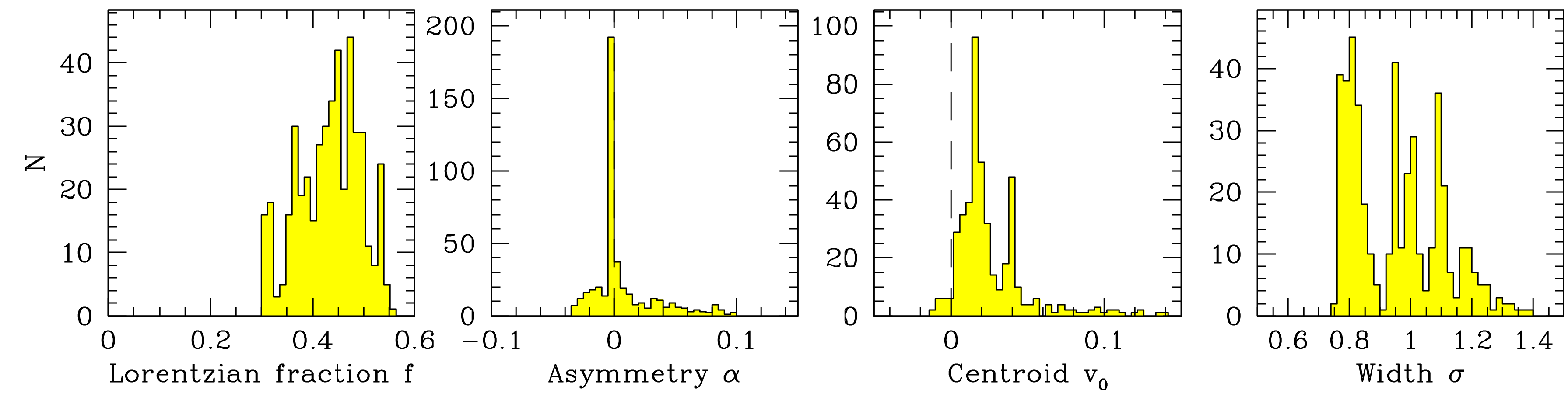}
  \includegraphics[width = \textwidth]{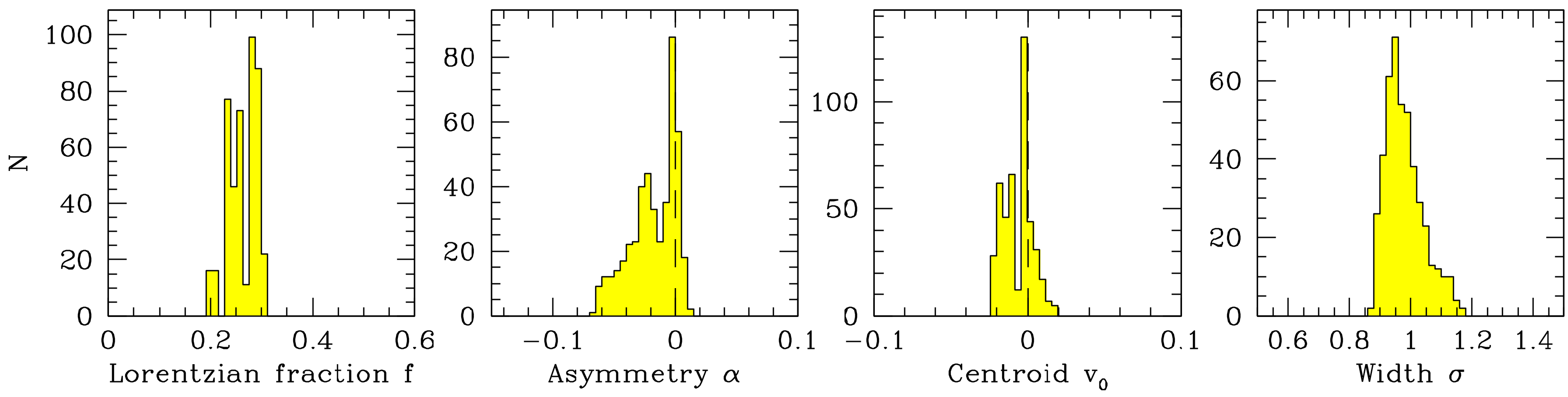}
  \caption{Histograms of $f$, $\alpha$, $v_0$, and $\sigma$ (with $\gamma_0 = 2\sqrt{2{\rm ln}2}\ \sigma$) obtained when fitting the along-scan (top) or across-scan (bottom) LSF model (Equation~\ref{eq:Gaussian_Lorentzian_def}) to the $448$ LSFs resulting from combining 14 SM CCDs, 16 stellar spectral-energy distributions, and two values of interstellar extinction (see Figure~\ref{fig:Stancik_and_Brauns_library_LSF}). The normalisation constant $A$ has been frozen to unity in all fits. The across-scan direction refers to average across-scan motion. Since the Gaia pixel ratio ($10 \times 30~\mu$m$^2$, i.e., $1:3$) effectively cancels the Gaia aperture ratio ($1.45 \times 0.50$~m$^2$, i.e., $3:1$), the along- and across-scan LSFs have roughly the same size ($\sigma \sim 1$~pixel).}\label{fig:parametrisation_histograms}
\end{figure*}

The LSF model in Equation~(\ref{eq:Gaussian_Lorentzian_def}) applies not only well to the along-scan direction but also to the across-scan direction. One peculiar aspect relevant (only) in the across-scan LSF is the fact that it varies in size (and shape) with time: stars, during their transit of the focal plane, have a small yet finite across-scan motion caused by the precession of the spin axis associated with the scanning law of the sky. The transverse speed of objects in the focal plane hence varies sinusoidally with a period equal to the satellite spin period (6 hours) and with an amplitude of $173$~mas~s$^{-1}$ (milli-arcsec~s$^{-1}$), corresponding to $2.80$~AC pixels over the 2900 integrating TDI lines in the SM CCDs.

Since we need to simulate and process hundreds of thousands of two-dimensional PSFs with random centre positions and noise configurations quickly, a parametrisation of the SM along- and across-scan LSFs using two sets of five parameters ($f$, $A$, $\alpha$, $v_0$, and $\gamma_0$; Figure~\ref{fig:parametrisation_histograms}) provides a convenient trade-off between realism and speed of our simulations. We thus simulate $750,000$ single-star images as follows:
\begin{enumerate}
\item parametrise the along-scan LSFs from the $448$-item full-fledged-simulation library by fitting, for each LSF, four free parameters ($f$, $\alpha$, $v_0$, and $\gamma_0$) to the LSF model from Equation~(\ref{eq:Gaussian_Lorentzian_def}); we freeze $A$ to unity in all fits to guarantee flux normalisation;
\item do the same but then across scan. We use three full-fledged PSF libraries with $448$ LSFs, (1) without across-scan motion ($0$~mas~s$^{-1}$), (2) with the average across-scan motion ($173 \cdot 2/\pi = 110$~mas~s$^{-1}$), and (3) with the maximum across-scan motion ($173$~mas~s$^{-1}$);
\item then repeat the following steps $750,000$ times:
\item select a random SM CCD, a random spectral type, and a random value of the interstellar extinction; in addition, select a random value of the across-scan motion with weights $\frac{1}{2}\cdot[\sin^{-1}(2/\pi)]/[\pi/2] = 0.2197$ for set (1), $0.5$ for set (2), and $\frac{1}{2}\cdot[(\pi/2)-\sin^{-1}(2/\pi)]/[\pi/2] = 0.2803$ for set (3);
\item get the five along-scan LSF fit parameters $f$, $A \equiv 1$, $\alpha$, $v_0$, and $\gamma_0$;
\item do the same but then across scan;
\item make a two-dimensional PSF, simply by multiplying the along-scan LSF with the across-scan LSF;
\item select a random sub-pixel position of the centre of the star, in two dimensions (along and across scan);
\item select a random magnitude between $G = 12.5$ and $20$~mag. In practice, we draw $100,000$ stars between $G = 12.5$ and $13.5$~mag, $100,000$ stars between $G = 13.5$ and $14.5$~mag, $\ldots$, and $50,000$ stars between $G = 19.5$ and $20$~mag. The total number of objects is hence $750,000$ exactly;
\item add sky background, corresponding to a typical surface brightness of $V = 22.5$~mag~arcsec$^{-2}$ (this corresponds to a background level of $0.63$~electrons per pixel after $2.85$~seconds of integration on the SM CCD);
\item add random Poisson noise, both on the object and on the sky-background counts;
\item project (``bin'') the PSF image on the SM samples (composed of $2 \times 2$ CCD pixels);
\item add a random total detection noise on each sample ($10.9$~electrons RMS per sample for the SM CCDs, based on ground-based payload-performance testing);
\item convert the electron counts to LSB units.
\end{enumerate}
We can ignore saturation, both at CCD-pixel-full-well and at CCD-charge-handling-capacity level, since our simulated stars, by construction, do not saturate (recall that saturared samples follow a different branch of the on-board detection software, ``extremity matching'' in Airbus Defence \& Space terminology). In the above process, to avoid border effects, we do not limit ourselves to $7 \times 7$~SM samples: each simulated image covers $40 \times 40$~samples ($80 \times 80$~pixels), which is then fed to the detection algorithm for object finding.

This recipe, clearly, does not provide a single-star library which is compatible with the astrophysical distribution of spectral types in the Gaia sky \citep[see, e.g.,][for a review of the expected spectral-type statistics and properties of the Gaia catalogue]{2012A&A...543A.100R}. But such a library is also not needed for our purposes: we aim to optimise the detection of all possible (CCD, spectral type, extinction) configurations, regardless of their existential probability, since we do not want Gaia's on-board detection to induce any biases in the selection of stars and hence in the final catalogue.

\subsection{Double stars}\label{subsec:double_star_dataset}

For double stars\footnote{
From now on, we will exclusively use the word ``double star'' to denote both optical double and (physical) binary stars; we do not treat higher-order multiple stars. In particular in dense areas, a significant fraction of Gaia double stars will not be binaries but optical doubles.
}, our requirements do not differ from those for single stars. We therefore follow the same recipe, except that we randomly select two objects (two PSFs) in each step (i.e., for each image). In practice, we simulate the primary component along the lines set out in Section~\ref{subsec:single_star_dataset}. The primary component is, by definition, the brightest and falls in the range $G = 12.5$ to $21$~mag; we go one magnitude fainter than for single stars since an unresolved, equal-brightness double star will be $0.75$~mag brighter than each component separately. Each simulated secondary component shares the CCD, the across-scan motion, and the interstellar extinction with its primary companion but has a random spectral type chosen among the 16 types listed in Section~\ref{subsec:single_star_dataset}, a random magnitude difference in the range $\Delta G = 0$--$5$~mag (with the added constraint that the secondary is brighter than $G = 21$ mag), a random orientation in the range $\alpha = 0^\circ$--$360^\circ$, a random separation in the range $\rho = 0$--$354$~mas, and a random sub-pixel centring. The maximum separation has been chosen to correspond to half of the (faint-star) along-scan window size in the astrometric field (i.e., $6$~AL pixels) since objects separated by larger angles will each receive their own window and can hence be considered as single stars.

As for single stars, we ignore saturation and avoid border effects by simulating oversized images covering $80 \times 80$ samples, which are then fed to the detection algorithm for object finding. In general, one double star simulated as described above can lead to either 1, 2, 3, or 4 local maxima:
\begin{itemize}
\item one local maximum typically results for double stars with small separations;
\item two local maxima typically result in cases of intermediate to large separations, allowing both components to be detected individually;
\item three and four local maxima can result if both components generate their own local maximum and if at least the primary component is bright and the separation is preferably not too large: the intersection(s) of the along-scan diffraction wing of one star with the across-scan diffraction wing of the other star (and/or vice versa) can yield a third (and/or fourth) local maximum.
\end{itemize}
We discriminate between double stars which generate one local maximum (symbolically $** \rightarrow *$) and double stars which generate two local maxima ($** \rightarrow **$). We construct two double-star data sets by simulating double stars in an open loop and assigning them either to the one-local-maximum or the two-local-maxima data set (or ignoring them in case of no local maximum) and repeating this exercise until both data sets have exactly $750,000$ entries (the $** \rightarrow **$ data set has thus $375,000$ underlying double stars whereas the $** \rightarrow *$ data set has $750,000$ underlying double stars).

Again, as for single stars, this recipe, clearly, does not provide a double-star library which is compatible with the (pairing) probability of physical binaries in the Gaia sky \cite[see, e.g.,][for a review of the expected binary-star statistics and properties of the Gaia double- and multiple-star catalogue]{2011AIPC.1346..107A}. But, for the same reasons as set out in Section~\ref{subsec:single_star_dataset} for single stars, this is also not required or desired for our purposes.

\subsection{Ghosts}\label{subsec:ghosts}

\begin{figure}[t]
  \centering
  \includegraphics[trim=52 20 25 25,clip,width = \columnwidth]{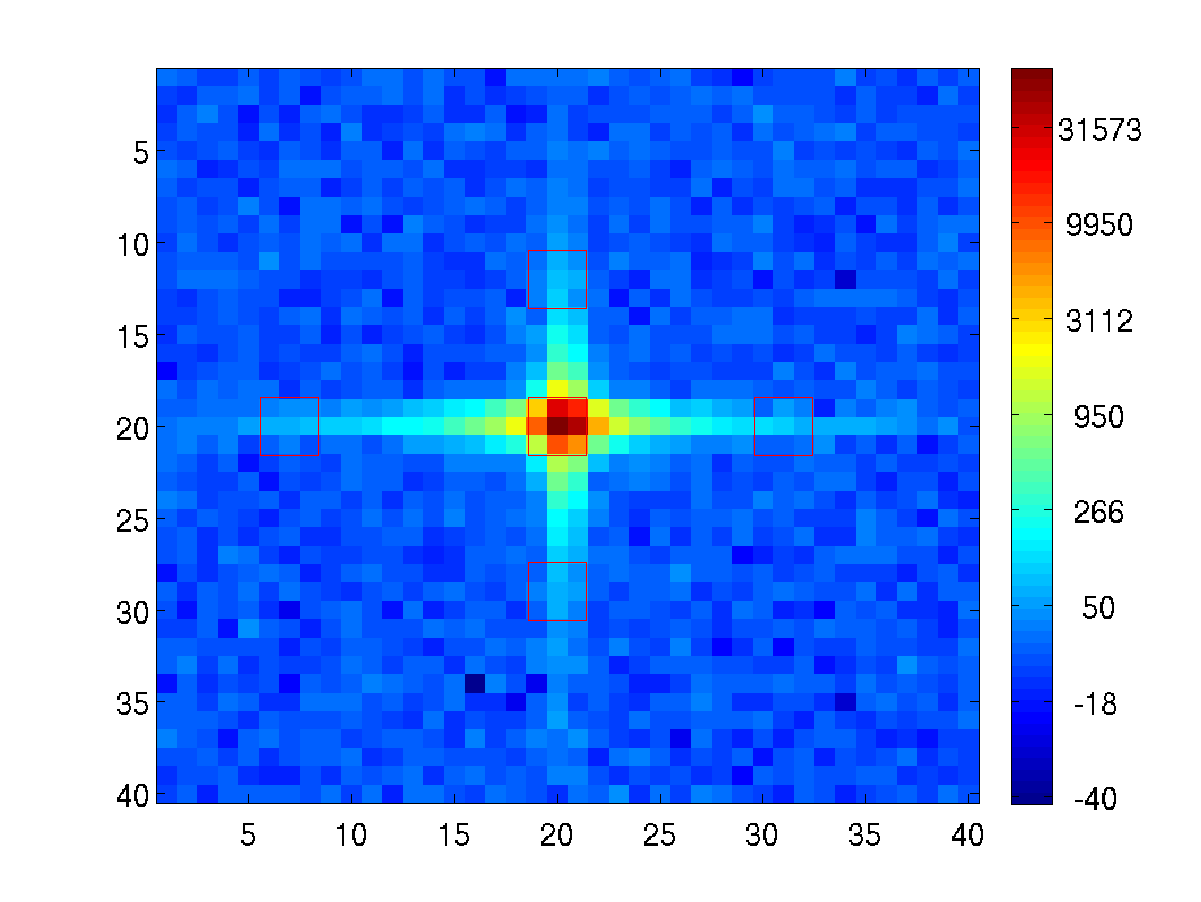}
  \caption{Example of the Sky-Mapper (SM) PSF of a single, bright star ($G \sim 13$~mag) which has five associated local maxima, one of the star itself and four spurious ghosts in the wings. The five red squares indicate the $3 \times 3$-samples VPA working windows of the five local maxima. The horizontal axis denotes along-scan (AL) SM sample while the vertical axis denotes across-scan (AC) SM sample. The colour coding is logarithmic and shows the sample flux in LSB after on-board background subtraction. Single stars can have associated ghosts out to a few dozen SM samples from the star centre (recall that our single-star simulations are based on a $40 \times 40$-samples grid).}\label{fig:example_PSF_bright_single_star}
\end{figure}

\begin{figure*}[t!]
  \centering
  \includegraphics[width=\textwidth]{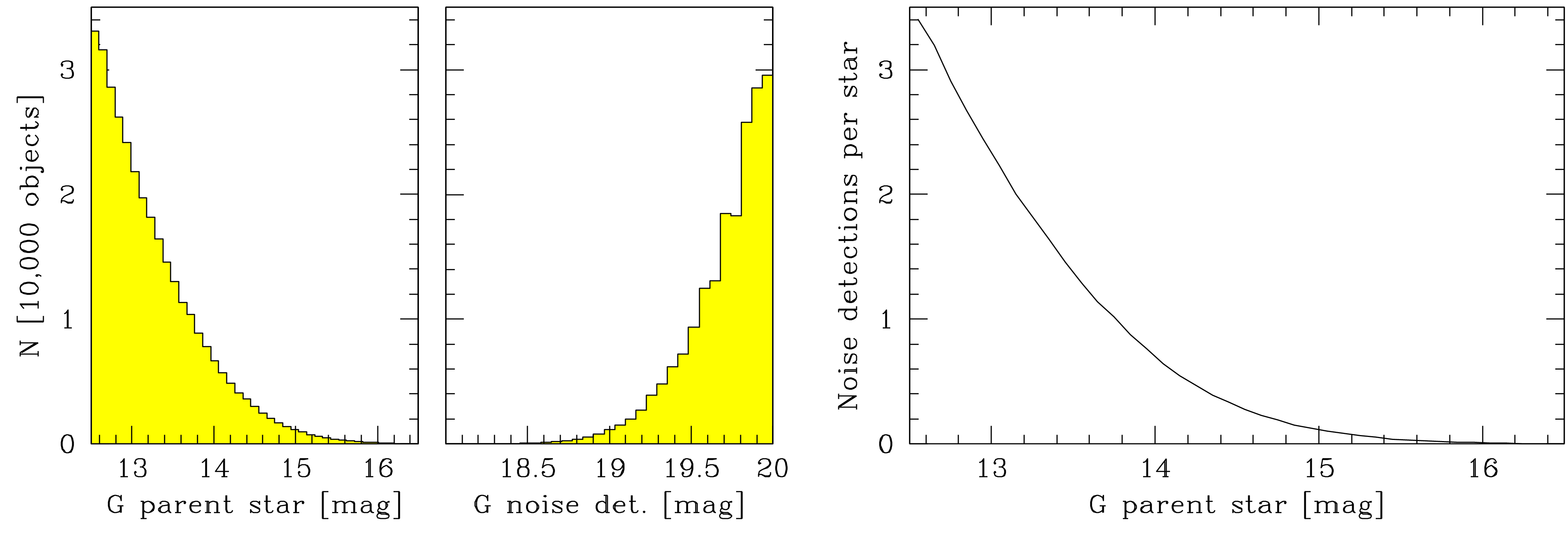}
  \caption{Properties of all $326,596$ ghosts brighter than $G = 20$~mag originating from single stars in the range $G = 12.5$--$20$ mag. {\it Panel 1}: histogram of the $G$ magnitude of the ``parent stars'' responsible for the ghosts. The faintest parent star has $G = 16.3$~mag. {\it Panel 2}: histogram of the $G$ magnitude of the ghosts. {\it Panel 3}: average number of ghosts that a star of magnitude $G$ generates. See also Figure~\ref{fig:ghosts2}.}
  \label{fig:ghosts1}
\end{figure*}

When feeding the single-star images described in Section~\ref{subsec:single_star_dataset} to the detection algorithm, it is not rare to retrieve multiple local maxima. Figure~\ref{fig:example_PSF_bright_single_star} shows an example of a single star which has five associated local maxima, one of the star core itself and four spurious ones in the (far) wings, from now on referred to as ghosts. This can happen since the (rather flat) PSF wing, some distance from the star centre, either along or across scan, can cause a local configuration of flux values in the $3 \times 3$-samples working window which satisfy the VPA local-maximum criteria on the PSF shape. Generally, such ghosts are found at some distance from the PSF core, where the PSF flattens out and where flux levels are low. They are hence typically faint. In our sample of $750,000$ single stars (Section~\ref{subsec:single_star_dataset}), we found $326,596$ ghosts. Figure~\ref{fig:ghosts1} shows their properties. The majority of the ghosts ($73$\%) is associated with the $100,000$ bright stars in the bin $G \in [12.5, 13.5]$~mag. The faintest star which has a ghost brighter than the VPA flux threshold at $G = 20$~mag is a $G = 16.3$-mag star. The ghosts vary in brightness from $G \approx 19$ to $20$~mag, with bright ones being (very) rare and faint ones being most common.

Ghosts which pass the thresholding stage are in principle harmful since they do compete in the window assignment (resource allocation) with real stars (Section~\ref{subsec:nomenclature}). We therefore follow what happens to ghosts when we optimise the rejection parameters by making a special object category labeled ``ghosts'', allowing to evaluate the performance of the optimised set of VPA parameters on this set of objects. This is further discussed in Section~\ref{subsec:discussion_ghosts}.

\subsection{Galactic cosmic rays and Solar protons}\label{subsec:PPEs}

\begin{figure}[t!]
  \centering
  \includegraphics[width=\columnwidth]{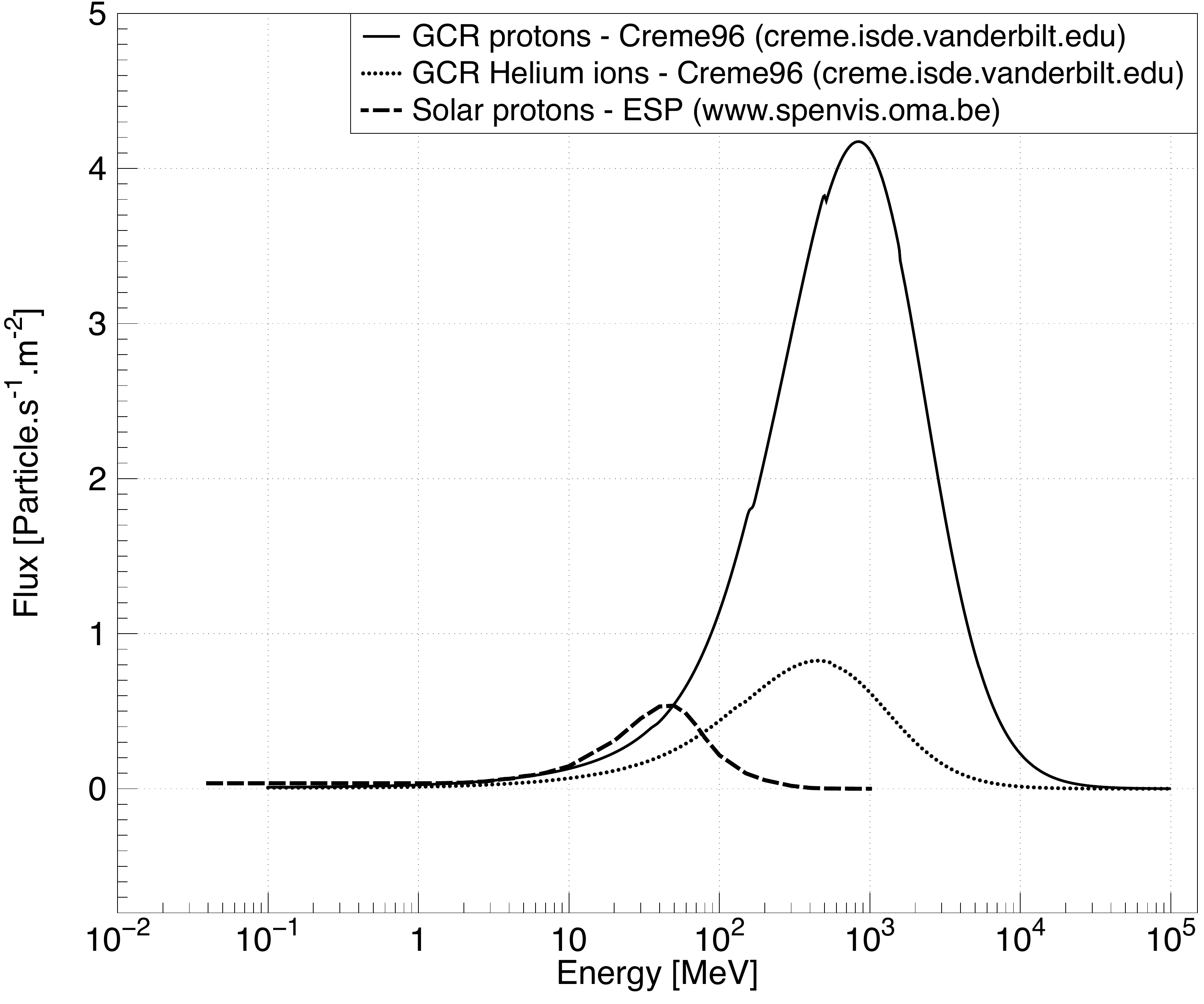}
  \caption{Energy distribution at L2 and at Solar maximum (after spacecraft shielding) for each considered type of incoming particle: Galactic-cosmic-ray proton (solid) and Helium nucleus (dotted), and Solar proton (dashed). The energy of each simulated prompt-particle event is randomly drawn from the respective distributions.}
  \label{fig:ppe_energy_distributions}
\end{figure}

\begin{figure}[t!]
  \centering
  \includegraphics[width=\columnwidth]{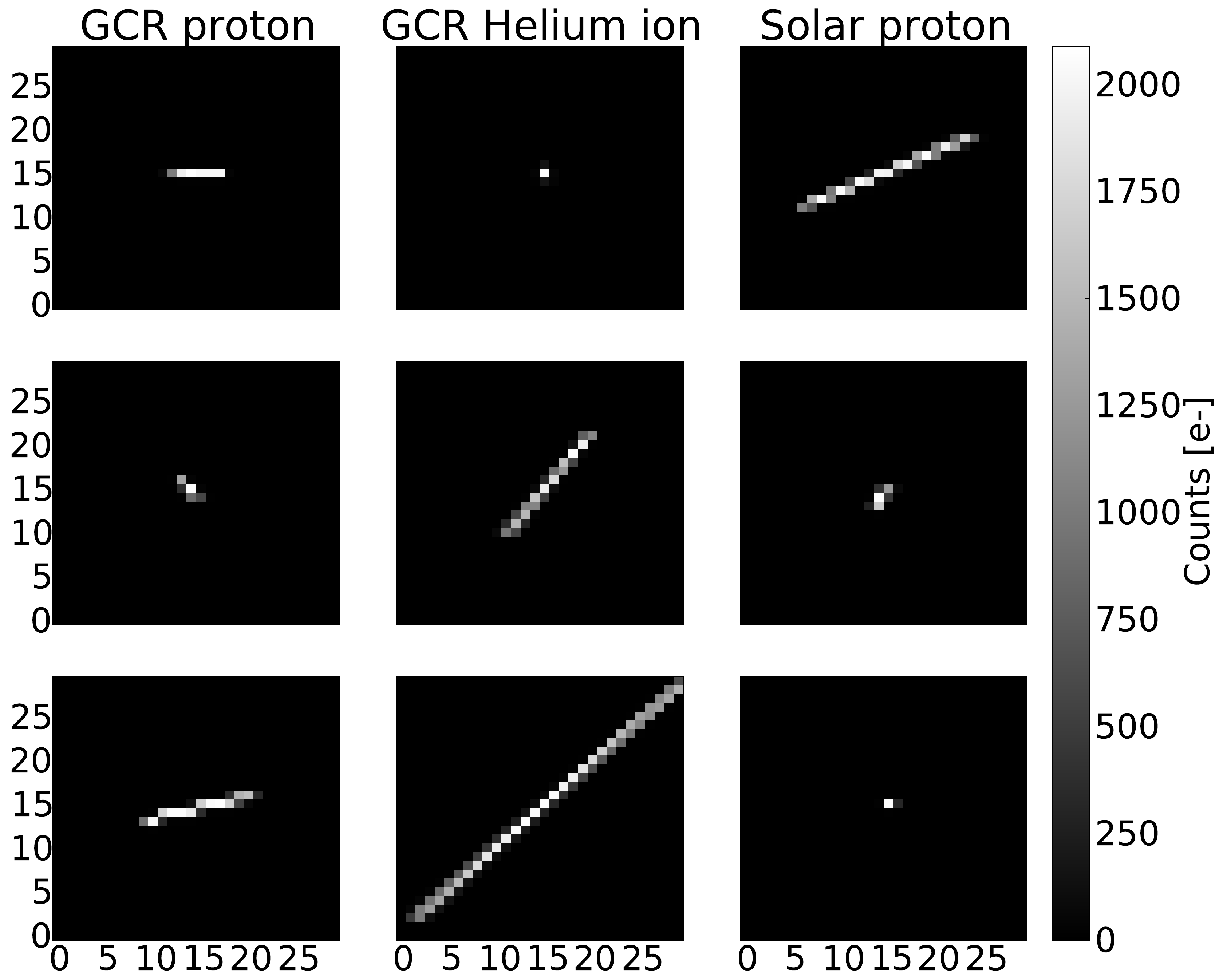}
  \caption{Examples of prompt-particle events for incoming particles of different nature and energy as generated by our simulator. The events are chosen arbitrarily to represent their diversity in orientation, size, and brightness. Elongated events, such as the one depicted in the bottom centre thumbnail, are less likely to occur since they need to pass through the CCD at a rather shallow angle. The most common events are circular (e.g., middle right thumbnail), with the incoming particle passing straight through the CCD.}
  \label{fig:ppe_example_images}
\end{figure}

Gaia's CCDs are not only sensitive to photons but also to energetic particles (radiation) that can lead to spurious events and, ultimately, unwanted detections\footnote{
Particles with energies less than $\sim$100~MeV are also responsible for displacement damage through generation of point defects (``traps'') in the CCD Silicon crystal lattice. These defects can trap and effectively delay electrons during their transfer from one pixel to the next, leading to an image distortion and decrease in signal-to-noise ratio. Implications of this charge-transfer inefficiency for the Gaia on-ground data processing are discussed in, e.g., \citet{2012MNRAS.419.2995P,2012MNRAS.422.2786H}.
}. As mentioned earlier, it is thus critical to discriminate prompt-particle events from astronomical sources at the detection stage. We hence simulate catalogues of prompt-particle events representative of the Gaia CCD architecture and the radiation environment of the spacecraft.

Gaia operates close to Solar maximum at the L2 Lagrangian point located 1.5~million~km beyond the Earth and its radiation belts. The L2 (interplanetary) radiation environment can be considered to be principally composed of Galactic Cosmic Rays (referred to as GCRs in Figures~\ref{fig:ppe_energy_distributions} and \ref{fig:ppe_example_images}) and Solar particles:
\begin{itemize}
\item Cosmic rays are high-energy particles (up to several GeV, cf.\ Figure~\ref{fig:ppe_energy_distributions}), generated mostly by supernovae, that are coincidentally passing through the Solar system. At the energies considered in this work, they are composed of approximately $90$\% protons, $9$\% Helium ions, and $1$\% heavier ions. The incoming flux of cosmic rays is rather continuous with a slight modulation by the Sun's activity (minimum at Solar maximum) and can be considered as a constant background of $5$~particles~cm$^{-2}$~s$^{-1}$.
\item Solar particles -- essentially protons -- are lower-energy particles (from several eV to a few hundred MeV, cf.\ Figure~\ref{fig:ppe_energy_distributions}) emitted by the Sun during discrete magnetic reconnection events occurring at the Solar surface. The Solar-proton flux hence varies from close to zero during Solar-quiet times to extremely high fluxes, up to millions of protons~cm$^{-2}$~s$^{-1}$, during Solar flares. 
\end{itemize}

Generating representative catalogues of prompt-particle events requires the energy spectrum for each type of incoming particle at L2 during Solar maximum, accounting for spacecraft shielding. This can be obtained using standard on-line models and tools. We use the CREME96 model \citep{1997ITNS...44.2150T} for cosmic rays and the SPace ENVironment Information System (SPENVIS) together with the Emission of Solar Protons (ESP) total-fluence model \citep{1999ITNS...46.1481X,2000ITNS...47..486X} for Solar protons. Spacecraft shielding stops a significant fraction of the lower-energy particles (i.e., mostly the Solar protons). To account for the impact of shielding on each spectrum, we use the particle-transport facility of each tool and an Aluminium thickness value of $11$~mm, corresponding to the average Al-equivalent shielding at the Gaia focal-plane assembly. The resulting spectra for each particle type are shown in Figure~\ref{fig:ppe_energy_distributions} and are used as input in our event simulation.

Each prompt-particle-event image in our catalogue is generated using code developed by A.~Short (2006, private communication) in support of GIBIS and validated against in-orbit XMM-EPIC MOS CCD data. To generate a single event, the main steps of the simulation consist of: 
\begin{enumerate}
\setcounter{enumi}{0}
\setlength{\itemsep}{1pt}
\item Random generation of the particle energy following the input energy spectrum, sub-pixel position, and angle of incidence;
\item Energy deposition (i.e., generation of free electrons) along the particle path through the CCD according to the Silicon stopping power applicable to the type of incident particle;
\item Electron diffusion in the field-free (and depleted) CCD region(s);
\item Mapping of the electrons to the CCD pixels and image generation.
\end{enumerate}
Our simulation takes into account the pixel architecture and geometry of the Gaia SM CCDs (normal-resistivity Silicon, $10 \times 30~\mu$m$^2$ pixels, $9~\mu$m depletion depth, and $7~\mu$m field-free thickness) and a nominal operating temperature of $163$~K.

We generate two catalogues, one for cosmic-ray events and one for Solar-proton events. Figure~\ref{fig:ppe_example_images} shows examples of simulated events for each particle type. One event can lead to multiple detections (including no detections): our $2,602,864$ cosmic-ray images lead to $3,884,976$ detections (i.e., local maxima in the VPA), which means the average multiplication factor is $1.49$, while our $1,195,992$ solar-proton images lead to $1,611,882$ detections (i.e., local maxima in the VPA), which means the average multiplication factor is $1.35$; this difference can be understood since cosmic rays are typically elongated while Solar protons are typically more point-like. For both event types, we ``only'' use $750,000$ randomly-selected local maxima in the VPA in our study (Section~\ref{sec:optimisation}).

 The statistical properties of our catalogues agree with the properties of similar catalogues which have been developed independently by Airbus Defence \& Space in 2008 in the frame of the Gaia project based on \cite{1979ITED...26.1742K,1990ITNS...37.1876L,1997SPIE.3063...77D}. One notable feature of both sets of prompt-particle-event catalogues is the lack of faint events: the faintest detected event has $G \sim 18.7$--$18.8$~mag ($\sim$$1,800$--$1,700$ electrons). This is not surprising, given the input energy distributions displayed in Figure~\ref{fig:ppe_energy_distributions}. In addition, one should realise that faint events come either from (very-)high-energy particles, which are hardly decelarated when they interact with the Silicon and hence deposit only few free electrons, or from low-energy particles, which are totally absorbed but which can only free a limited number of electrons. In addition, particles ineracting with CCDs deposit most energy just before they come to a stop, which gives a hard cut-off at low energies.

\subsection{Unresolved galaxies}\label{subsec:galaxies}

Gaia will not only observe stars but will also encounter millions of poorly-to-unresolved galaxies all over the sky \citep{2014A&A...568A.124D}. This unique dataset is a valuable by-product of the mission, and specific groups in the Gaia Data Processing and Analysis Consortium (DPAC) are in charge of developing strategies and the necessary software implementation for spectral \citep{Tsalmantza2009} and morphological \citep{KroneMartins2013} studies of these objects.

\begin{figure}[t!]
  \centering
  \includegraphics[width=\columnwidth]{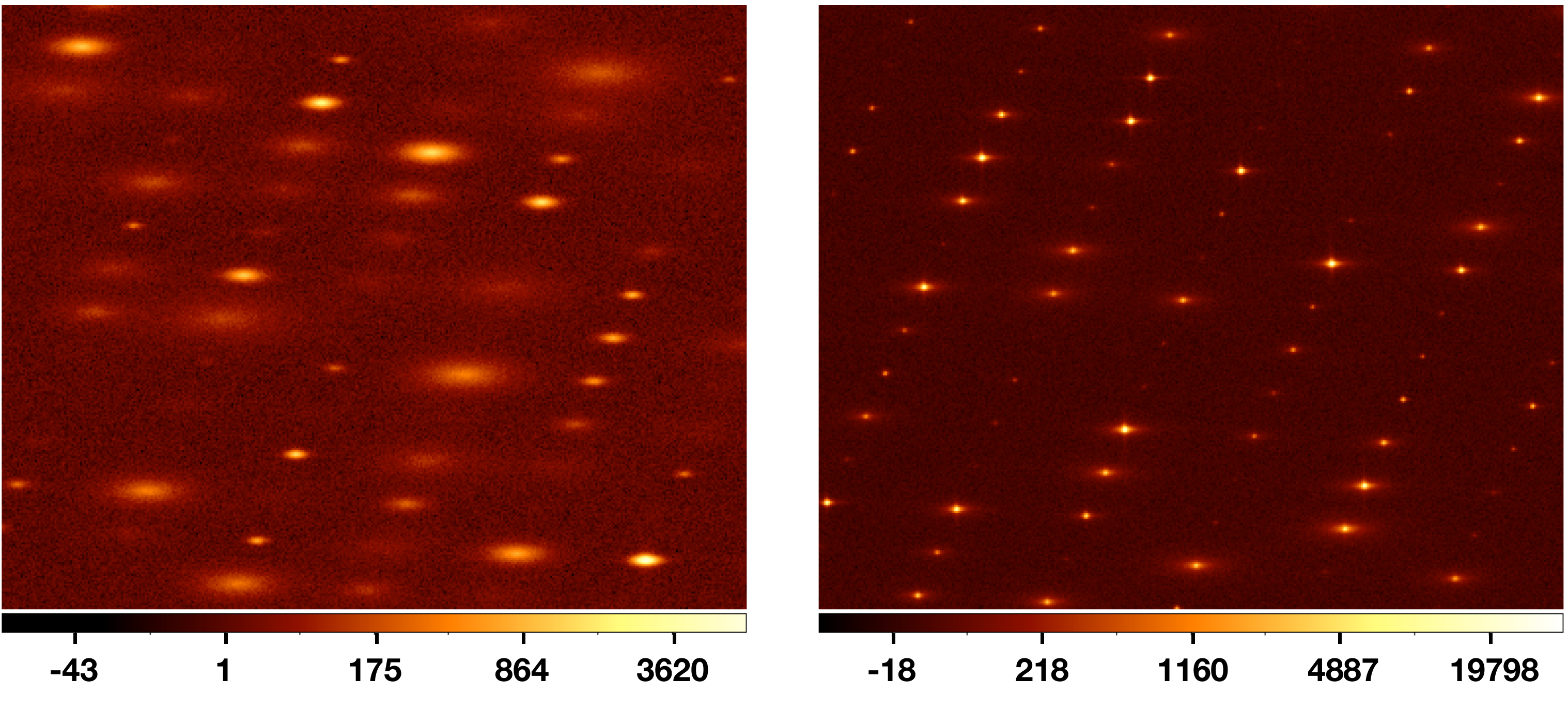}
  \caption{Examples of extreme galaxy profiles in SM CCDs simulated with GIBIS. Exponential disk profiles are shown in the top panel, while de Vaucouleurs profiles are shown in the bottom panel. The colour map is logarithmic and encodes the flux in each pixel in electrons. The profiles do not appear circularly symmetric since the pixels in this representation are square while Gaia's pixels are rectangular. In our detection-performance assessment (Section~\ref{subsec:scientific_results_galaxies}), individual images of all objects, at the correct angle for each transit, are generated and analysed.}
  \label{Fig:GalProf-FP-SM-GB13}
\end{figure}

As Gaia is primarily a Galactic astrometry mission, we do not take galaxies into account for the optimisation of the rejection parameters (Section~\ref{sec:optimisation}). However, it is important to study the impact of this optimisation on the detection of such objects, as this may have a direct impact on the scientific outcome of their study as well as on the strategies to be adopted for their analysis during the data processing. Thus, to assess the detection of unresolved galaxies, we create a catalogue of synthetic galaxy profiles covering two extreme cases: (i) pure de Vaucouleurs profiles, representing pure classical galaxy bulges or elliptical objects, and (ii) pure exponential profiles, representing pure galaxy disks. We have deliberately chosen not to include the most extreme case of galaxy profiles, representing active galactic nuclei (AGNs), as their point-source-like profiles will be naturally detected by Gaia. The simulations have been performed with GIBIS, which simulates the de Vaucouleurs profiles using the effective radius $R_{V}$, corresponding to:
\begin{equation}
I_{V}(r) \propto \exp{ \left( -7.67 \left( \frac{r}{R_{V}} \right)^{1/4} \right)}
\end{equation}
and the exponential profile using the disk scale length $R_{E}$:
\begin{equation}
I_{E}(r) \propto \exp{\left(-\frac{r}{R_{E}}\right)}.
\end{equation}
The simulated profiles are circularly symmetric, as elliptical profiles are equivalent to a circular profile of a smaller radius for detection purposes. They uniformly cover the parameter space with radii between $0.2$ and $2.0$~arcsec and integrated magnitudes from $V = 14$ to $20$~mag, regardless of the physical relevance of each parameter combination \citep[e.g., a fraction of this parameter space is not expected to be occupied by real galaxies; see][]{2014A&A...568A.124D}. As generating GIBIS simulations is time consuming, the simulations have been performed arranging several profiles in the same image. The profiles have been arranged on a regular grid around galactic coordinates $(l, b) = (40^\circ, 52^\circ)$. These coordinates have been chosen since -- due to Gaia's scanning law used in GIBIS -- the satellite will perform $152$ observations with different transit angles around this position, making the analysis of the results less prone to statistical fluctuations. Considering each transit as an independent observation, a total of $179,056$ observations have been simulated. Figure~\ref{Fig:GalProf-FP-SM-GB13} shows two examples of the resulting SM images.

\subsection{Asteroids}\label{subsec:asteroids}

\begin{figure}
  \centering
  \includegraphics[width=\columnwidth]{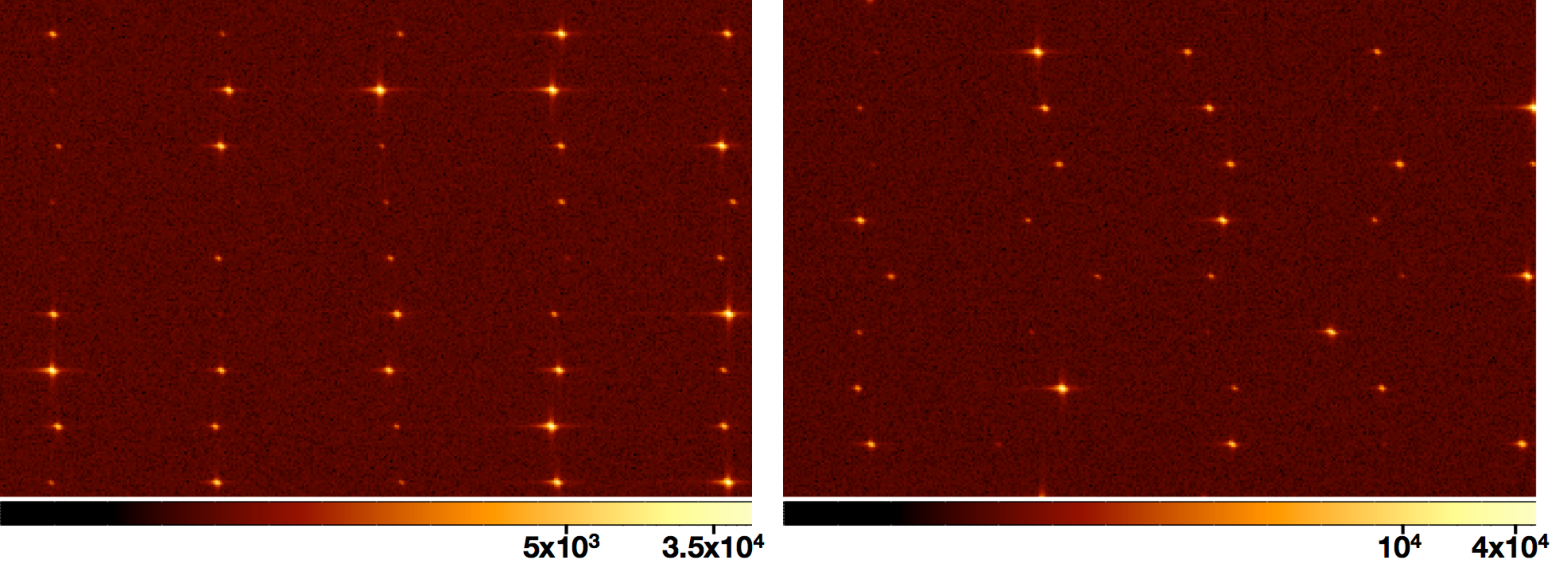}
  \caption{Example GIBIS images of main-belt asteroids (left) and near-Earth objects (right) in SM CCDs. The colour map is logarithmic and encodes the flux in each pixel in electrons.}
  \label{Fig:AstProf-FP-SM-GB13}
\end{figure}

Besides stars (Sections~\ref{subsec:single_star_dataset}--\ref{subsec:double_star_dataset}) and unresolved galaxies (Section~\ref{subsec:galaxies}), Gaia will also observe a few hundred thousand Solar-system bodies, mainly asteroids \citep[e.g.,][]{hestro10_lnp,hestro14_cosp}. A specific data-reduction pipeline with customised identification and centroiding algorithms has been implemented in DPAC for these moving, generally unresolved objects. Like for unresolved galaxies, we do not take asteroids into account for the optimisation of the rejection parameters (Section~\ref{sec:optimisation}) albeit we do assess their detection performance using GIBIS simulations. Compared to current and upcoming ground-based surveys, Gaia's limiting magnitude is modest. However, Gaia has the unique capability to discover new near-Earth objects (NEOs) at low Solar elongation, i.e., the faint end of the detected population is of particular interest and important for the science-alerts-driven ground-based follow-up network Gaia-FUN-SSO \citep{thuillot14_sf2a}. We hence distinguish two groups, the main-belt asteroids (MBAs) and NEOs; the latter are generally fainter and have larger apparent motion. The asteroid velocity vectors are randomly sampled from the distributions from \citet{Mignard07}. Since the motion of asteroids around the Sun is within some tens of degrees from the Laplacean plane, their motion relative to the Gaia focal plane is not uniformly distributed: speeds are on average larger in the across-scan direction. To produce statistics for the detection analysis for each type of asteroid, ten independent simulation grids (across-scan speed versus along-scan speed versus magnitude between $V = 14$ and $21$~mag) have been created, resulting in $4640$ MBAs and $4640$ NEOs. The asteroids have been shuffled around at random positions in the focal plane between the different simulations to average out any possible positional dependency. Figure~\ref{Fig:AstProf-FP-SM-GB13} shows two examples of asteroid images.

\section{Optimising the free parameters}\label{sec:optimisation}

\subsection{Defining the merit function}

In order to optimise the 20 free parameters of the low- and high-frequency rejection curves, we need to define a merit function. First, it is important to realise that the low- and high-frequency curves are independent. The 20-dimensional problem hence reduces to two 10-dimensional problems. After some experimenting, we settled -- for both the low- and the high-frequency optimisation -- on the functional form:
\begin{eqnarray}
P(\mathbf{u}) & = & P_{*}(\mathbf{u})\ \cdot P_{** \rightarrow *}(\mathbf{u})\ \cdot P_{** \rightarrow **}(\mathbf{u})\ \cdot\nonumber \\
&& [1 - P_{\rm CR}(\mathbf{u})]\ \cdot [1 - P_{\rm SP}(\mathbf{u})],\label{eq:merit_function}
\end{eqnarray}
where the 10-dimensional vector $\mathbf{u} = (a_\rightarrow, b_\rightarrow, c_\rightarrow, d_\rightarrow, e_\rightarrow, \break a_\uparrow, b_\uparrow, c_\uparrow, d_\uparrow, e_\uparrow)$ is the vector of unknowns (free parameters) of either the low- or the high-frequency problem; the subscript $\rightarrow$ denotes the along-scan parameters whereas the subscript $\uparrow$ denotes the across-scan parameters. The subscript $*$ stands for a single star, $** \rightarrow *$ for a double star inducing a single detection, $** \rightarrow **$ for a double star inducing two detections, CR for cosmic ray, and SP for Solar proton. The general symbol $P$ denotes detection probability, i.e., the fraction of objects which fall above the high-frequency curve in the high-frequency case or below the low-frequency curve in the low-frequency case. In essence, the merit function from Equation~(\ref{eq:merit_function}) defines a balance between single- and double-star detection versus cosmic-ray and Solar-proton rejection: the higher $P$, the better Gaia's (stellar) science return. We do not consider the detection performance of external galaxies and/or asteroids in the merit function since these objects are not a core science product: Gaia is a Galactic astrometry mission and the on-board detection should be optimised for stars.

The detection probability of single stars, $P_{*}(\mathbf{u})$, is calculated as:
\begin{equation}
P_{*}(\mathbf{u}) = \sum_{G = 13}^{20} w_G \cdot P_{G*}(\mathbf{u}),
\end{equation}
where the summation is over the $G$-magnitude range of interest, $w_G$ denotes the weight of each magnitude bin, i.e., the fractional number of stars in that bin from the standard Gaia Galaxy model (Table~\ref{tab:weights}), and $P_{G*}(\mathbf{u})$ denotes the average detection probability of the $N_G$ simulated stars in each magnitude bin (recall that $N_G = 100,000$ for $G = 13,\ldots,19$, while $N_{G = 20} = 50,000$):
\begin{eqnarray}
P_{G*} &=& {{{1}\over{N_G}}}\sum_{i=1}^{N_G}
\left\{
  \begin{array}{l l}
    \rm{\ } & \rm{\ }\\
    \rm{\ } & \rm{\ }\\
    1 & \quad \rm{if\ }
\left\{\begin{array}{c}
  \left[\left(\left([v_{0,i} + a_{\rightarrow}]_{18} \cdot [v_{2,i} + b_{\rightarrow}]_{18}\right)_{4} \cdot c_{\rightarrow}\right)_{8}\right]_{32} \\
\begin{array}{c}
< \rm{\ for\ low\ frequency}\\
> \rm{\ for\ high\ frequency}
\end{array} \\
 \left[\left(\left[(F_i)_2+d_{\rightarrow}\right]_{18}^2 + e_{\rightarrow}\right)_{4}\right]_{32} \\
     {\rm \  } \\
     {\wedge } \\
     {\rm \  } \\
  \left[\left(\left([h_{0,i} + a_{\uparrow}]_{18} \cdot [h_{2,i} + b_{\uparrow}]_{18}\right)_{4} \cdot c_{\uparrow}\right)_{8}\right]_{32} \\
\begin{array}{c}
< \rm{\ for\ low\ frequency}\\
> \rm{\ for\ high\ frequency}
\end{array} \\
 \left[\left(\left[(F_i)_2+d_{\uparrow}\right]_{18}^2 + e_{\uparrow}\right)_{4}\right]_{32} \\
 \end{array}\right.\\
    \rm{\ } & \rm{\ }\\
    0 & \quad \rm{otherwise,}\\
  \end{array} \right.\label{eq:define_P_G*}
\end{eqnarray}
where $F_i = v_{0,i} + v_{1,i} + v_{2,i} = h_{0,i} + h_{1,i} + h_{2,i}$ is the (background-subtracted) LSB flux of star $i$ in the $3 \times 3$-samples working window, and $v_{j,i}$ and $h_{j,i}$ denote the LSB flux sums of the $j^{\rm th}$ vertical (across-scan) and horizontal (along-scan) vectors of the $3 \times 3$-samples working window of star $i$ (see Section~\ref{sec:vpa}, Equation~\ref{eq:h_and_v}). The saturation and truncation operators $[\ldots]_{n}$ and $(\ldots)_{n}$ are defined in Section~\ref{subsec:detection}.

The detection probabilities of double stars, $P_{** \rightarrow *}(\mathbf{u})$ and $P_{** \rightarrow **}(\mathbf{u})$, are calculated along the same line as the detection probability for single stars. The detection probabilities of cosmic rays and Solar protons, $P_{\rm CR}(\mathbf{u})$ and $P_{\rm SP}(\mathbf{u})$, are calculated nearly the same, the only difference being that the weights $w_G$ are all identical to 1 since the probability of occurrence of a particular event with a certain energy (i.e., magnitude) is already covered in the creation of the event catalogues (see Section~\ref{subsec:PPEs}).

\subsection{Regularising the merit function}

With the choice made above to link the weights $w_G$ to the frequency of occurrence of stars in the sky, bright stars ($G\sim 13$--$16$~mag) implicitly receive reduced weight compared to faint stars since the latter are (far more) numerous. This is desirable to some extent but risks not detecting a disproportionate fraction of bright stars, which generally have high scientific importance and small astrometric errors. We therefore introduce regularisation factors $R_{*}$ and $R_{**}$ in the merit function $P(\mathbf{u})$ as defined in Equation~(\ref{eq:merit_function}) enforcing a minimum detection performance for single and double stars which varies as function of~magnitude:
\begin{equation}
R_{*} = \prod_{G = 13}^{20} R_{G*} \quad{\rm with}\quad R_{G*} =
\left\{ \begin{array}{l l}
    1 & \quad \rm{if\ } P_{G*} \geq P_{G*,{\rm min}}\\
    0 & \quad \rm{otherwise,}\\
\end{array} \right.\label{eq:regularisation_function}
\end{equation}
and similar for double stars ($R_{**}$).

Gaia's scientific mission requirements entail at least $95$\% on-board observation efficiency for single and double stars over the full magnitude range, down to the faint limit $G = 20$~mag. This implies that the detection probability shall be even higher than $95$\% since other losses exist (for example, there is a finite confirmation probability in AF1, $0.2$\% of faint-object transits is lost as a result of prioritised allocation of windows to bright stars, $0.1$\% of transits is lost as a result of focal-plane ``blinding'' caused by nearby bright stars or planets, etc.; Section~\ref{subsec:nomenclature}). Since in early industrial software verification tests $>$$98$\% detection performance on single stars has been reached, and since experiments with our software indicate that single-star detection percentages of $99.99$\% can be reached, we adopt threshold values (Table~\ref{tab:weights}) $P_{G*,{\rm min}} = \sqrt{0.9999}$ and $P_{G**,{\rm min}} = \sqrt{0.99}$ for $12.5 < G~{\rm [mag]} < 16.5$ (bins $G = 13,\ldots,16$) and $P_{G*,{\rm min}} = \sqrt{0.9999}$ and $P_{G**,{\rm min}} = \sqrt{0.97}$ for $16.5 < G~{\rm [mag]} < 20$ (bins $G = 17,\ldots,20$). The square roots refer to the fact that $P$ defines either the high- or the low-frequency detection probability; the total detection probability is the ``logical AND'' (i.e., the ``product'') of these probabilities.

\begin{table}[t]
\caption{Statistics of the Gaia Universe Model Snapshot GUMS \citep{2012A&A...543A.100R}. $N$ denotes the number of objects in the model in each magnitude bin (not to be confused with $N_G$ which denotes the number of simulated objects in magnitude bin $G$); $w_G$ denotes the relative, normalised weight of each bin, such that $\sum_{G = 13}^{20} w_G = 1$; $P_{*,{\rm min}}$ and $P_{**,{\rm min}}$ denote the minimum-required detection probabilities for single and double stars, respectively (Equation~\ref{eq:regularisation_function}). The square root refers to the fact that $P$ refers to either the high- or the low-frequency detection probability; the final detection probability is the ``logical AND'' (i.e., the ``product'') of these probabilities.}\label{tab:weights}
\begin{center}
\begin{tabular}[h]{cccccc}
\hline\hline
\\[-8pt]
$G$ & $G$ range & $N$ & $w_G$ & $P_{G*,{\rm min}}$ & $P_{G**,{\rm min}}$\\
$[$mag$]$ & $[$mag$]$ & $[$$10^6$~stars$]$ & &\\
\hline
\\[-8pt]
13& 12.5--13.5&  10& 0.0092& $\sqrt{0.9999}$& $\sqrt{0.99}$\\
14& 13.5--14.5&  24& 0.0223& $\sqrt{0.9999}$& $\sqrt{0.99}$\\
15& 14.5--15.5&  38& 0.0351& $\sqrt{0.9999}$& $\sqrt{0.99}$\\
16& 15.5--16.5&  71& 0.0660& $\sqrt{0.9999}$& $\sqrt{0.99}$\\
17& 16.5--17.5& 125& 0.1167& $\sqrt{0.9999}$& $\sqrt{0.97}$\\
18& 17.5--18.5& 183& 0.1713& $\sqrt{0.9999}$& $\sqrt{0.97}$\\
19& 18.5--19.5& 377& 0.3526& $\sqrt{0.9999}$& $\sqrt{0.97}$\\
20& 19.5--20.0& 243& 0.2268& $\sqrt{0.9999}$& $\sqrt{0.97}$\\
\hline
\end{tabular}
\end{center}
\end{table}

\subsection{Optimising the merit function}

To optimise the regularised merit function ($P(\mathbf{u}) \cdot R_{*} \cdot R_{**}$ from Equations~\ref{eq:merit_function} and \ref{eq:regularisation_function}), we use the downhill-simplex minimisation method \citep[][in practice, since we want $P$ to be maximised, we minimise $1 - P(\mathbf{u}) \cdot R_{*} \cdot R_{**}$]{1965CJ......7..308N,Press:2007:NRE:1403886}. For both the low- and high-frequency problems, we adopt a three-step minimisation approach:
\begin{enumerate}
\item We first explore the full parameter space ($-32,768$ to $+32,767$ for each parameter) in a coarse manner, using randomly-placed starting simplices with large characteristic length scales ($10,000$) and a reduced set of data ($10$\% of all objects, randomly selected from our object/event catalogues). These settings allow to repeat the optimisation a large number of times within a reasonable time (e.g., $12$ days for $\sim$$50,000$ repeats on a normal workstation), enabling deep exploration of the full parameter space.
\item We then zoom in on the minimum found in the previous step and start the optimisation again in that area -- still allowing the starting simplex to vary from run to run over the characteristic length scale -- but now with reduced characteristic length scales (typically $\sim$$100$ for $a$, $b$, and $c$ and $\sim$$1000$ for $d$ and $e$) and with the full set of objects ($750,000$ single stars, $750,000$ double stars generating one local maximum, $375,000$ double stars generating two local maxima, $750,000$ Solar-proton-induced local maxima, and $750,000$ cosmic-ray-induced local maxima). We repeat this minimisation $1,000$ times.
\item We finally restart the optimisation from the minimum found in the previous step, but now with further-reduced characteristic length scales (typically by a factor $10$ compared to the previous step). We repeat this minimisation $100$ times. The outcome of this step yields the optimised vector $\mathbf{u}$ of unknowns as well as the achieved detection performance of stars and rejection performance of cosmic rays and Solar protons. These are discussed further in Section~\ref{sec:results}.
\end{enumerate}

\section{Results}\label{sec:results}

After optimisation, the merit function (Equation~\ref{eq:merit_function}) reaches $P = 14.51$ for the low-frequency case, with -- by construction -- regularisation factors $R_{*} = R_{**} = 1$, compared to $P = 12.04$ for the baseline parameters. In the latter case, however, the minimum detection percentages defined in Table~\ref{tab:weights} are not met, neither for single nor for double stars, i.e., $R_{*} = R_{**} = 0$. All low-frequency star-detection probabilities have improved: $P_{*,{\rm LF}}$ went from $99.964$\% to $99.999$\%, $P_{** \rightarrow *,{\rm LF}}$ from $98.417$\% to $99.867$\%, and $P_{** \rightarrow **,{\rm LF}}$ from $98.308$\% to $99.961$\%. At the same time, the low-frequency cosmic-ray and solar-proton detections also improved: $P_{\rm CR, LF}$ went from $65.843$\% to $63.123$\% and $P_{\rm SP, LF}$ from $63.560$\% to $60.587$\%. For the high-frequency optimisation, we reached $P = 79.91$ (with $R_{*} = R_{**} = 1$), compared to $P = 80.39$ for the default settings; again, the functional baseline does not meet the minimum detection percentages defined in Table~\ref{tab:weights}, neither for single nor for double stars, i.e., $R_{*} = R_{**} = 0$. As for the low-frequency case, all high-frequency star-detection probabilities improved: $P_{*,{\rm HF}}$ went from $99.997$\% to $99.998$\%, $P_{** \rightarrow *,{\rm HF}}$ from $99.999$\% to $100.000$\%, and $P_{** \rightarrow **,{\rm HF}}$ from $99.963$\% to $99.968$\%; the prompt-particle-event performance slightly degraded, from $11.668$\% to $11.717$\% for $P_{\rm CR, HF}$ and from $8.951$\% to $9.453$\% for~$P_{\rm SP, HF}$.

After combining the low- and high-frequency results, the following situation emerges: the single-star (faint-star) detection probability $P_{*}$ increases from $99.961$\% to $99.997$\%; the probability $P_{** \rightarrow *}$ to detect a double star as one detection (``unresolved double star'') increases from $98.417$\% to $99.866$\%; the probability $P_{** \rightarrow **}$ to detect a double star as two detections (``resolved double star'') increases from $98.271$\% to $99.928$\%; the probability $P_{\rm CR}$ to detect a cosmic ray decreases from $6.349$\% to $5.276$\%; and the probability $P_{\rm SP}$ to detect a Solar proton decreases from $3.401$\% to $3.064$\%. The magnitude dependence of these results is provided in Table~\ref{tab:optimised_results}; for comparison, Table~\ref{tab:baseline_results} presents the magnitude dependence of the functional baseline. One can immediately conclude that the functional baseline for the rejection parameters provides a starting point which meets the single-star scientific requirements of the mission (albeit not the more stringent minimum detection percentages defined in Table~\ref{tab:weights}). Nonetheless, we have found room for optimisation, the main reason being that we have no constraint beyond $G \sim 18.5$~mag to reject cosmic rays and/or Solar protons, simply because such events do not exist in significant quantities (see the discussion in Section~\ref{subsec:PPEs}). So, the flux-dependence freedom of the rejection curves for faint objects has been used in the optimisation to select virtually all detections (local maxima). This is, clearly, beneficial for extended objects, in particular unresolved galaxies and asteroids (see Sections~\ref{subsec:scientific_results_galaxies} and \ref{subsec:scientific_results_asteroids}). The price to pay is, of course, that also ghosts (Section~\ref{subsec:ghosts}) are now frequently detected: whereas the functional baseline only lets $1.800$\% of the ghosts through, this increases to $99.866$\% for the optimised parameters. This side effect is discussed further in Section~\ref{subsec:discussion_ghosts}.

\begin{figure}[t!]
  \centering
  \includegraphics[width=\columnwidth]{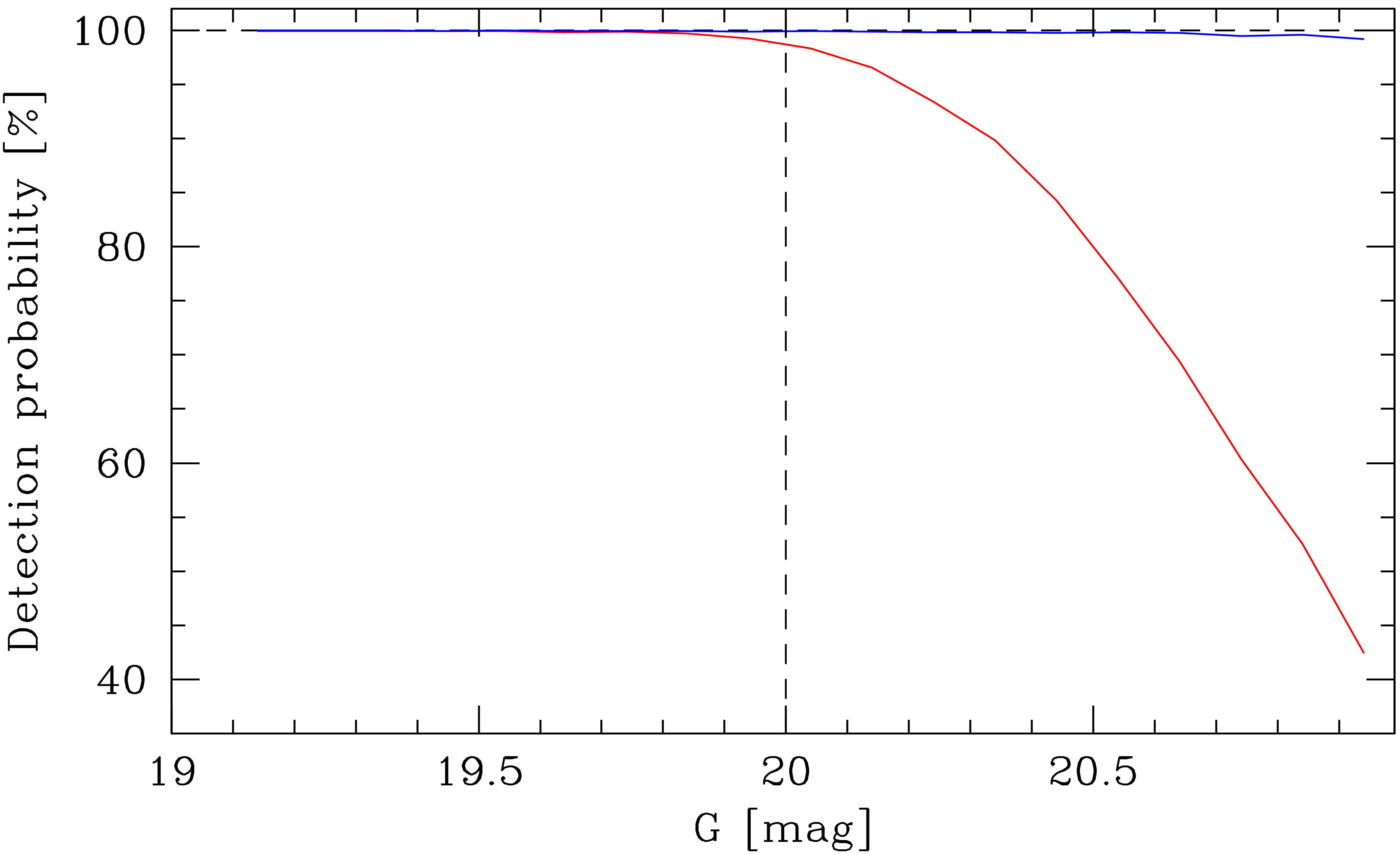}
  \caption{Single-star detection probability -- without any flux thresholding -- as function of $G$ magnitude for both the functional-baseline (red) and the optimised (blue) rejection parameters.}
  \label{fig:performance@G=20mag}
\end{figure}

Figure~\ref{fig:performance@G=20mag} shows the single-star detection probability as function of $G$ magnitude for both the functional-baseline and the optimised rejection parameters. These results do not involve a flux thresholding: they purely reflect the intrinsic detection performance of Gaia, including the effect of the rejection parameters. Surprisingly, therefore, the baseline parameters already show the start of a downgoing trend in the detection probability of stars brighter than the nominal threshold of $G = 20$~mag. The optimised parameters, on the other hand, show a constant probability, close to $100$\%, up to $G = 21$~mag (compared with $\sim$$40$\% for the functional-baseline parameters reached at $G = 21$~mag).

\begin{table}[t]
\caption{Magnitude dependence of object-detection probabilities for the functional-baseline rejection parameters. The symbol '--' indicates the absence of faint cosmic rays and Solar protons in our prompt-particle-event catalogues, as explained in Section~\ref{subsec:PPEs}. The magnitude-averaged star-detection probabilities $P_{G*}$, $P_{G** \rightarrow *}$, and $P_{G** \rightarrow **}$ in the last line are weighted with the Galaxy-model weights $w_G$ from Table~\ref{tab:weights}. The functional-baseline detection performance is not compatible with the minimum detection percentages defined in Table~\ref{tab:weights}.}\label{tab:baseline_results}
\begin{center}
\begin{tabular}[h]{cccccc}
\hline\hline
\\[-8pt]
$G$ range & $P_{G*}$ & $P_{G** \rightarrow *}$ & $P_{G** \rightarrow **}$ & $P_{G,\rm CR}$ & $P_{G,\rm SP}$\\
$[$mag$]$ & [\%] & [\%] & [\%] & [\%] & [\%] \\
\hline
\\[-8pt]
12.5--13.5& 100.000 &  99.917 &  99.188 &   8.584 &  15.781\\
13.5--14.5& 100.000 &  99.929 &  99.302 &  11.218 &  12.091\\
14.5--15.5& 100.000 &  99.936 &  99.263 &   4.589 &   6.346\\
15.5--16.5& 100.000 &  99.946 &  98.763 &  11.957 &   1.048\\
16.5--17.5& 100.000 &  99.886 &  99.027 &   2.780 &   0.454\\
17.5--18.5& 100.000 &  99.808 &  99.121 &   7.646 &   2.651\\
18.5--19.5&  99.999 &  99.326 &  98.660 &   3.871 &      --\\
19.5--20.0&  99.831 &  94.306 &  96.200 &      -- &      --\\
\hline
\\[-8pt]
12.5--20.0&  99.961 &  98.417 &  98.271 &   6.349 &   3.401\\
\hline
\end{tabular}
\end{center}
\end{table}

\begin{table}[t]
\caption{As Table~\ref{tab:baseline_results}, but for the optimised rejection parameters.}\label{tab:optimised_results}
\begin{center}
\begin{tabular}[h]{cccccc}
\hline\hline
\\[-8pt]
$G$ range & $P_{G*}$ & $P_{G** \rightarrow *}$ & $P_{G** \rightarrow **}$ & $P_{G,\rm CR}$ & $P_{G,\rm SP}$\\
$[$mag$]$ & [\%] & [\%] & [\%] & [\%] & [\%] \\
\hline
\\[-8pt]
12.5--13.5& 100.000 &  99.726 &  98.998 &   5.722 &  12.672\\
13.5--14.5& 100.000 &  99.713 &  99.303 &   7.387 &   8.692\\
14.5--15.5& 100.000 &  99.579 &  99.641 &   3.212 &   5.232\\
15.5--16.5&  99.997 &  99.550 &  99.775 &   9.645 &   1.424\\
16.5--17.5&  99.995 &  99.505 &  99.896 &   2.318 &   0.889\\
17.5--18.5&  99.997 &  99.869 &  99.978 &   6.543 &   3.011\\
18.5--19.5&  99.999 & 100.000 &  99.994 &   1.864 &      --\\
19.5--20.0&  99.995 & 100.000 &  99.994 &      -- &      --\\
\hline
\\[-8pt]
12.5--20.0&  99.997 &  99.866 &  99.928 &   5.276 &   3.064\\
\hline
\end{tabular}
\end{center}
\end{table}

\begin{figure*}[p!]
  \centering
  \includegraphics[width=0.92\textwidth]{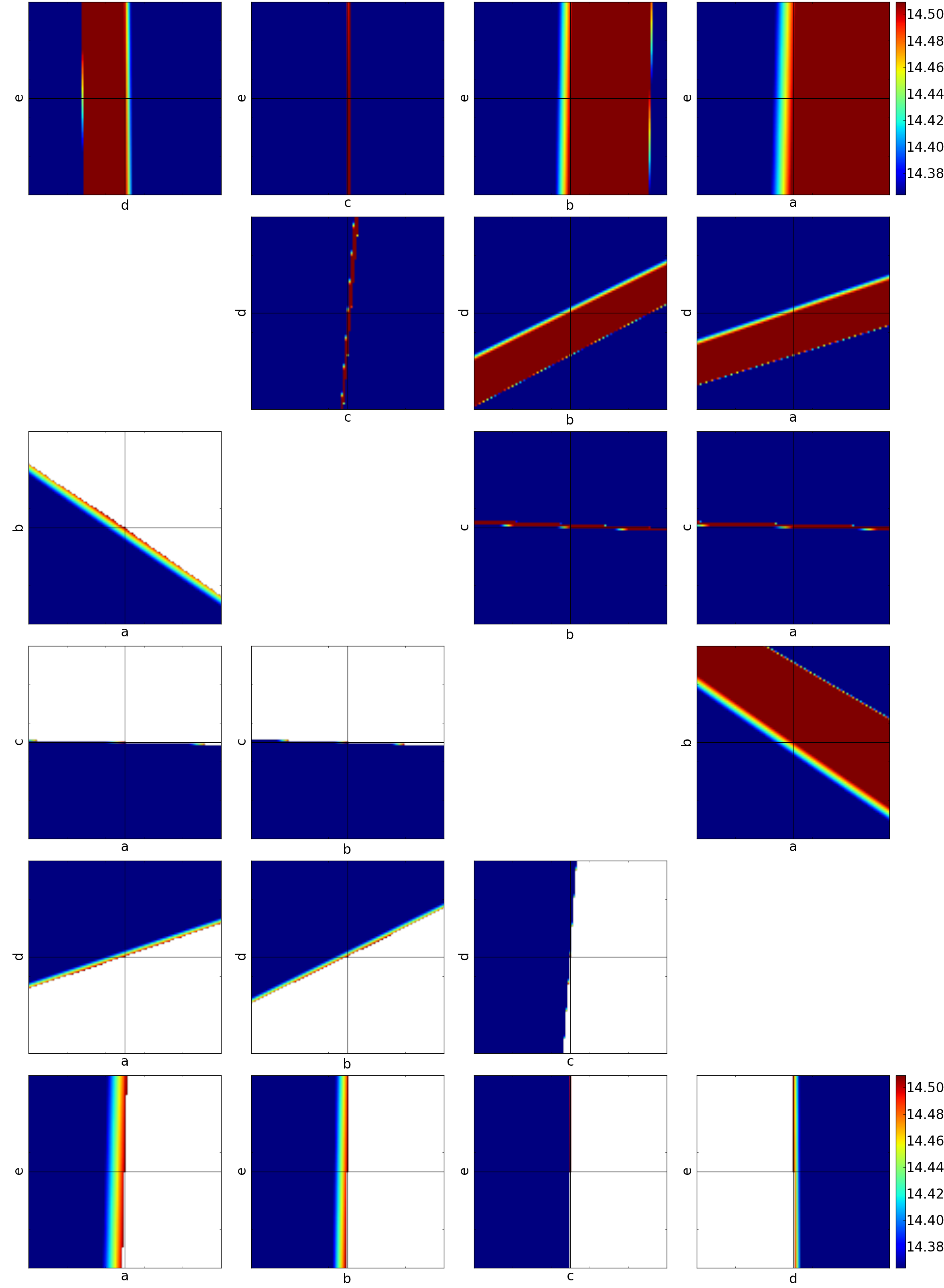}
  \caption{Contour plots, for the low-frequency, along-scan case, of all ten parameter combinations. The panels above the diagonal refer to the merit function $P(\mathbf{u})$ while the panels below the diagonal refer to the regularised merit function $P(\mathbf{u}) \cdot R_{*} \cdot R_{**}$. The panels are centred on the optimised parameter values (intersection of the black lines) and cover a range of $100$ for $a$, $b$, $c$, and $d$ and $500$ for $e$. White areas refer to parameter combinations which violate the minimum detection percentages defined in Table~\ref{tab:weights}.}
  \label{fig:contour_LFAL}
\end{figure*}

\onlfig{14}{
\begin{figure*}[p!]
  \centering
  \includegraphics[width=0.92\textwidth]{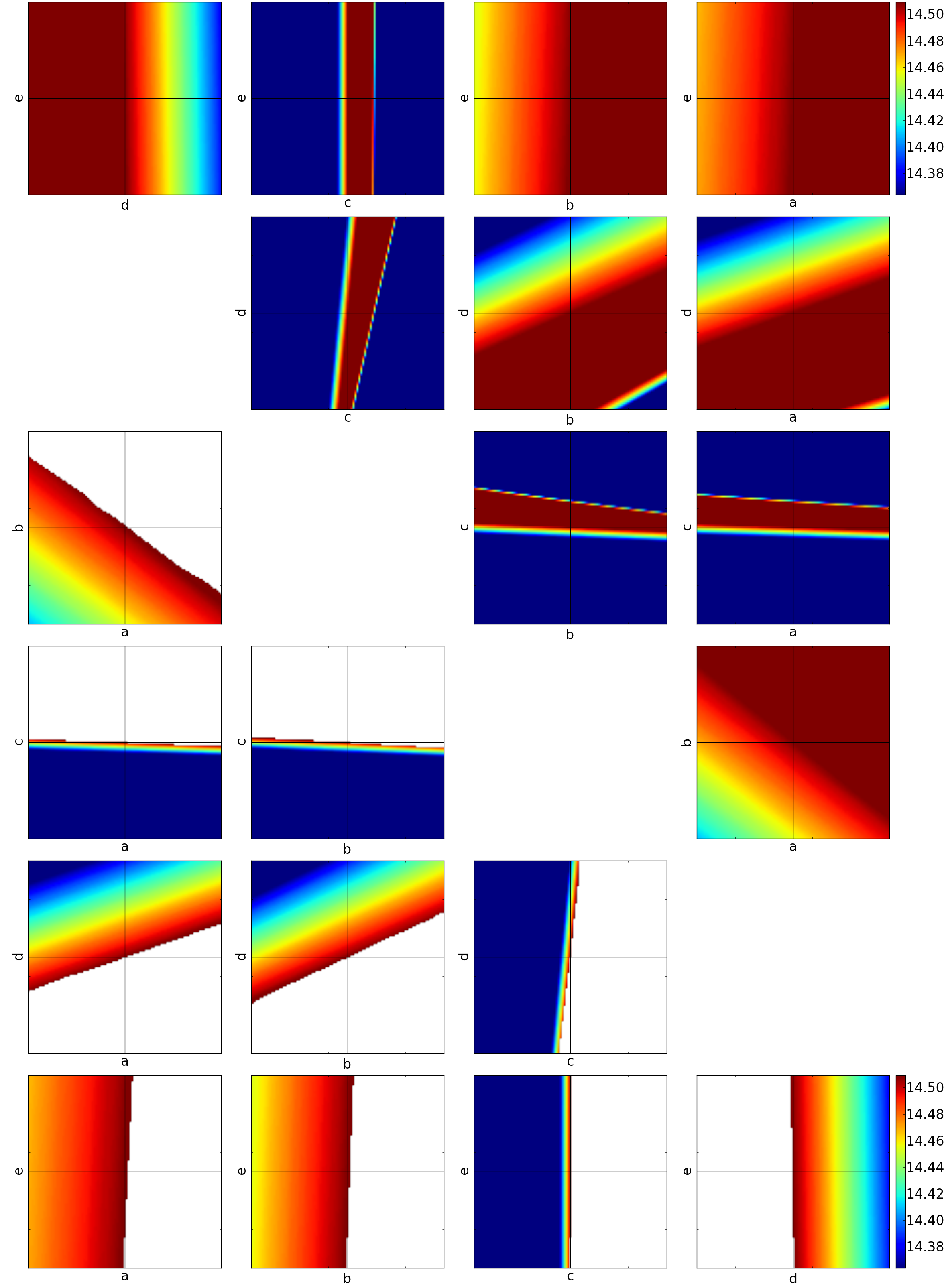}
  \caption{As Figure~\ref{fig:contour_LFAL}, but for the low-frequency, across-scan case.}
  \label{fig:contour_LFAC}
\end{figure*}}

\onlfig{15}{
\begin{figure*}[p!]
  \centering
  \includegraphics[width=0.92\textwidth]{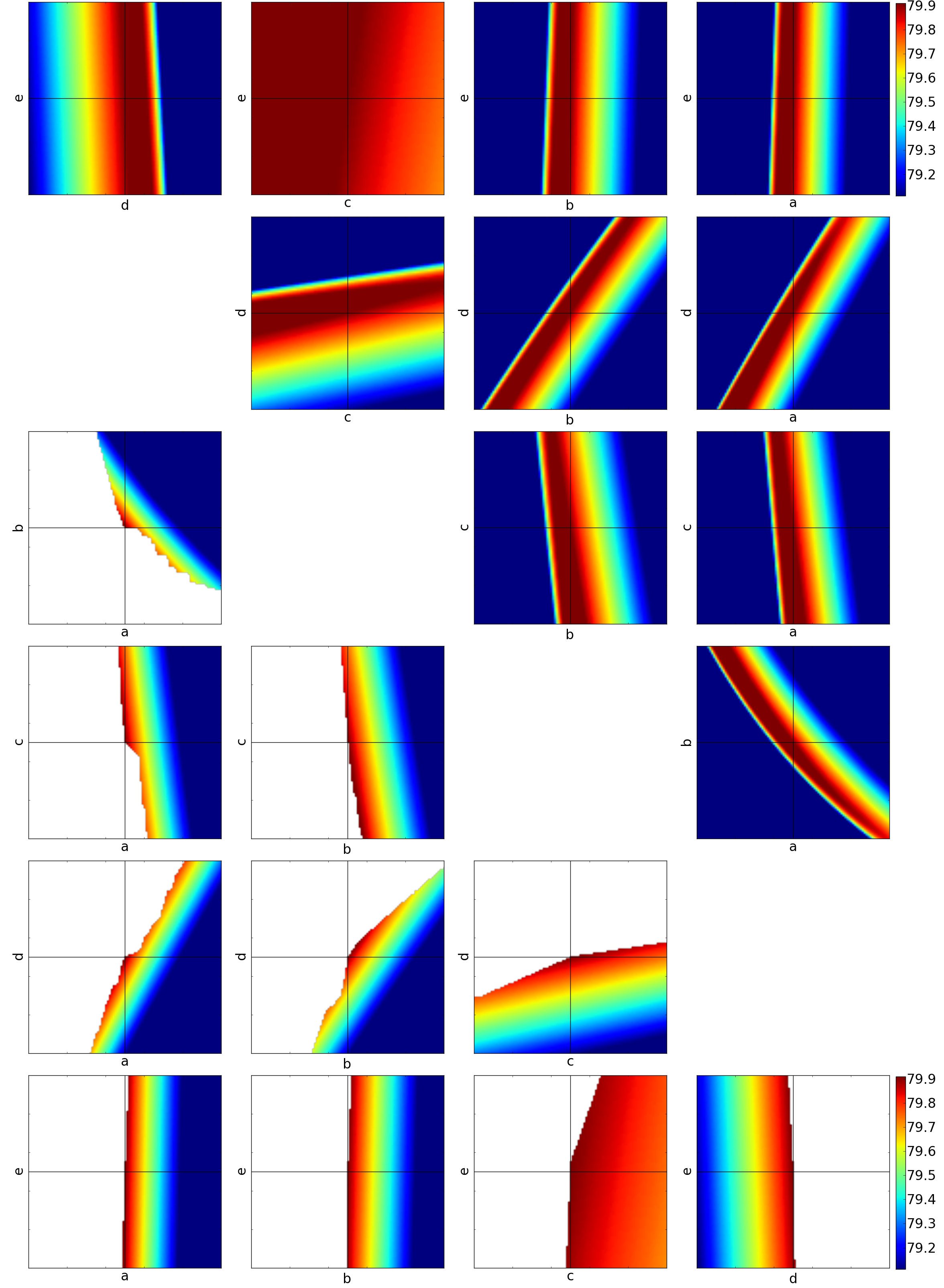}
  \caption{As Figure~\ref{fig:contour_LFAL}, but for the high-frequency, along-scan case.}
  \label{fig:contour_HFAL}
\end{figure*}
}

\onlfig{16}{
\begin{figure*}[p!]
  \centering
  \includegraphics[width=0.92\textwidth]{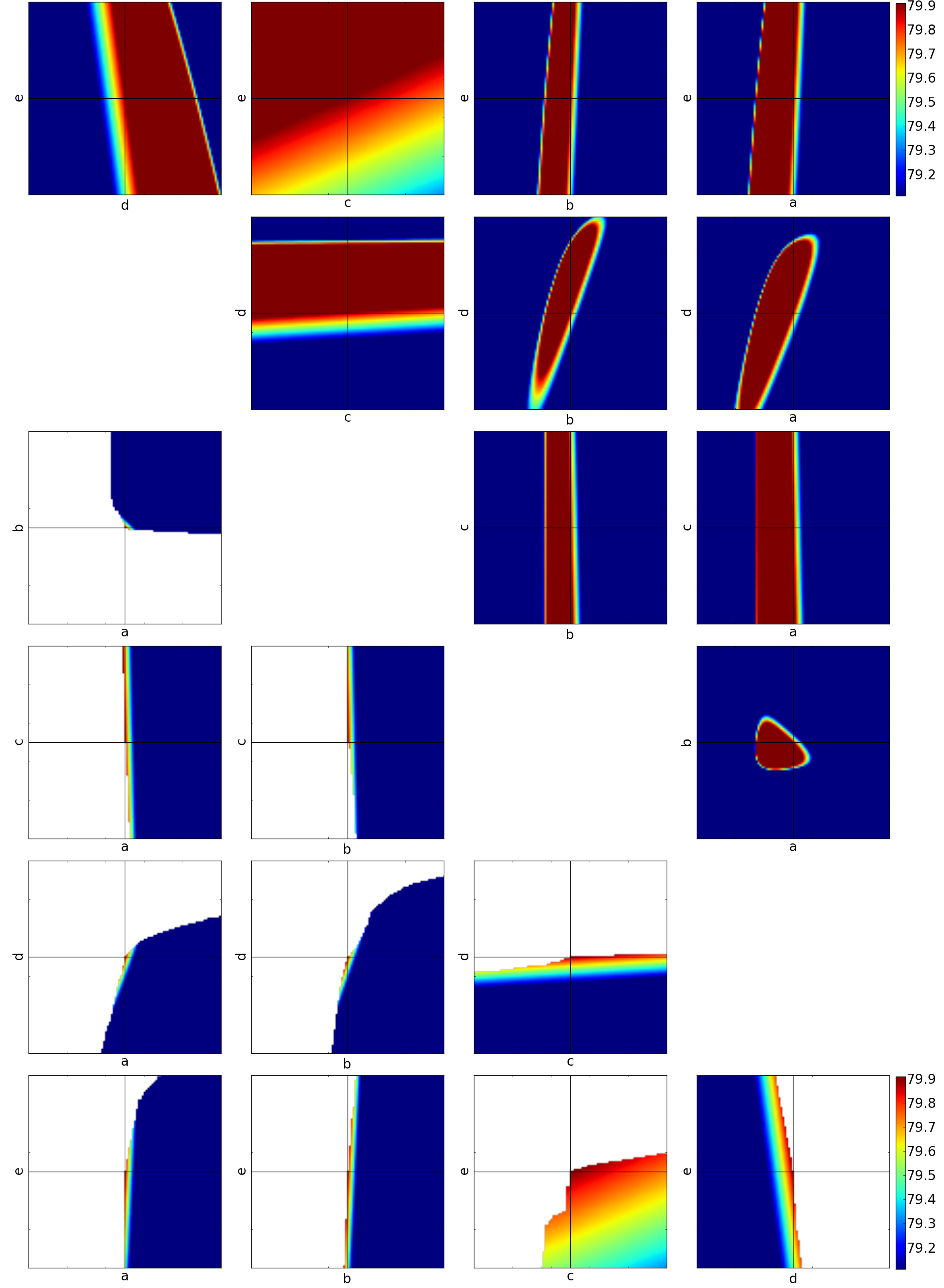}
  \caption{As Figure~\ref{fig:contour_LFAL}, but for the high-frequency, across-scan case.}
  \label{fig:contour_HFAC}
\end{figure*}
}

Figures~\ref{fig:contour_LFAL}--\ref{fig:contour_HFAC} provide two-dimensional contour plots of the merit function $P(\mathbf{u})$ and the regularised merit function $P(\mathbf{u}) \cdot R_{*} \cdot R_{**}$ for the various frequency-direction combinations. As one can clearly see by the presence of sharp boundaries, the regularisation -- introduced to maintain the star-detection probabilities above some minimum-acceptable thresholds (Table~\ref{tab:weights}) -- does influence the results. Without regularisation, better figures of merit could be obtained but such solutions would sacrifice either too many star detections to reduce prompt-particle-event detections or too many bright-star detections ($G \la 16$~mag) to improve faint-star performance ($G \ga 18$~mag). One can also see from the various panels that many parameters are correlated: the contour regions are often (strongly) elongated. This is not surprising since the rejection equations have been designed to offer coarse and fine adjustment \citep{LL:GAIA.ASF.MEM.PLM.00259}: roughly speaking, for a given flux level $F$, parameters $a$ and $b$ determine the values of the vertical and horizontal asymptotes, parameter $c$ determines the coarse position of the vertex of the rejection curve (Figure~\ref{fig:example_rejection_plot}), and parameters $d$ and $e$ can be used to fine-tune the vertex position. It is hence not surprising to see that the optimum values of $c$ are not too different from the functional baseline (low-frequency: $170$ and $160$ versus $155$ and $160$ for along and across scan, respectively; high-frequency: $2447$ and $2569$ versus $2667$ and $2667$ for along and across scan, respectively).

\section{Discussion}\label{sec:discussion}

\subsection{Solar protons}\label{subsec:protons}

As already explained in Section~\ref{subsec:PPEs}, the Solar-proton rate varies with time from essentially zero during Solar-quiet times to extremely high fluxes during Solar flares. In practice, however, the Sun behaves bi-modally: it is either ``quiet'', i.e., not emitting protons, or ``bursting'', i.e., emitting such a high proton flux that Gaia's star trackers are blinded, the spacecraft goes into transition mode, and scientific-data collection is suspended. In-between states do not really exist, except for the very short, intermittent states corresponding to the rise and fall (onset and offset) of Solar flares. One might therefore argue that Solar protons (i.e., the factor $[1 - P_{\rm SP}(\mathbf{u})]$) should not be included in the merit function, Equation~(\ref{eq:merit_function}). In practice, however, the inclusion or exclusion of protons in the merit function does not significantly affect the results of the optimisation since the shape and magnitude distributions of protons resembles those of cosmic rays. We, somewhat arbitrarily, decided to include a Solar-proton factor in the merit function.

\subsection{Secondary particles}

Whilst shielding a CCD will stop a fraction of the (low-energy) prompt-particle events, excessively thick shielding will introduce a flux (``shower'') of secondary particles created by the electromagnetic interaction at nuclear level between the primary particles (i.e., protons and Helium nuclei) and the shielding. However, these secondaries only become significant for shield thicknesses in excess of $\sim$$10$~cm of Aluminium \citep[e.g.,][]{1993ITNS...40.1628D}. For Gaia, the effective shielding thickness is at the level of $11$~mm of Aluminium, implying that secondary particles are (likely) not significant. Nonetheless, an exploratory study for Gaia has been made by the Space Environments and Effects Analysis Section in the Technical Directorate of the European Space Agency (G.~Santin, 2009, private communication), assuming, as a very worst case, CCD shieldings of $2.0$~cm of Aluminium plus $1.5$~cm of SiC from the back and $3.5$~cm of glass from the front. Various nuclear collision processes (hadronic models, quark gluon strings, and binary cascades) have been considered and the results indicate that secondaries will not be present in large numbers. Our prompt-particle-event catalogues are hence adequate for Gaia.

\subsection{AF1 confirmation}\label{subsec:AF1_confirmation}

As explained in Section~\ref{subsec:nomenclature}, the confirmation criterion is purely flux-based: the ``PSF shape'' of the confirmed object is not tested. In the frame of this investigation, we can thus ignore the AF1 confirmation stage. Basically, this stage is only useful to prevent Solar protons and cosmic rays which accidentally pass the SM detection stage (as well as the pre-selection and resource-allocation stages) from being windowed throughout the focal plane. This is beneficial only for the volume of science data transmitted to ground: prompt-particle events which pass the SM filter and are only uncovered in AF1 as spurious detection do already occupy a window at that stage and this resource allocation is irreversible, i.e., this window can be dropped but cannot be given back to observe one more star. In other words: the number of prompt-particle-event detections shall really be minimised at SM-detection level.

\subsection{The impact of non-rejected ghosts}\label{subsec:discussion_ghosts}

\begin{figure*}[t!]
  \centering
  \includegraphics[width=\textwidth]{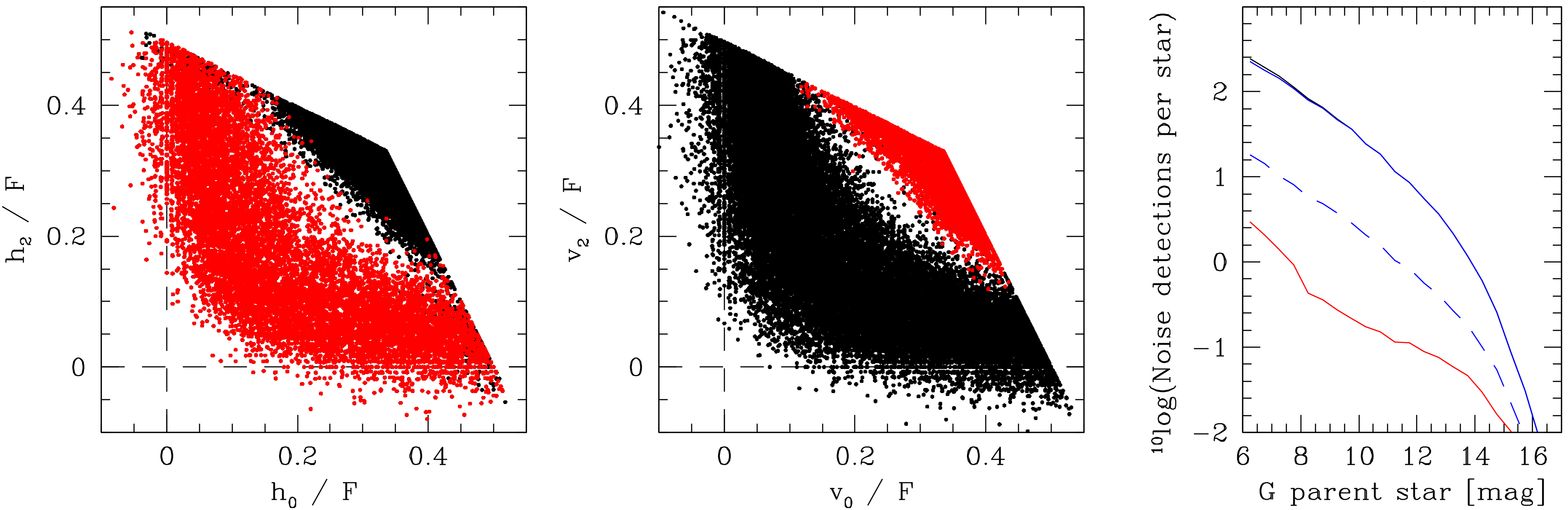}
  \caption{Properties of ghosts brighter than $G = 20$~mag originating from single stars in the range $G = 6$--$20$ mag. {\it Panel 1}: across-scan rejection plot of ghosts, showing ghosts in the across-scan wings as red symbols and ghosts in the along-scan wings as black symbols. {\it Panel 2}: as {\it Panel 1}, but for the along-scan direction. {\it Panel 3}: average number of ghosts that a (parent) star of magnitude $G$ generates. This is essentially the same plot as {\it Panel 3} in Figure~\ref{fig:ghosts1} but for an extended magnitude range of parent stars and displayed on a logarithmic scale. The black curve denotes all local maxima associated with ghosts, the blue curve -- mostly overlying the black curve -- denotes those local maxima passing the optimised rejection parameters defined in Section~\ref{sec:results}, and the red curve denotes those local maxima passing the functional-baseline rejection parameters. The dashed blue curve represents the alternative set of optimised parameters with reduced ghost sensitivity, as discussed in Section~\ref{subsec:discussion_ghosts}.}\label{fig:ghosts2}
\end{figure*}

Section~\ref{sec:results} already concluded that a byproduct of the optimised rejection parameters is the increased sensitivity to ghosts: whereas the functional baseline only lets $1.800$\% of the ghosts generated by parent stars in the range $G = 12.5$--$20$~mag through, this increases to $99.866$\% for the optimised parameters. To put these numbers in perspective, we extend the analysis presented in Section~\ref{subsec:ghosts} (and Figure~\ref{fig:ghosts1}), which is based on parent stars in the range $G = 12.5$--$20$~mag, to parent stars covering the range $G = 6$--$20$ mag. Panel~3 in Figure~\ref{fig:ghosts2} shows that stars as bright as $G = 6$~mag typically generate more than $250$ ghosts on an SM-CCD transit; their magnitude distribution is essentially the same as found in Section~\ref{subsec:ghosts}, strongly peaking at $G = 20$~mag. Panels~1 and 2 show that ghosts occupy very distinct areas in the rejection plots: ghosts in the across-scan wings resemble stars in the across-scan rejection plot and resemble ripples in the along-scan rejection plot. Similarly, ghosts in the along-scan wings resemble stars in the along-scan rejection plot and resemble ripples in the across-scan rejection plot. Whereas the functional-baseline rejection parameters stop the vast majority of ghosts, the optimised rejection parameters do the opposite and let most of them through.

After folding the average number of ghosts that a parent star generates with the total number of stars of that magnitude in the sky, it is easy to compute the increase in the number of detected ``sources'' caused by ghosts: it is $+12.98$\% for the optimised parameters, versus $+0.22$\% for the functional baseline. These numbers, however, are by far not the full story:
\begin{itemize}
\item As explained in Section~\ref{subsec:AF1_confirmation}, ghost detections made in SM need to pass the (flux-based) confirmation stage in AF1 before being assigned a window in AF and before being telemetered to ground. Since the noise pattern in AF1 will differ from that in SM, not all ghost detections will be confirmed in AF1. GIBIS simulations suggest, however, that the majority of ghosts detections are confirmed in AF1.
\item In addition to increasing telemetry, ghost detections can be harmful for the faint-end transit (and catalogue) completeness -- in particular in dense areas -- since they occupy resources (windows), the number of which in AF is limited to $W = 20$ at each TDI line (see Section~\ref{subsec:nomenclature}). One particular aspect with the ghosts is that ghost detections in the across-scan wings have the same along-scan (TDI) coordinate, and hence compete mutually -- as well as with the parent star and with other stars that happen to be present at that along-scan position in the same CCD -- for the $W = 20$~resources available per TDI line (in addition, there is the pre-selection limitation of $5$~detections per TDI line that can enter the resource allocation; Section~\ref{subsec:nomenclature}). Ghost detections in the along-scan wings, on the other hand, do typically not mutually compete for resources but ``only'' compete with other stars. Nonetheless, the overall conclusion is that the number of ghost detections shall preferably be minimised at SM-detection level.
\end{itemize}

As a result, we performed some experiments to find solutions for the rejection parameters which improve upon the functional-baseline results for what regards single- and double-star detections but which do not let through a large percentage of ghosts. This proved possible but not without penalty (see Table~\ref{tab:alternative_results}): the ghost-detection probability dropped from $99.866$\% to $10.841$\% (cf.\ $1.800$\% for the functional baseline) while, at the same time, the star detection probabilities improved from $99.961$\% to $99.986$\% for $P_{*}$ (optimised: $99.997$\%), from $98.417$\% to $99.221$\% for $P_{** \rightarrow *}$ (optimised: $99.866$\%), and from $98.271$\% to $99.119$\% for $P_{** \rightarrow **}$ (optimised: $99.928$\%); however, this performance increase was achieved at the expense of reduced prompt-particle-event rejection capabilities: $P_{\rm CR}$ increased from $6.403$\% to $11.186$\% (optimised: $5.276$\%) while $P_{\rm SP}$ increased from $3.401$\% to $5.219$\% from (optimised: $3.064$\%).

\begin{table}[t]
\caption{As Table~\ref{tab:optimised_results}, but for the alternative set of optimised rejection parameters which sacrifice object-detection and prompt-particle-event rejection performance to improve the ghost rejection performance (see Section~\ref{subsec:discussion_ghosts}).}\label{tab:alternative_results}
\begin{center}
\begin{tabular}[h]{cccccc}
\hline\hline
\\[-8pt]
$G$ range & $P_{G*}$ & $P_{G** \rightarrow *}$ & $P_{G** \rightarrow **}$ & $P_{G,\rm CR}$ & $P_{G,\rm SP}$\\
$[$mag$]$ & [\%] & [\%] & [\%] & [\%] & [\%] \\
\hline
\\[-8pt]
12.5--13.5& 100.000 &  99.951 &  99.528 & 10.443  & 18.943 \\
13.5--14.5& 100.000 &  99.972 &  99.679 & 13.523  & 14.012 \\
14.5--15.5& 100.000 &  99.970 &  99.703 &  5.231  &  6.951 \\
15.5--16.5& 100.000 &  99.982 &  99.373 & 15.586  &  2.139 \\
16.5--17.5& 100.000 &  99.945 &  99.537 &  9.784  &  4.706 \\
17.5--18.5& 100.000 &  99.905 &  99.552 & 11.339  &  5.421 \\
18.5--19.5& 100.000 &  99.664 &  99.327 &  7.312  &     -- \\
19.5--20.0&  99.939 &  97.202 &  98.018 &     --  &     -- \\
\hline
\\[-8pt]
12.5--20.0&  99.986 &  99.221 &  99.119 &  11.186 &   5.219\\
\hline
\end{tabular}
\end{center}
\end{table}

\subsection{Robustness}\label{subsec:robustness}

One may ask how robust the optimised parameters are to, for instance, radiation-damage effects, noise, PSF-shape (changes), etc. It is important in this respect to recall the following:
\begin{enumerate}
\item We have deliberately chosen to define one set of rejection parameters applicable to all VPUs, despite the available degree of freedom, which could improve the detection performance further, to define optimised parameter sets for each VPU separately, or even separately for SM1 and SM2 within each VPU. By design, therefore, the optimised parameters cover the large(st possible) variety of wave-front errors and PSF shapes, which clearly improves their robustness;
\item The figure-of-merit contour plots (Figures~\ref{fig:contour_LFAL}, \ref{fig:contour_LFAC}, \ref{fig:contour_HFAL}, and \ref{fig:contour_HFAC}) show extended ``good regimes'' and gradual changes, suggesting that the detection / rejection performance is not sensitive to modest changes in the parameter values, provided that the correlations between parameters are~considered;
\item The detection algorithm is not strongly sensitive to sky-background or straylight, which manifest themselves as a constant background which is eliminated through the background-subtraction step. Clearly, enhanced sky-background or stray-light levels would induce extra noise. However, Figure~\ref{fig:performance@G=20mag} shows that the optimised parameters provide excellent performance not only at $G = 20$~mag (where the signal-to-noise ratio is $21$) but at least down to $G = 21$~mag, i.e., when the signal-to-noise ratio has dropped to $13$. Such a drop would also correspond, for instance, to a straylight level of 14 electrons per pixel per second;
\item The detection algorithm works on integrated fluxes contained in samples, consisting of $2\times2$-binned pixels, of size $0.12 \times 0.36$~arcsec. It merely inspects the PSF shape based on gross, zeroth-order quantities, namely the flux-vector elements $h_0$, $h_1$, and $h_2$ (and $v_0$, $v_1$, and $v_2$) and the total flux $F$. There is hence no strong sensitivity of the on-board detection to fine (milli-arcsecond-level) features in the PSF;
\item The robustness of the detection algorithm to CCD degradation from non-ionising radiation damage (see Footnote~4) has been evaluated by Airbus Defence \& Space in a dedicated laboratory test campaign using a CCD connected to the VPAs \citep{LL:GAIA.ASF.TCN.PLM.00709}. These tests have demonstrated that (long-term) PSF-shape changes induced by radiation damage, even for a worst-case, end-of-mission-accumulated radiation dose, are much more subtle and negligible compared to (short-term) PSF ``changes'' caused by, for instance, across-scan motion, sub-pixel location, etc.
\end{enumerate}
In short, we believe the optimised parameters are robust against (instrumental) effects not included in our assessment.

\subsection{The real Gaia mission}\label{subsec:real_mission}

The study presented in this work shows that, compared to the functional-baseline rejection parameters, room for improvement exists if a different trade-off is applied: a better detection of celestial objects and rejection of cosmic rays and Solar protons at the expense of more ghosts. The alternative parameters represent an intermediate option. These results will be validated in orbit during the commissioning phase of the mission by means of a four-day test in which the functional-baseline rejection parameters will be used during $24$ hours, followed by $24$ hours of operations with the optimised parameters from Section~\ref{sec:results}, followed by $24$ hours of operations with the alternative set of optimised parameters from Section~\ref{subsec:discussion_ghosts}, followed by $24$ hours of operations with a set of rejection parameters which effectively do not filter any local maximum. During this test, the spacecraft will be operated in ecliptic-pole-scanning mode; this guarantees that each telescopes scans each of the ecliptic poles on each 6-hour revolution, with only a precession in ecliptic longitude. The ``continuously-observed'' ecliptic-pole regions, the stellar content of which has been carefully observed from the ground prior to the launch through dedicated efforts by DPAC, constitute the best-available benchmark against which the Gaia-detection performance can be assessed, although the ground-based observations are limited in/by spatial resolution, bandpass-transformation errors, (unrecognised) variable stars, (unresolved) double stars, star-galaxy classification errors, etc. In addition, one should keep in mind that the ecliptic poles just represent two particular density regimes (low density for the north and average density for the south pole) which are not representative of dense areas such as the Galactic bulge. Both prompt-particle events and ghosts can be harmful, in particular in dense areas where all resources are needed by stars, albeit prompt-particle events and ghosts differ in the sense that cosmic rays and Solar protons are always brighter than $G \sim 19$~mag whereas ghosts are always fainter than $\sim$$19$~mag: prompt-particle events hence compete with bright stars whereas ghosts compete with faint stars. In areas with sufficient resources (the vast majority of the sky in terms of area), such that additional ghosts or prompt-particle events can be supported, the preference is clearly to have a good star-detection efficiency. Spurious detections are, and only if confirmed in AF1, ``just'' a nuisance for the data processing in such cases. This, however, is not necessarily true in dense areas. Based among others on the outcome of the in-orbit test, a decision will be made on which parameter set will be flown during nominal operations. The final trade-off and decision, however, is beyond the scope of this work, which mainly aims to put the various elements that go into the trade-off onto the table. Further calibration and adjustment of the rejection parameters remains always possible during the~mission.

\section{Scientific implications}\label{sec:implications}

After having established optimised rejection parameters, it is interesting to assess their benefit for the science return of Gaia. Since it goes without saying that the improved performance of the single-star detection, in particular around $G = 20$~mag, will be beneficial to science, we focus on three other areas, namely double stars (Section~\ref{subsec:scientific_results_double_stars}), unresolved galaxies (Section~\ref{subsec:scientific_results_galaxies}), and asteroids (Section~\ref{subsec:scientific_results_asteroids}).

\subsection{Double stars}\label{subsec:scientific_results_double_stars}

\begin{figure}[t!]
  \centering
  \includegraphics[trim=45 13 77 39,clip,width = \columnwidth]{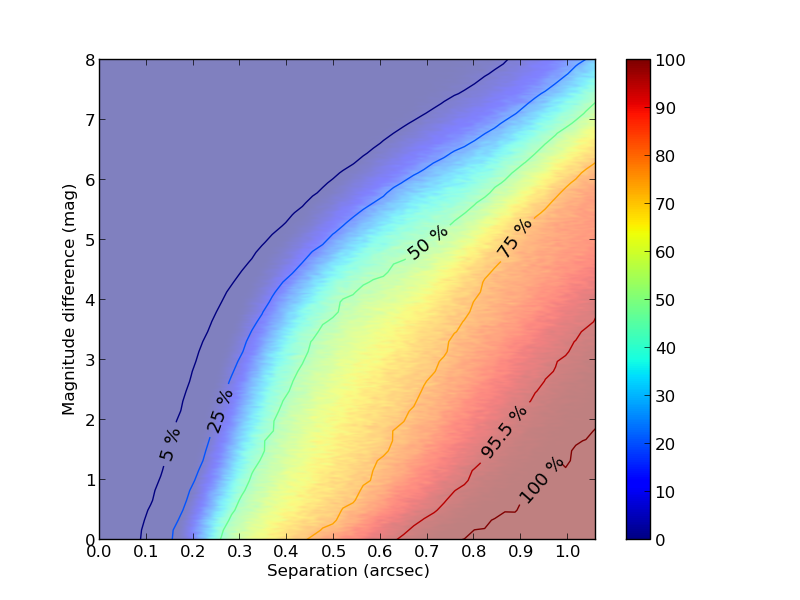}
  \caption{Probability (in \%) of a double star with a primary component with $G=13$~mag to be resolved into two local maxima as function of separation $\rho$, in units of arcsec, and magnitude difference $\Delta G$ (averaged over all orientation angles). This probability does not take the rejection curves into account; in practice, however, the results do not significantly change when applying the optimised rejection parameters. Contours at $5$\%, $25$\%, $50$\%, $75$\%, $95.5$\%, $98$\%, and $100$\% have been labeled.}\label{fig:double_stars_resolution}
\end{figure}

\begin{figure}[t!]
  \centering
  \includegraphics[trim=38 11 89 39,clip,width = \columnwidth]{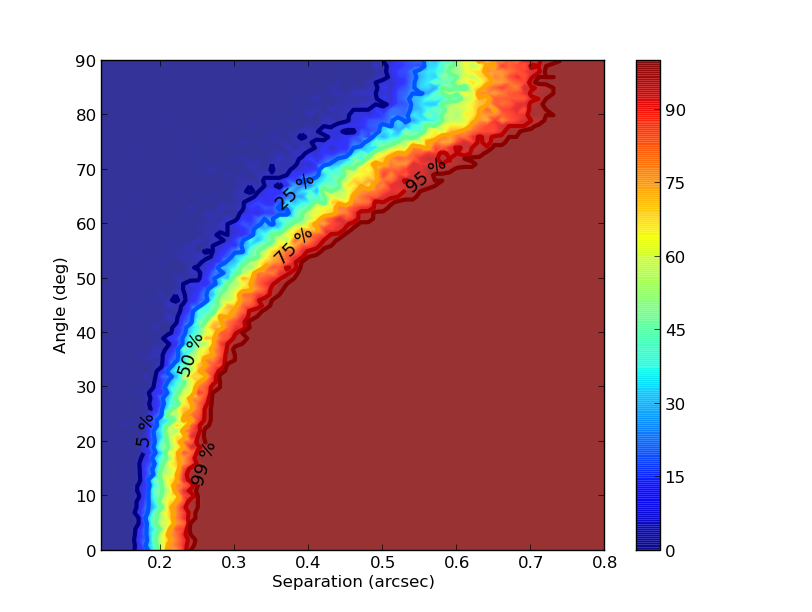}
  \caption{Probability of an equal-brightness double star ($\Delta G = 0$~mag) to be resolved into two local maxima as function of separation, in units of arcsec, and orientation angle, in units of degrees. The $G$ magnitude of the primary has been fixed at $G = 13$~mag. This probability does not take the rejection curves into account; in practice, however, the results do not significantly change when applying the optimised rejection parameters. Contours at $5$\%, $25$\%, $50$\%, $75$\%, $95$\%, and $99$\% have been labeled.}\label{fig:double_stars_probability}
\end{figure}

\begin{figure*}[t]
  \centering
  \includegraphics[width = 0.49\textwidth]{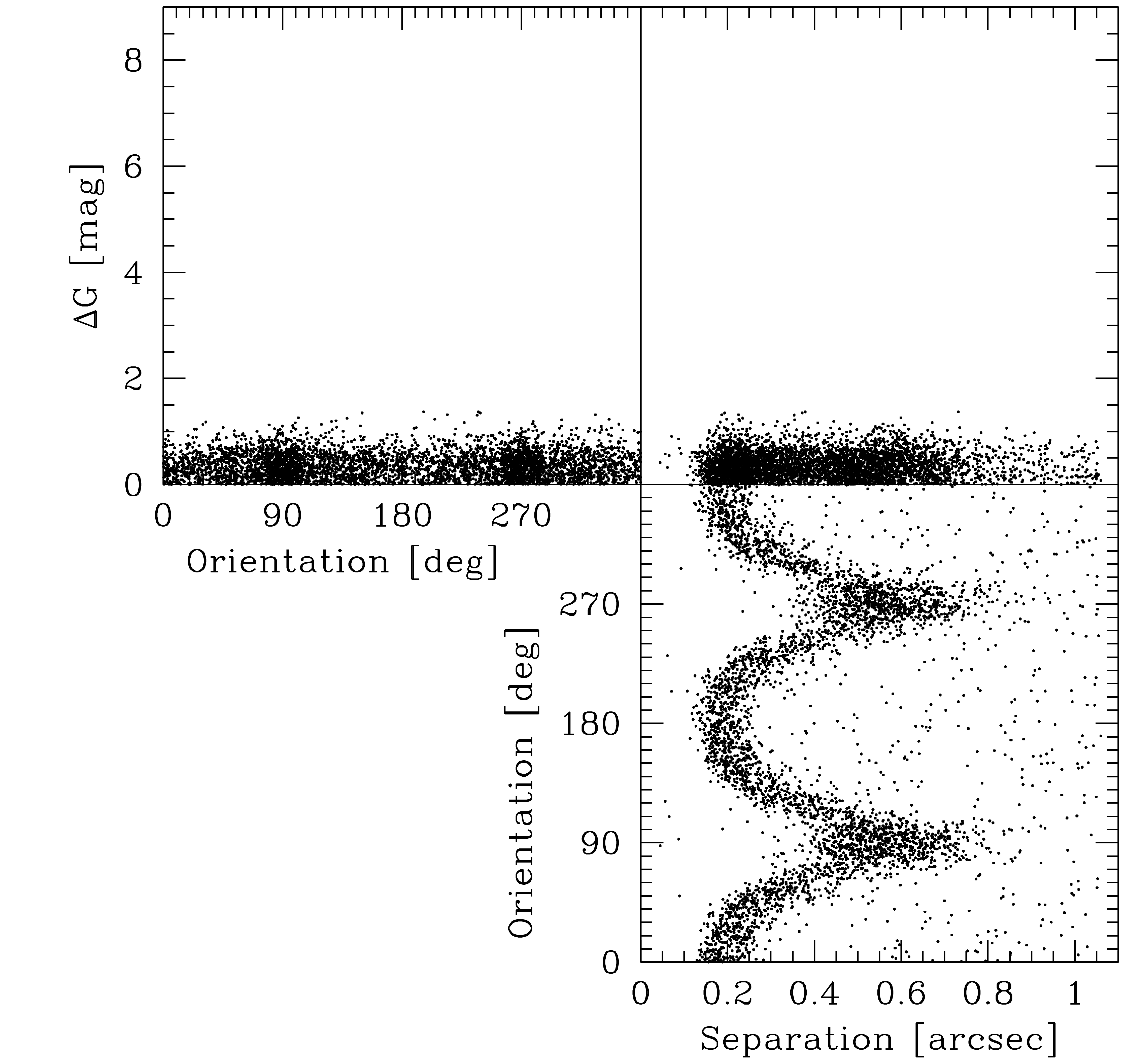}
  \includegraphics[width = 0.49\textwidth]{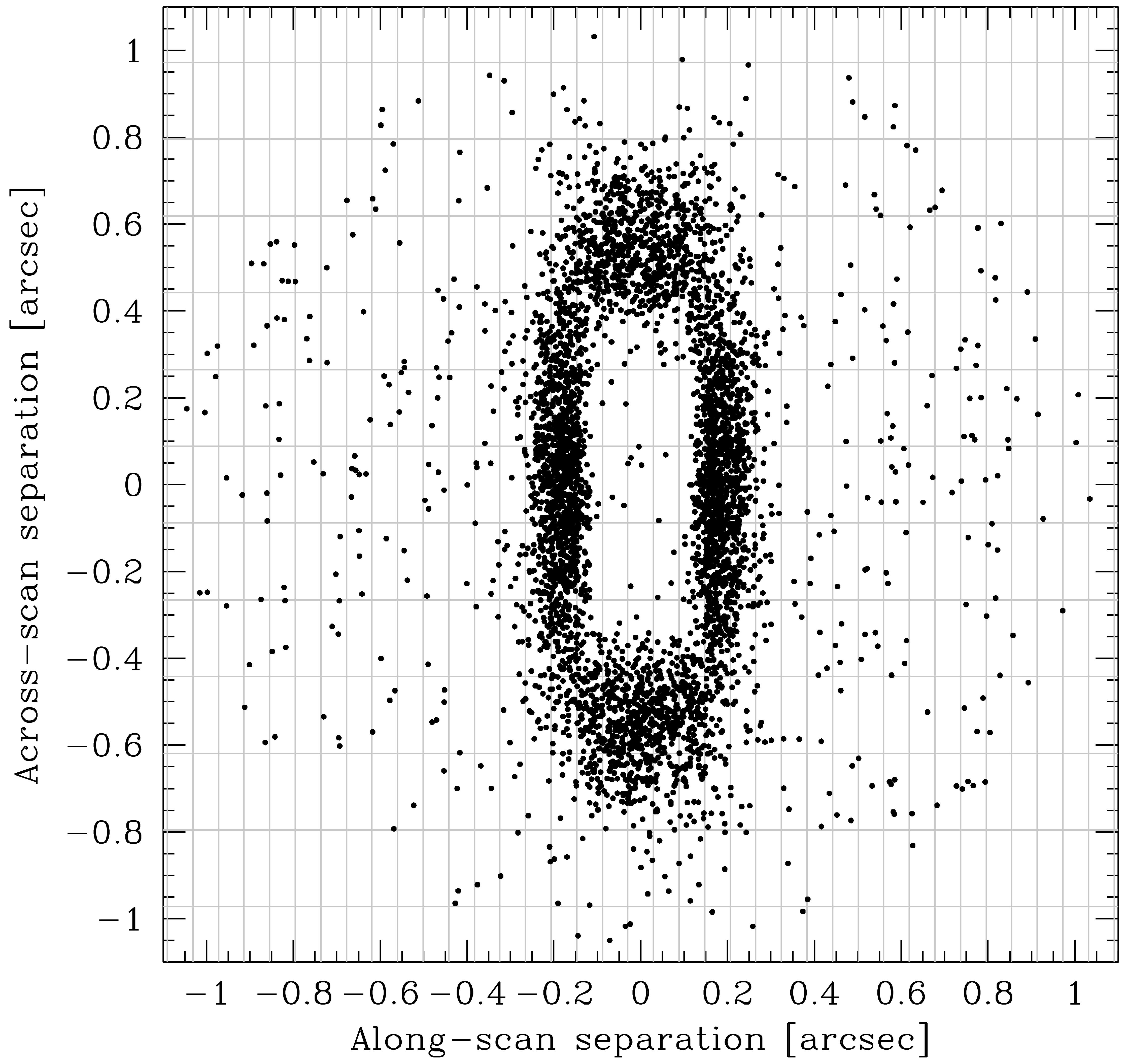}
  \includegraphics[width = 0.49\textwidth]{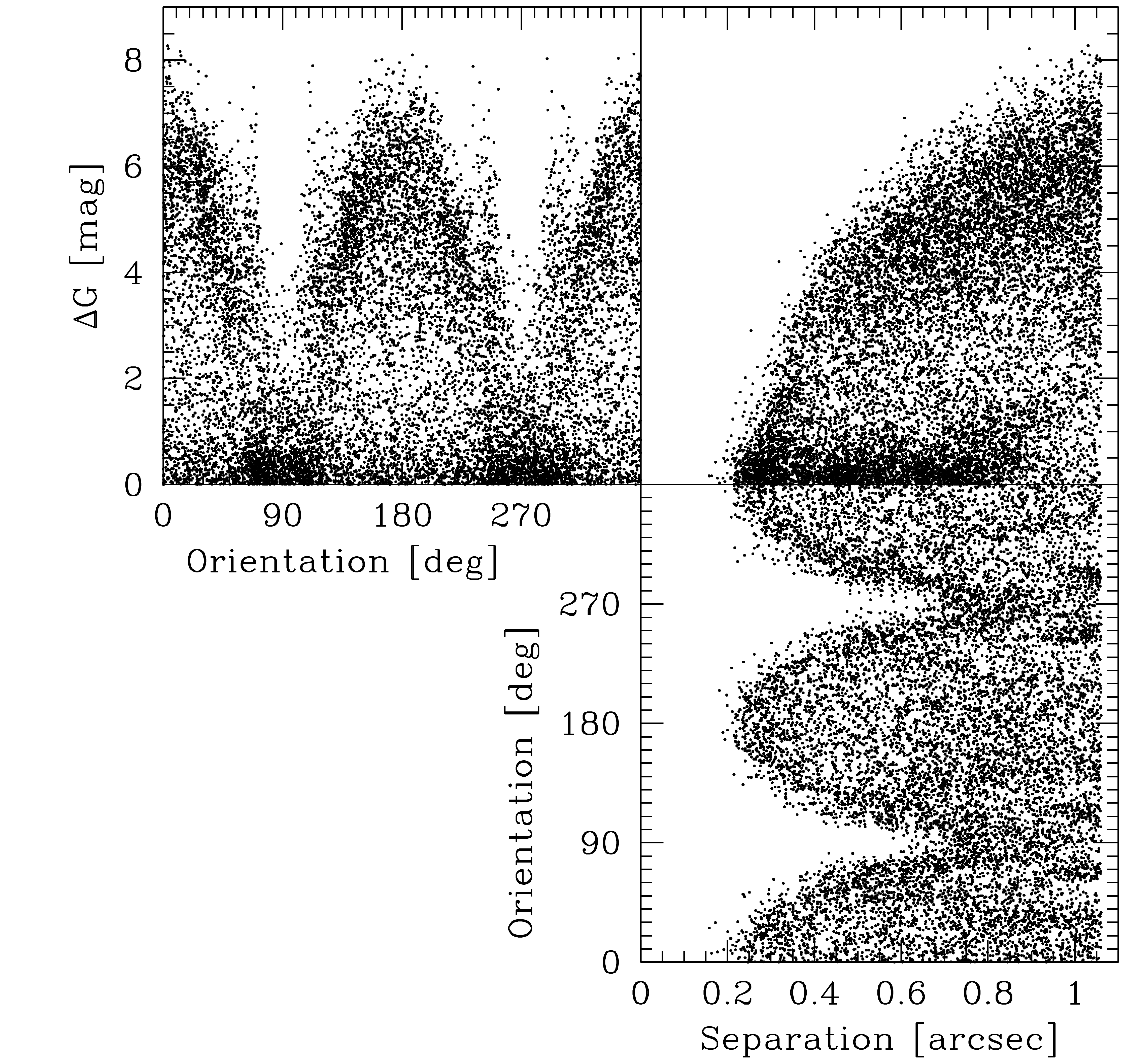}
  \includegraphics[width = 0.49\textwidth]{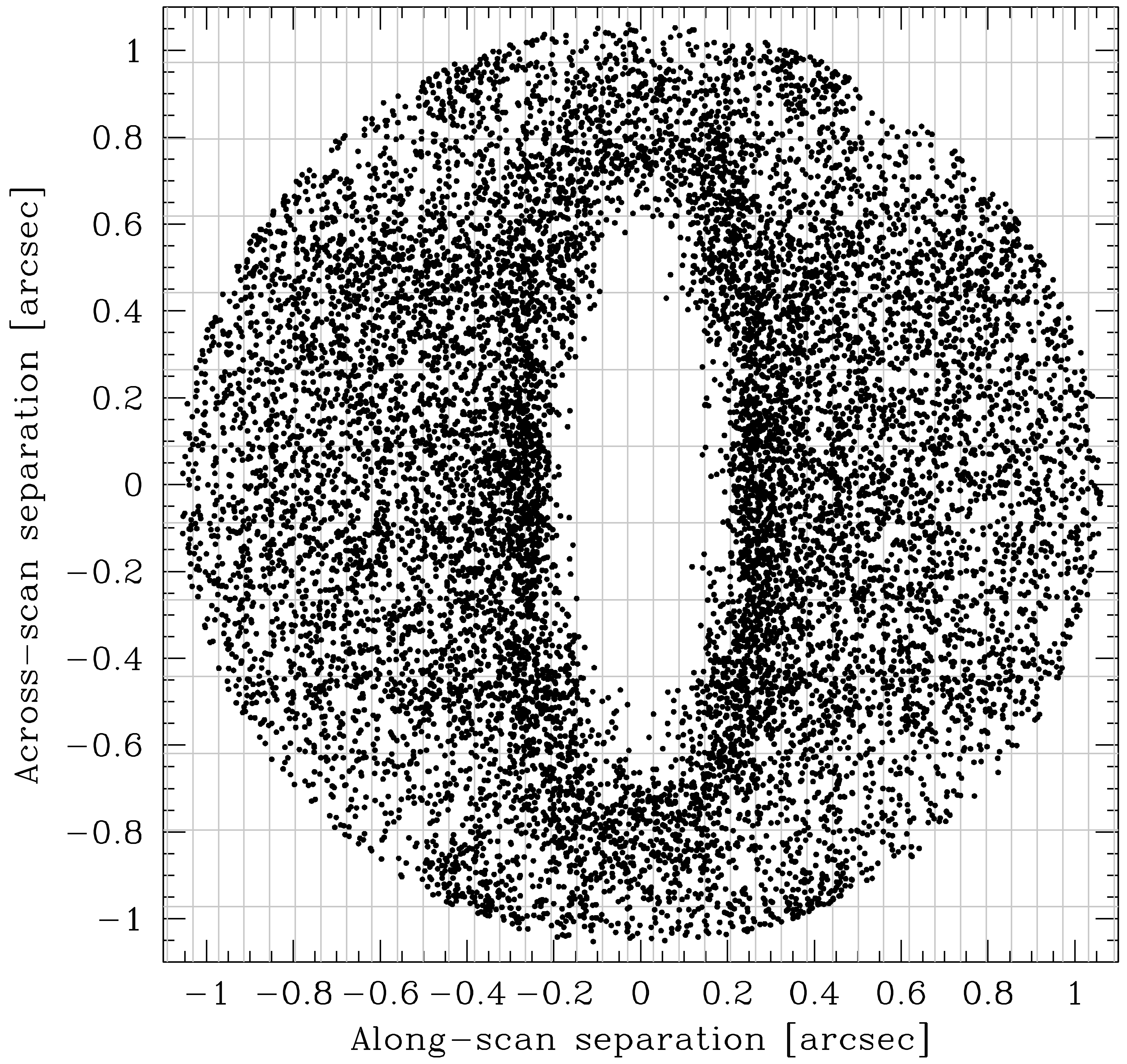}
  \caption{{\it Left-hand-side panels:} configuration of rejected double stars leading either to a single local maximum (i.e., rejected unresolved double stars; top panel; $4631$ objects), or to a local maximum for both components individually (i.e., rejected secondary components of resolved double stars; bottom panel; $13,376$ objects) for the functional-baseline rejection parameters. {\it Right-hand-side panels:} orientation, in along- versus across-scan ``coordinates'', of rejected objects leading either to a single local maximum (top panel), or to two local maxima (bottom panel). The dots represent the orientation of the secondary components with respect to their primary components, which are situated at the origin. The lines denote the CCD pixel grid; recall that Gaia's detection is based on SM samples, composed of $2 \times 2$-binned pixels.}\label{fig:double_stars_baseline}
\end{figure*}

\begin{figure*}[t]
  \centering
  \includegraphics[width = 0.49\textwidth]{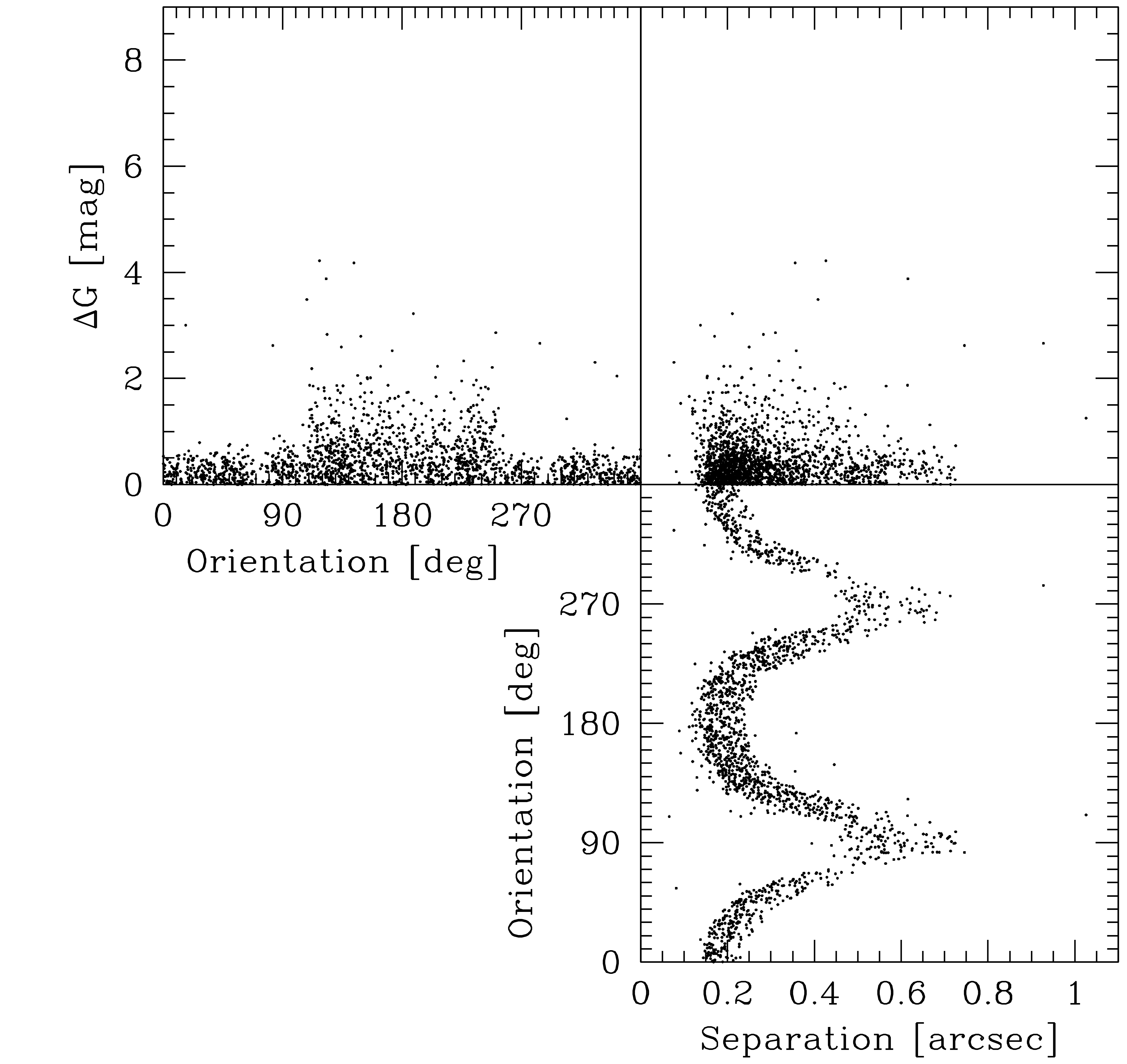}
  \includegraphics[width = 0.49\textwidth]{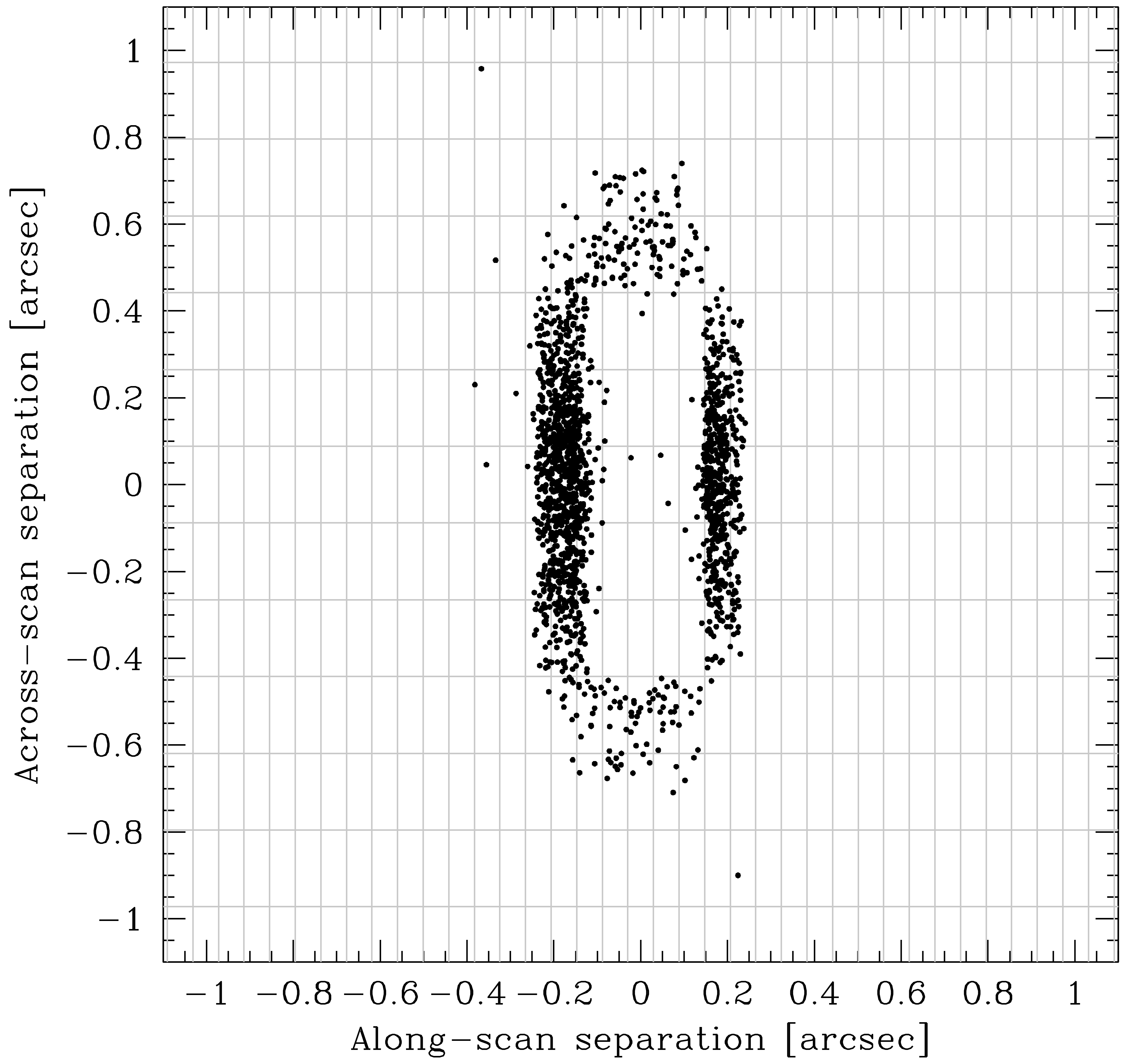}
  \includegraphics[width = 0.49\textwidth]{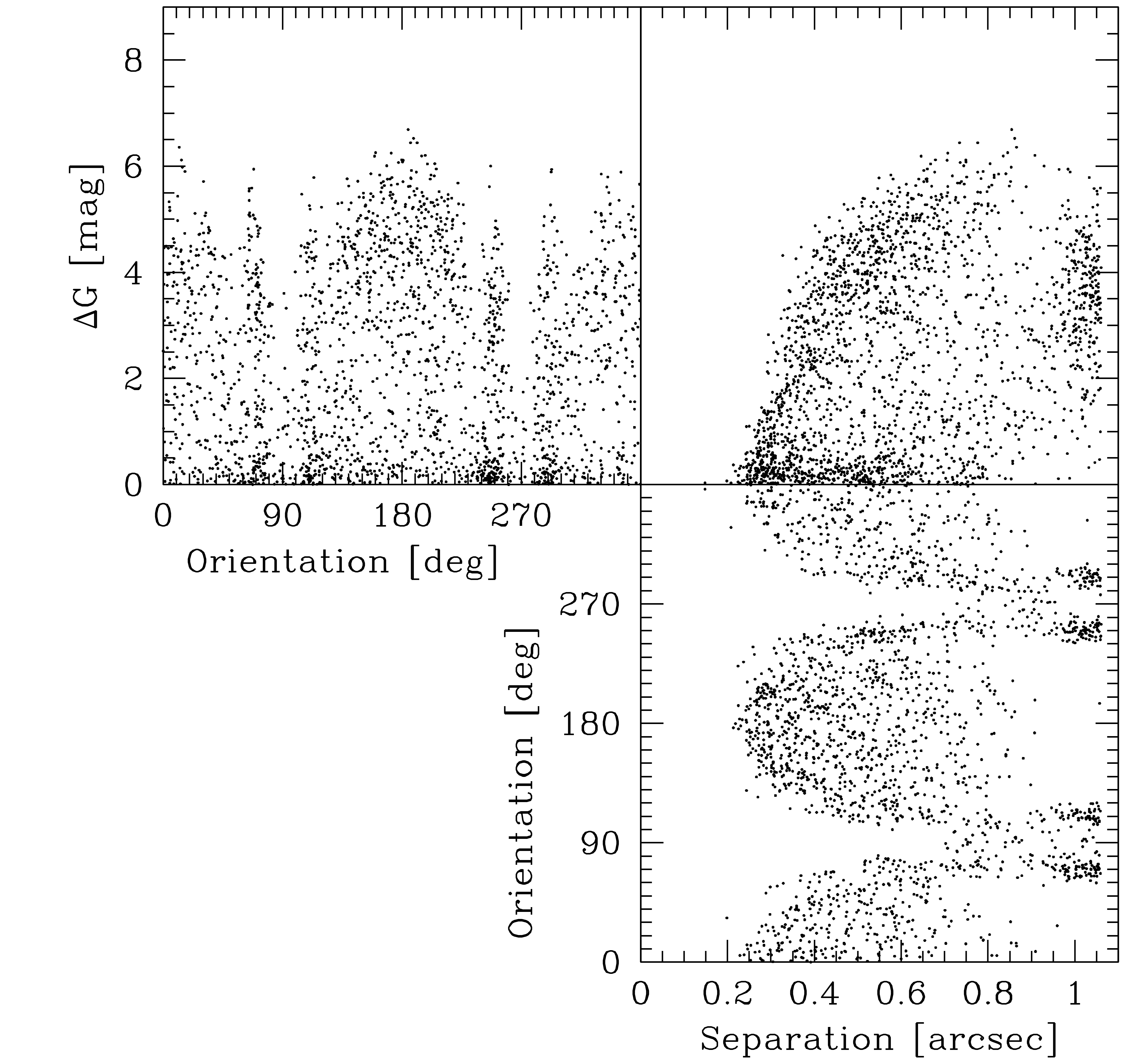}
  \includegraphics[width = 0.49\textwidth]{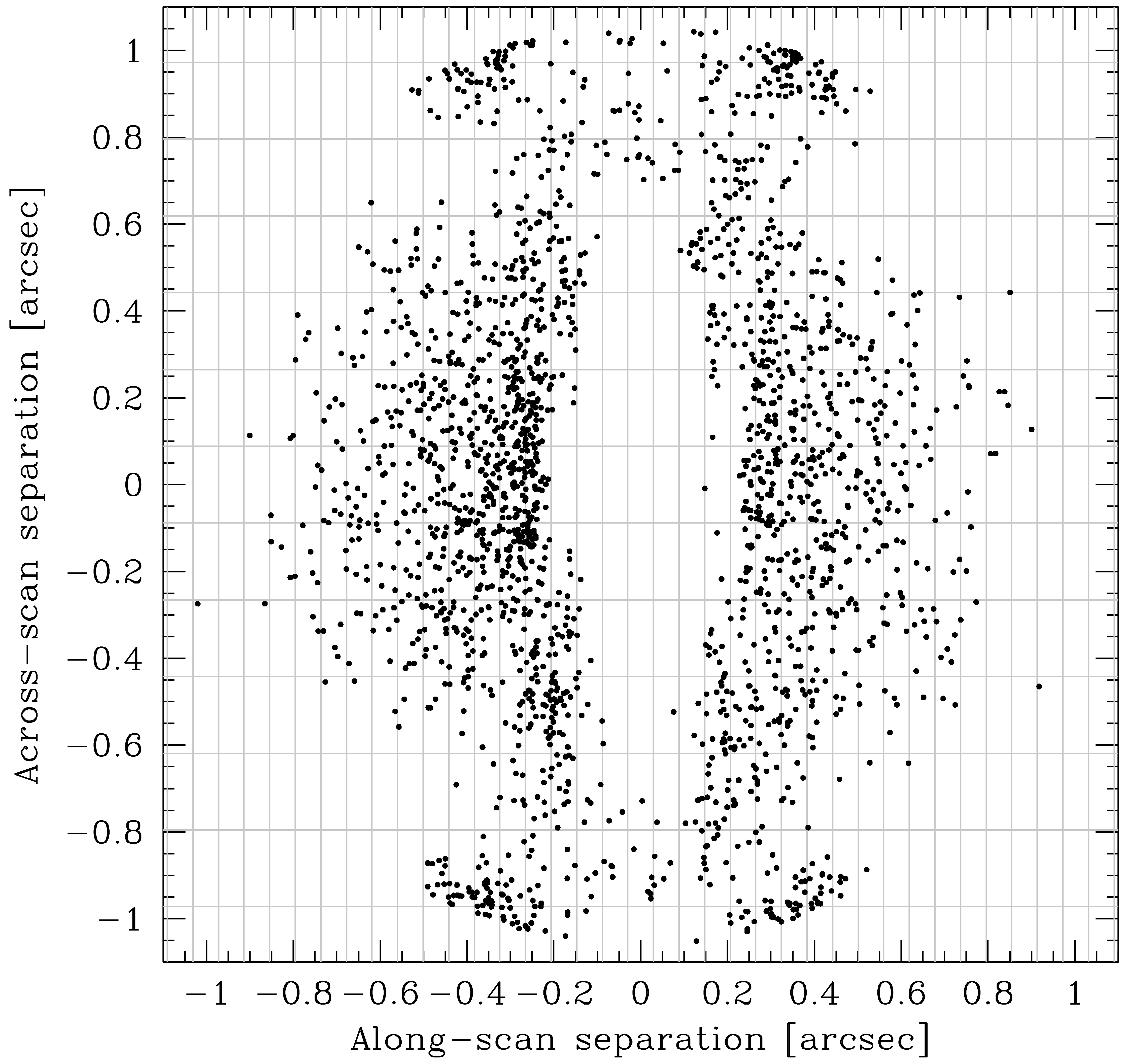}
  \caption{As Figure~\ref{fig:double_stars_baseline}, but for the optimised rejection parameters. The top and bottom panels contain $1746$ and $2140$ objects, respectively.}\label{fig:double_stars_optimised}
\end{figure*}

Section~\ref{subsec:double_star_dataset} summarises the contents of our double-star data set. We differentiate unresolved double stars, which only lead to one local maximum in the detection (symbolically $** \rightarrow *$), from resolved double stars, which lead to two local maxima in the detection (symbolically $** \rightarrow **$). Our resolved data set has $375,000$ underlying double-star systems, each with two resolved components, whereas the unresolved data set has $750,000$ underlying double-star systems with only unresolved components. Compared to the $\Delta G$ and $\rho$ ranges $0$--$5$~mag and $0$--$354$~mas used in the optimisation (Sections~\ref{sec:optimisation} and \ref{sec:results}), the simulated configuration space of double stars discussed here is limited to primaries in the range $G = 12.5$--$21$~mag and secondaries with magnitude difference $\Delta G = 0$--$8.5$~mag, separation $\rho = 0$--$1061$~mas ($0$--$18$~AL pixels), and orientation $\alpha = 0^\circ$--$360^\circ$ ($\alpha=0^\circ$ denotes the along-scan axis whereas $\alpha= 90^\circ$ denotes the across-scan axis). We only consider detections brighter than $G = 20$~mag.

Actually, the double-star data set allows to quantify Gaia's capability to resolve (close) double stars. The spatial distribution of the {\it resolved} secondary components with respect to their primary companions is composed of a semi-uniform background with an ellipsoidal hole centred around the primary. The spatial distribution of the {\it unresolved} secondary components with respect to their primary companions displays a fully complementary behaviour and shows a strongly-peaked distribution around the origin (``the primary'') which fits into the hole. The hole boundary represents the transition between resolved and unresolved double stars, refering to Gaia's detection capability of (close) double stars. The axis-ratio of this hole ($1:3$) reflects the rectangular pixel size ($10 \times 30$~$\mu$m$^{2}$, or $0.06 \times 0.18$~arcsec$^{2}$); this asymmetric sensitivity is not expected to introduce biases in the final Gaia catalogue since objects are typically observed $70$ times over the mission, with ``random'' scanning orientations.

Figure~\ref{fig:double_stars_resolution} shows the probability of a double star to be resolved into two local maxima as function of separation $\rho$ and magnitude difference $\Delta G$ when averaged over orientation angle $\alpha$, the idea being that even a few-percent probability means that the system will be resolved in at least one of the few dozen transits acquired during the mission. Gaia's resolving power does degrade with $\Delta G$ but does not vary with the primary's magnitude for a given $\Delta G$ value. Our results are consistent with those of \cite{LL:ASP-006}. Figure~\ref{fig:double_stars_probability} shows -- for bright, equal-brightness double stars ($G = 13$~mag and $\Delta G = 0$~mag) -- how Gaia's detection probability depends on orientation and separation. The best performance, $0.23$~arcsec with $95$\% confidence level, is obviously found for double stars with a pure along-scan separation. For systems with a separation purely in the across-scan direction, this number is a factor three worse, while for double stars with a random orientation, this number is a factor two~worse\footnote{
$$
(2\pi)^{-1} \cdot \int_{0}^{2\pi}{\rm d}\alpha \left([1\cdot\cos\alpha]^{2} + [3\cdot\sin\alpha]^{2}\right)^{1/2} = 2 \cdot E(-8) / \pi \approx 2.13,
$$
where $E(m)$ is the complete elliptic integral of the second kind.
}.

Figure~\ref{fig:double_stars_baseline} focuses on the impact of the rejection equations on close double stars: the top panels refer to unresolved systems, i.e., systems which lead to one local maximum, whereas the bottom panels refer to resolved systems, i.e., systems which lead to two local maxima. Figure~\ref{fig:double_stars_baseline} only shows the properties of the secondary component of those systems which have been rejected by the functional-baseline rejection parameters. The spatial distribution of the secondary components with respect to their primary companions for the $4631$ rejected unresolved double stars (top panels) is composed of a uniform background plus an ellipsoidal ring centred around the primary, of radius $\sim$$0.2$~arcsec ($\sim$$3$~AL pixels) in the along-scan direction and $\sim$$0.6$~arcsec ($\sim$$3$~AC pixels) in the across-scan direction, corresponding to marginally unresolved systems. The vast majority of rejected unresolved objects have $G \ga 19.5$~mag (see column $P_{G** \rightarrow *}$ in Table~\ref{tab:baseline_results}), meaning that their companions have $\Delta G \la 1$~mag (since Gaia's limit is $G = 20$~mag). This explains the rather sharp cut-off seen in $\Delta G$-space. The spatial distribution of the $13,376$ rejected secondary components with respect to their primary companions for the resolved systems (bottom panels) also follows a uniform background with an ellipsoidal ring superimposed, which is larger and more diffuse than for the unresolved double stars. The magnitude distribution of the primary components in these systems is rather uniformly spread between $G = 12.5$ and $20$~mag (see also column $P_{G** \rightarrow **}$ in Table~\ref{tab:baseline_results}).

Figure~\ref{fig:double_stars_optimised} shows the double-star-rejection results for the optimised rejection parameters from Section~\ref{sec:results}. The main structures from Figure~\ref{fig:double_stars_baseline} persist but are significantly thinned out, i.e., Gaia's rejection performance of (close) double stars with the optimised rejection parameters is significantly better than with the functional baseline ($4631$ rejected unresolved double stars reduce to $1746$ systems while $13,376$ rejected resolved double stars reduce to $2140$ objects). Striking is the absence of the uniform background of rejected unresolved double stars achieved with the functional-baseline parameters, i.e., unresolved, faint double stars with large(r) separations are correctly classified with the optimised rejection parameters. One should, however, not over-interpret the importance of detecting both components of close double stars since what ultimately matters is whether the secondary component is observed or not. This condition can be claimed to be met if the secondary is sufficiently-well contained in the $0.4 \times 2.1$-arcsec$^2$ window of the primary, so for instance when the along-scan separation is less than $0.2$~arcsec and/or when the across-scan separation is less than $1$~arcsec. Nonetheless, the observation always improves with a dedicated window centred on the secondary, allowing to improve the on-ground deconvolution of the (across-scan-binned) data.

\subsection{Unresolved galaxies}\label{subsec:scientific_results_galaxies}

\begin{figure}[t!]
  \centering
  \includegraphics[trim=20 20 20 20,clip,width=\columnwidth]{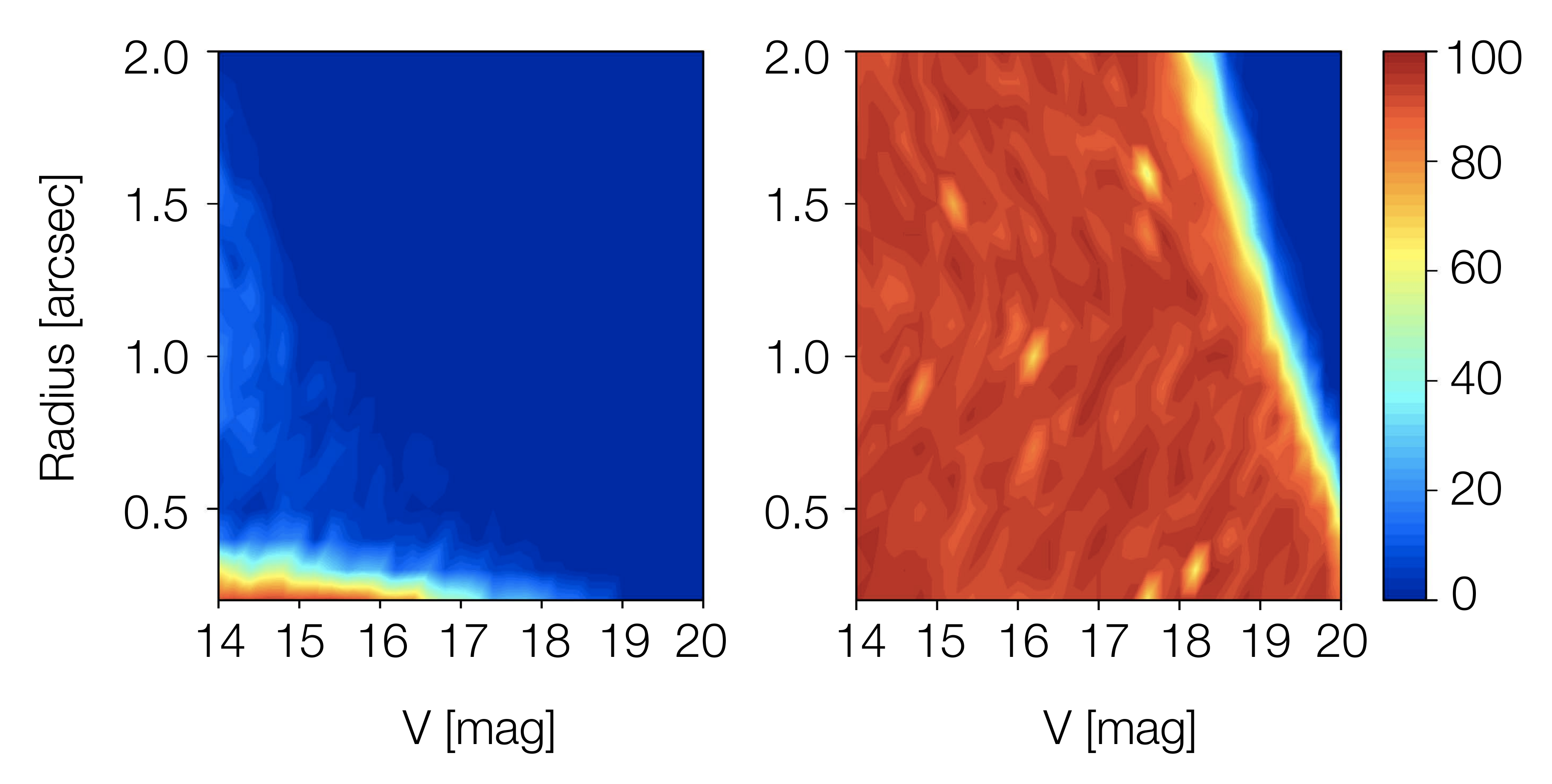}
  \includegraphics[trim=20 20 20 20,clip,width=\columnwidth]{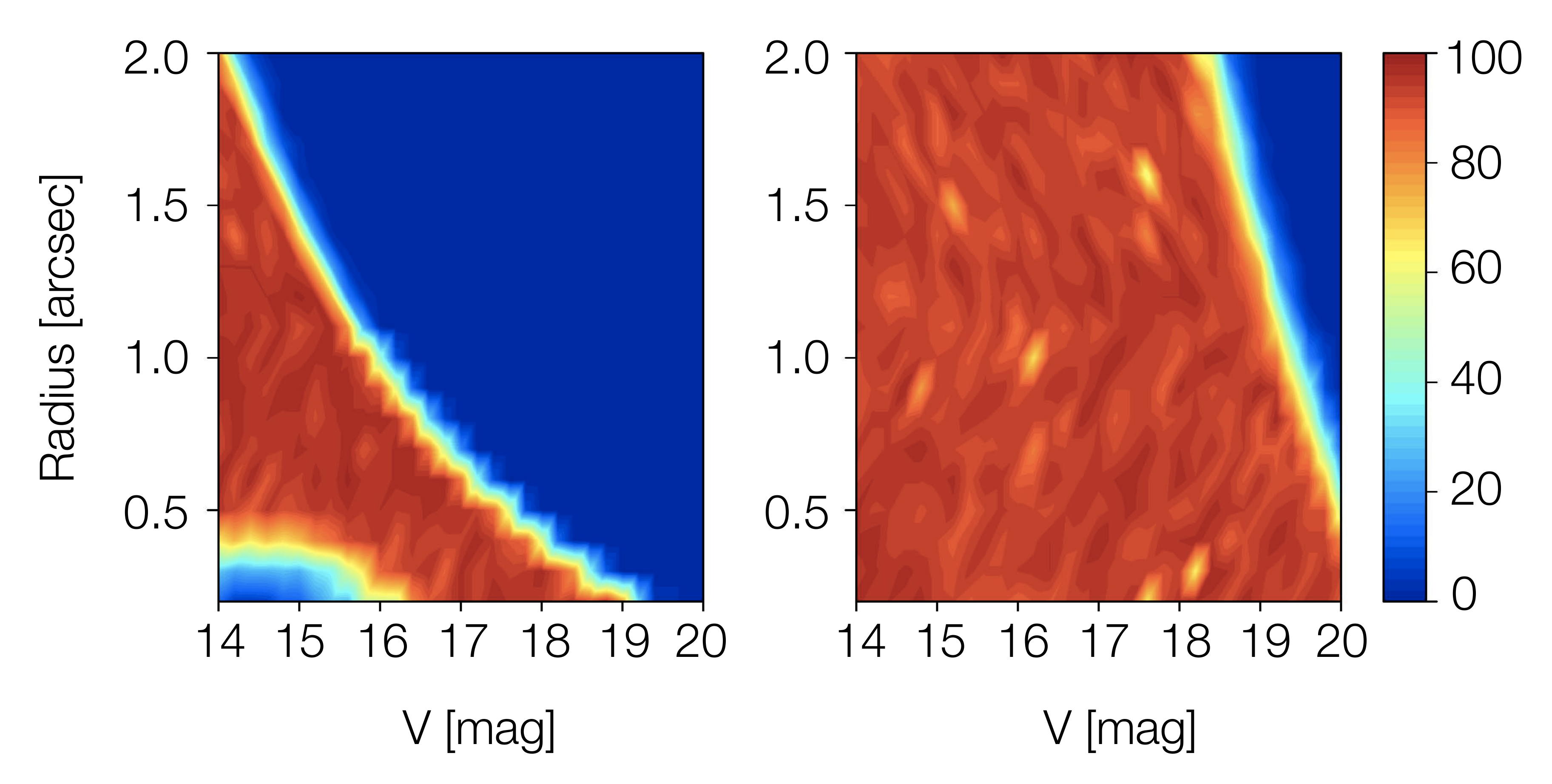}
  \caption{Galaxy detection probability (in \%) for the functional-baseline rejection parameters (top panels) and the optimised parameters (bottom panels). The left-hand-side panels represent exponential disk profiles while the right-hand-side panels represent de Vaucouleurs profiles. Noise structure shared between the upper and lower panels is due to the GIBIS simulation, for instance gaps between CCD~rows.}
  \label{Fig:GalProf-DetectionMap} 
\end{figure}

\begin{figure}[t!]
  \centering
  \includegraphics[trim=20 20 20 10,clip,width=\columnwidth]{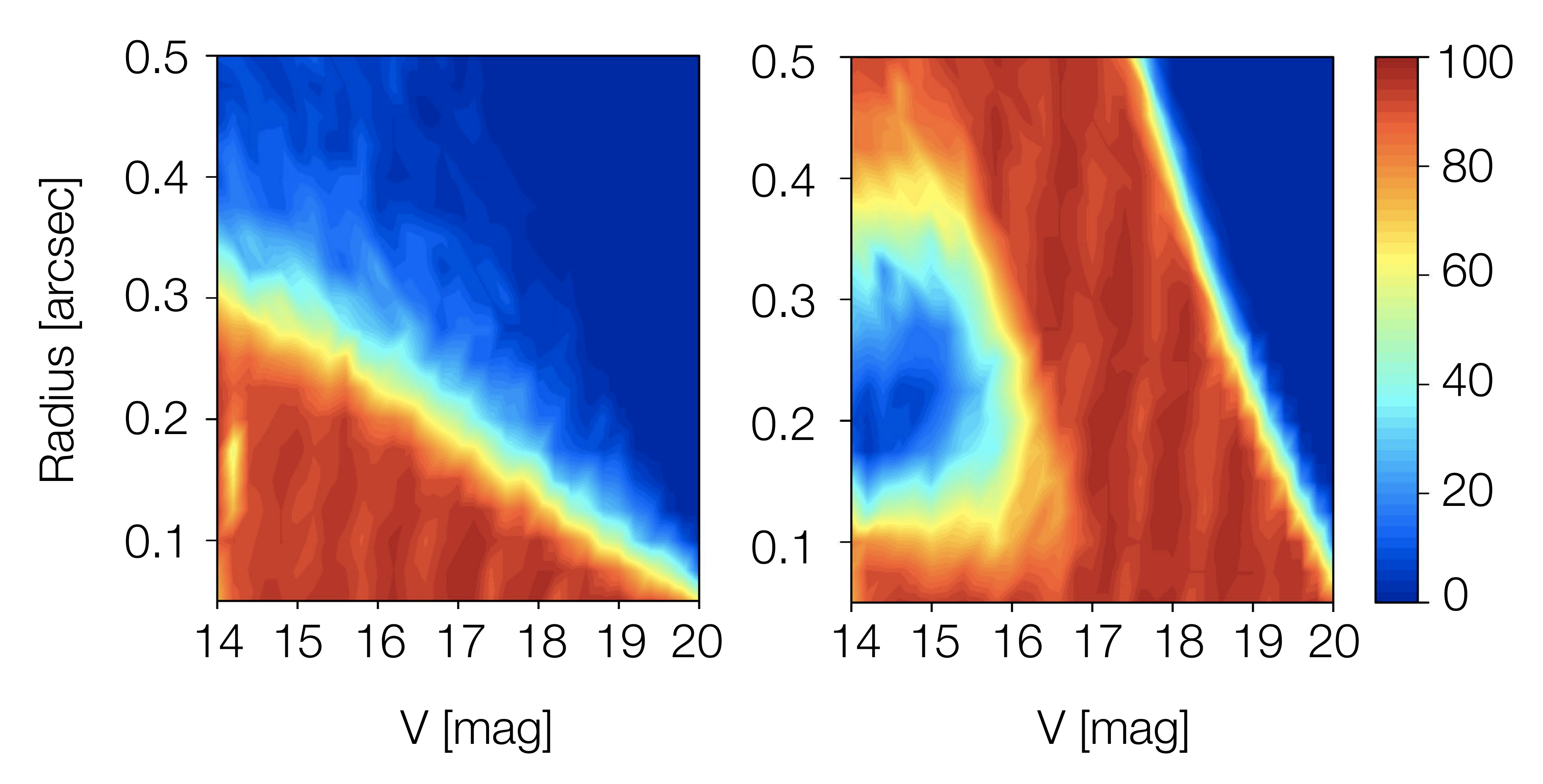}
  \caption{Detection probability (in \%) for exponential disk profiles with radius $<$$0.5$~arcsec for the functional-baseline rejection parameters (left) and the optimised parameters (right).}
  \label{Fig:GalProf-DetectionMap-verysmall} 
\end{figure}

To study the impact of the optimised parameters on the detectability of unresolved galaxies, we adopt the GIBIS simulations of exponential and de Vaucouleurs galaxy profiles summarised in Section~\ref{subsec:galaxies}. Figure~\ref{Fig:GalProf-DetectionMap} shows the resulting detection probabilities of the two extreme-profile cases as function of integrated brightness and radius. Two well-defined regions with a sharp transition can be seen: a region where the profiles are almost always detected, in red, and another where the profiles are not detected, in blue. Small-scale (noise) variations are due to the position of the galaxies in the focal plane and their position in the sky: since there are gaps between the CCD rows, and since some profiles may even fall outside the focal plane in some transits, the simulated profiles are {\it a priori} not expected to be observed in all transits and some random variation is expected.

The detection maps obtained with the functional-baseline rejection parameters, in the upper pannels, indicate that exponential disk profiles are not detected except for the most compact ones. On the other hand, most simulated de Vaucouleurs profiles are detected, with the most compact ones (with radius $<$$0.5$~arcsec) reaching the faint end of the simulated sources. The optimised rejection parameters significantly improve the detection of exponential disk profiles, as can be seen in the lower-left panel of Figure \ref{Fig:GalProf-DetectionMap}. The diagonal transition between the red (bottom-left) and blue (upper-right) area in that panel does {\it not} mark the intrinsic detection performance of Gaia, i.e., the fact that a detection is necessarily linked to the presence of a local maximum (see Section~\ref{sec:vpa} and Equation~\ref{eq:local_maximum} in particular) but rather marks the faint-end threshold at $G = 20$~mag: for large(r) effective radii, the galaxy flux is severely underestimated on board as a result of an over-subtraction of the background induced by the presence of galaxy light in the samples from which the background is estimated. As a result, for disk profiles, the region of the parameter space with successful detections is concentrated around the most compact or bright exponential profiles. The detection map shows an unexpected feature in the region of the brightest simulated compact profiles, with radius $\lesssim$$0.3$~arcsec and integrated magnitude $V \lesssim 16$~mag: profiles lying therein are rejected as, not surprisingly, ripples. In order to assess this effect, we performed additional GIBIS simulations, including a set of simulations with no-filter rejection parameters (see Section~\ref{subsec:real_mission}), for exponential profiles with radii varying from $50$ to $500$~mas. The resulting detection maps in Figure~\ref{Fig:GalProf-DetectionMap-verysmall} confirm (i) that the region of lower detections is a feature introduced by the optimised parameters, and (ii) that for radii $\lesssim$$100$~mas, the profiles are detected again for the entire magnitude range, up to the faint end of the simulations. The latter result is expected, as such profiles are almost indistinguishable from single stars. The ``detection hole'' is caused by the interplay between the low-frequency, along-scan rejection curves (which -- for $G \la 18$~mag -- are more strict for the optimised parameters than for the functional-baseline parameters), the size and brightness of the galaxy (which influence the [core] size of the image), the on-board background subtraction (which operates on the $5 \times 5$-samples [$0.6 \times 1.8$~arcsec$^{2}$] ring around the sample of interest -- see Section~\ref{sec:vpa}), and the on-board magnitude estimation (which is severely biased -- up to a few magnitudes for large(r) effective radii -- towards fainter magnitudes).

The results presented in Figure~\ref{Fig:GalProf-DetectionMap} for de Vaucouleurs profiles (right panels) show that the optimised parameters have no significant impact on their detectability. As for disk profiles, the diagonal transition between the red (bottom-left) and blue (upper-right) area marks the faint-end threshold of Gaia. The same region of parameter space that was covered using the baseline rejection parameters is also covered with the optimised ones. However, the optimised parameters do improve the detectability of intermediate profiles that are between the exponential and the de Vaucouleurs profiles.

Taking into account these results, we expect mostly elliptical galaxies and galaxy bulges to be detected by Gaia, and thus to be present in the catalogue, while late-type spiral galaxies -- even those with weak bulges -- will be mostly absent. One should keep in mind, however, that the window data transmitted to ground are of limited extent -- typically $4.7 \times 2.1$~arcsec$^2$ (along-scan $\times$ across-scan) in SM and $0.4 \times 2.1$~arcsec$^2$ in AF -- and of limited sampling -- typically $0.2 \times 0.7$~arcsec$^2$ per sample in SM and $0.06 \times 2.1$~arcsec$^2$ per sample in AF -- and hence not comparable to classical imaging data. These conclusions are a par with those by \citet{2014A&A...568A.124D}. Nevertheless, even though few real galaxies are expected to populate the region of the parameter space opened up by the optimised parameters, the adoption of this configuration will enable the exploration of a morphological regime that is exclusively available from space observations. Even the confirmation of no-detections due to the absence of real objects populating this region of the parameter space will provide important constraints. Moreover, as Gaia is the only space-based all-sky survey in the visible wavelength domain and this is a regime that has never been explored so systematically and extensively as Gaia will be able to do, this prospective is scientifically invaluable and deserves to be pursued.

\subsection{Asteroids}\label{subsec:scientific_results_asteroids}

\begin{figure}
  \centering
  \includegraphics[trim=12 10 10 20,clip,width=\columnwidth]{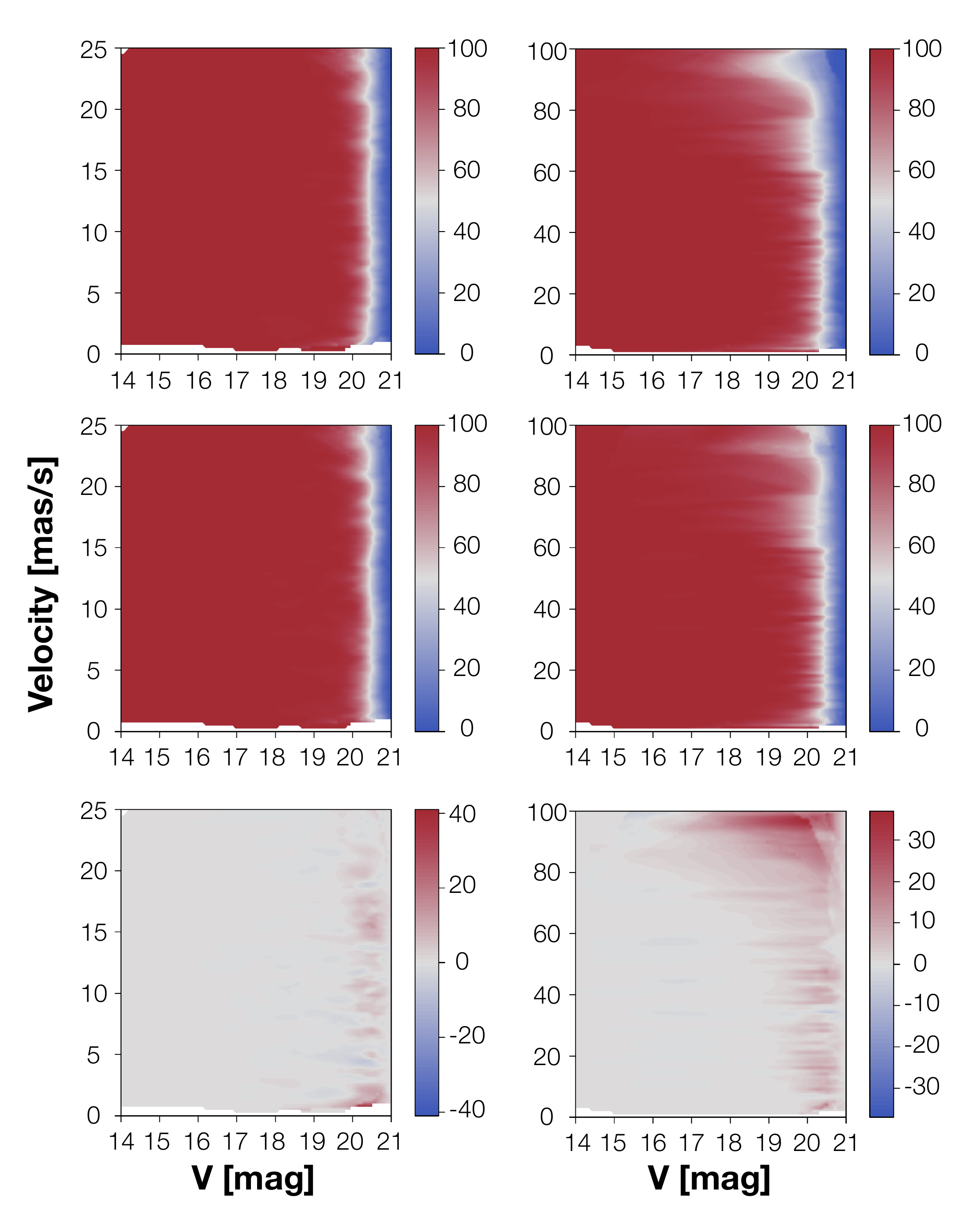}
  \caption{Detection-probability maps (in \%) for asteroids (left: MBAs; right: NEOs), using the functional-baseline rejection parameters (top) and the optimised parameters (middle), as function of $V$ magnitude and velocity modulus. The bottom panels represent the resulting gain (optimised minus functional baseline).}
  \label{Fig:AstProf-DetectionMap}
\end{figure}

\begin{figure}
  \centering
  \includegraphics[trim=15 1 10 21,clip,width=0.92\columnwidth]{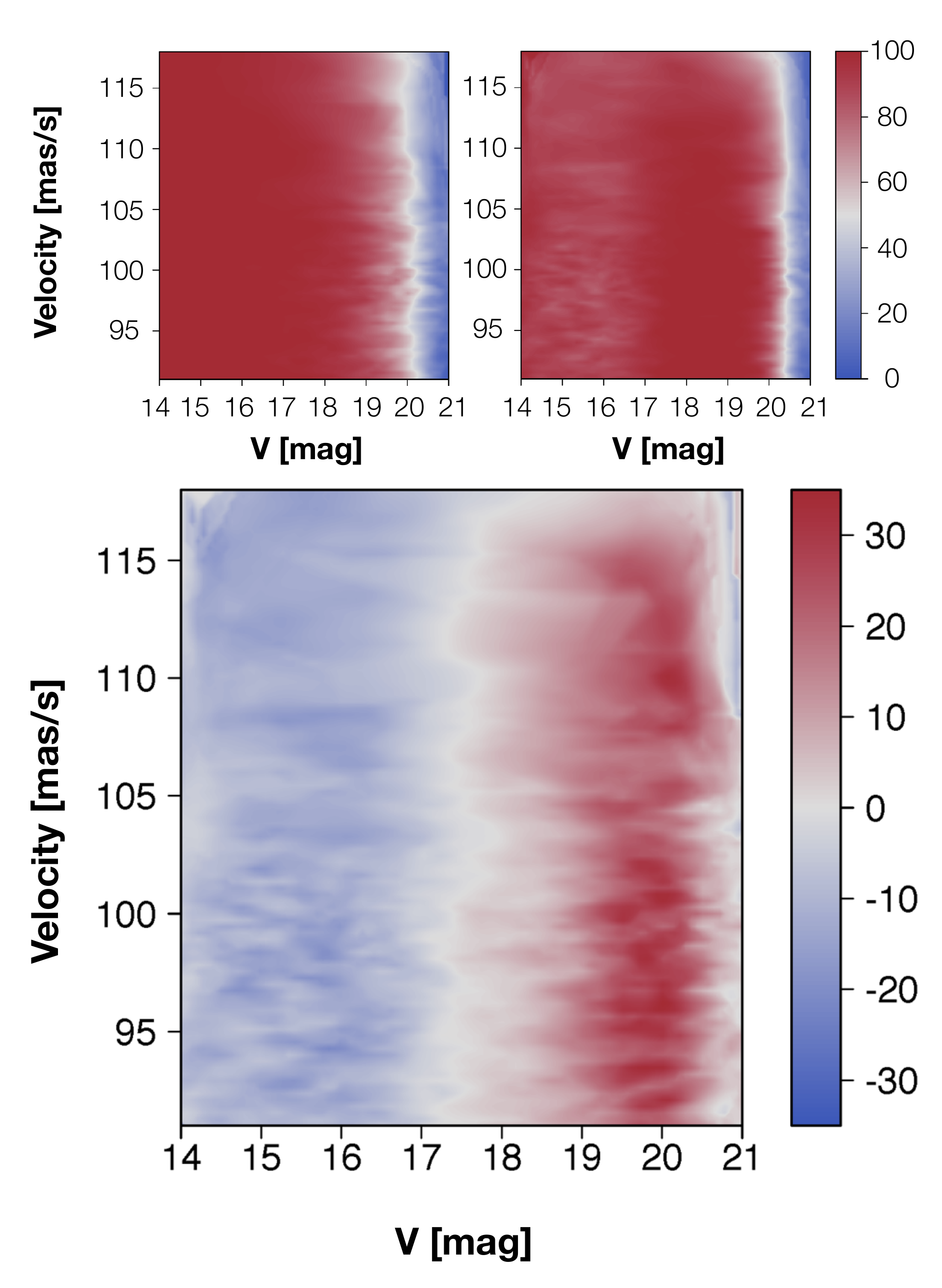}
  \caption{Detection-probability maps (in \%) for NEOs faster than $90$~mas~s$^{-1}$, using the functional-baseline rejection parameters (top, left) and the optimised parameters (top, right), as function of $V$ magnitude and velocity modulus. The bottom panel represents the resulting gain.}
  \label{Fig:AstProf-DetectionMap2} 
\end{figure}

To investigate how the adoption of the optimised parameters impacts the detection of asteroids, we use the GIBIS simulations of main-belt asteroids (MBAs) and near-Earth objects (NEOs) described in Section~\ref{subsec:asteroids}. The resulting detection maps for both types of objects, after averaging over the ten simulation grids and after taking the different projections of the velocity vector into account, are presented in Figure~\ref{Fig:AstProf-DetectionMap}. They show that, in both cases, the optimised rejection parameters provide detection gains for objects fainter than $20$~mag, in line with Figure~\ref{fig:performance@G=20mag}. Also, the results indicate that for high-velocity NEOs, with speed modulus exceeding $\sim$$80$~mas~s$^{-1}$, there is a $\sim$$30\%$ gain. To assess this effect, we perform additional simulations of $4640$ fast NEOs, with velocity modulus greater than $90$~mas~s$^{-1}$. The results are represented in Figure~\ref{Fig:AstProf-DetectionMap2}, and show a significant detection gain for objects at the faint end. On the other hand, these simulations also show that fast, bright NEOs have slightly lower detection rates when the optimised parameters are adopted.

The optimised rejection parameters allow a better coverage and completeness at the faint end of the sample population, in particular for fast-moving NEOs, albeit this gain comes at the expense of a $\sim$$10$--$20$\% detection probability loss for bright ($V < 17$~mag), fast-moving ($>$$80$~mas~s$^{-1}$) NEOs. However, the asteroid population is far from uniformly distributed in brightness-velocity space: bright and fast-moving bolides are much less frequent than faint and slower ones, implying that the gain and loss give a net science improvement. The optimised parameters allow in particular to find new faint objects ($V > 20$~mag for MBAs and slow NEOs and $V > 18$~mag for fast NEOs). Unfortunately, fast-moving objects, once detected, may not be observed in all subsequent astrometric CCDs because the window propagation is the same for all detections and based on 'fixed stars'. Science alerts in this respect can confirm the detection and bring valuable, complementary (astrometric) data.

\section{Conclusions}\label{sec:conclusion}

We present a study of Gaia's detection capability of objects, in particular (non-saturated) stars, double stars, and unresolved galaxies. We have developed an emulation of the on-board detection software which has been carefully validated against the real software. The algorithm has $20$ free, so-called rejection parameters governing the boundaries between stars on the one hand and point-like or elongated (high-frequency) prompt-particle events or extended (low-frequency) ripples on the other hand. We evaluate the detection and rejection performance of the algorithm using catalogues of simulated single stars, resolved and unresolved double stars, Galactic cosmic rays, and Solar protons. The functional-baseline rejection parameters allow to detect $99.961$\% of single stars, $98.417$\% of unresolved double stars, and $98.271$\% of resolved double stars (Table~\ref{tab:baseline_results}). At the same time, $6.349$\% and $3.401$\% of the cosmic rays and Solar protons, respectively, do not get rejected.

After optimisation, we managed to improve these performances to $99.997$\% for single stars, $99.866$\% for unresolved double stars, $99.928$\% for resolved double stars, $5.276$\% for cosmic rays, and $3.064$\% for Solar protons (Table~\ref{tab:optimised_results}). The optimised rejection parameters also remove the artefact of the functional-baseline parameters that the reduction of the detection probabililty of faint stars as function of $G$ magnitude already sets in before the nominal threshold at $G = 20$~mag.

We not-surprisingly find, as a result of the rectangular pixel size (along-scan $:$ across-scan $= 1:3$), that Gaia's intrinsic power to resolve close double stars -- i.e., before the application of PSF-shape criteria through the rejection equations and parameters in the detection process -- is better in the along- than in the across-scan direction: the minimum separation to resolve -- with $95$\% confidence level -- a close double star is $0.23$~arcsec in the along-scan and $0.70$~arcsec in the across-scan direction, which applies to equal-brightness double stars ($\Delta G = 0$~mag), regardless of the brightness of the primary. To resolve double stars with $\Delta G > 0$~mag, larger separations are required. For a given value of $\Delta G$, Gaia's resolving power does not vary with the magnitude of the primary.

Whereas the optimised rejection parameters have no significant (beneficial) impact on the detectability of pure de Vaucouleurs profiles, they do significantly improve the detection of pure exponential-disk profiles, and hence also the detection of unresolved external galaxies with intermediate profiles. The optimised rejection parameters also improve the detection of faint asteroids and high-velocity near-Earth objects, albeit at the expense of a modest detection-probability reduction of fast, bright near-Earth objects.

The only major side effect of the optimised parameters is that spurious ghosts in the wings of bright stars also pass unfiltered ($99.866$\% detection efficiency, versus $1.800$\% with the baseline). We have identified an alternative set of optimised parameters, which sacrifices some object-detection and prompt-particle-event rejection performance to reduce the ghost-detection sensitivity from $99.866$\% to $10.841$\% (Table~\ref{tab:alternative_results}). An in-orbit test during commissioning with the functional-baseline and the optimised parameters will provide input for the ultimate decision which parameter set to operate the spacecraft with during its operational~life.

\begin{acknowledgements}

It is a pleasure to thank Benjamin Massart, Juanma Fleitas, Alcione Mora, Timo Prusti, and Matthias Erdmann for constructive feedback received during this research project. Alex Short kindly provided his prompt-particle-event generation code. The Video Processing Algorithms (VPAs) have been developed by Airbus Defence \& Space (formerly known as EADS Astrium SAS), Toulouse, under the contract awarded by the European Space Agency (ESA) for the development of the Gaia spacecraft. Juanma Fleitas kindly performed some tests with the VPA prototype code. We thank the anonymous referee for providing detailed feedback which improved the presentation of the paper. This research has made use of the SIMBAD database and VizieR catalogue access tool, both operated at the Centre de Donn\'{e}es astronomiques de Strasbourg (CDS), of NASA's Astrophysics Data System (ADS), of SPace ENVironment Information System (SPENVIS), initiated by the Space Environment and Effects Section (TEC-EES) of ESA and developed by the Belgian Institute for Space Aeronomy (BIRA-IASB) under ESA contract through ESA's General Support Technologies Programme (GSTP), administered by the Belgian Federal Science Policy Office. This research has also made use of the Gaia Instrument and Basic Image Simulator (GIBIS), developed by Carine Babusiaux and collaborators within the Gaia Data Processing and Analysis Consortium (DPAC) and deployed by the Centre National d'\'Etudes Spatiales (CNES). AKM acknowledges the Portuguese agency \emph{Funda\c c\~ao para Ci\^encia e Tecnologia} \emph{(FCT)} for financial support (SFRH/BPD/74697/2010 \& PTDC/CTE-SPA/118692/2010).

\end{acknowledgements}


\begin{thebibliography}{57}
\expandafter\ifx\csname natexlab\endcsname\relax\def\natexlab#1{#1}\fi

\bibitem[{{Arenou}(2011)}]{2011AIPC.1346..107A}
{Arenou}, F. 2011, in American Institute of Physics Conference Series, Vol.
  1346, American Institute of Physics Conference Series, ed. J.~A. {Docobo},
  V.~S. {Tamazian}, \& Y.~Y. {Balega}, 107--121

\bibitem[{{Babusiaux}(2005)}]{2005ESASP.576..417B}
{Babusiaux}, C. 2005, in ESA Special Publication, Vol. 576, The
  Three-Dimensional Universe with Gaia, ed. C.~{Turon}, K.~S. {O'Flaherty}, \&
  M.~A.~C. {Perryman}, 417

\bibitem[{{Babusiaux} {et~al.}(2011){Babusiaux}, {Sartoretti}, {Leclerc}, \&
  {Ch{\'e}reau}}]{2011ascl.soft07002B}
{Babusiaux}, C., {Sartoretti}, P., {Leclerc}, N., \& {Ch{\'e}reau}, F. 2011,
  Astrophysics Source Code Library, 7002

\bibitem[{{Bailer-Jones}(2010)}]{2010MNRAS.403...96B}
{Bailer-Jones}, C.~A.~L. 2010, \mnras, 403, 96

\bibitem[{{Bertin} \& {Arnouts}(1996)}]{1996A&AS..117..393B}
{Bertin}, E. \& {Arnouts}, S. 1996, \aaps, 117, 393

\bibitem[{{Dale} {et~al.}(1993){Dale}, {Marshall}, {Cummings}, {Shamey}, \&
  {Holland}}]{1993ITNS...40.1628D}
{Dale}, C., {Marshall}, P., {Cummings}, B., {Shamey}, L., \& {Holland}, A.
  1993, IEEE Transactions on Nuclear Science, 40, 1628

\bibitem[{{de Bruijne}(2012)}]{2012Ap&SS.341...31D}
{de Bruijne}, J.~H.~J. 2012, \apss, 341, 31

\bibitem[{{de Bruijne} {et~al.}(2010){de Bruijne}, {Kohley}, \&
  {Prusti}}]{2010SPIE.7731E..35D}
{de Bruijne}, J.~H.~J., {Kohley}, R., \& {Prusti}, T. 2010, in Society of
  Photo-Optical Instrumentation Engineers (SPIE) Conference Series, Vol. 7731,
  Society of Photo-Optical Instrumentation Engineers (SPIE) Conference Series

\bibitem[{de~Keijser {et~al.}(1982)de~Keijser, Langford, Mittemeijer, \&
  Vogels}]{deKeijser:a21783}
de~Keijser, T.~H., Langford, J.~I., Mittemeijer, E.~J., \& Vogels, A. B.~P.
  1982, Journal of Applied Crystallography, 15, 308

\bibitem[{{de Souza} {et~al.}(2014){de Souza}, {Krone-Martins}, {dos Anjos},
  {Ducourant}, \& {Teixeira}}]{2014A&A...568A.124D}
{de Souza}, R.~E., {Krone-Martins}, A., {dos Anjos}, S., {Ducourant}, C., \&
  {Teixeira}, R. 2014, \aap, 568, A124

\bibitem[{{Dutton} {et~al.}(1997){Dutton}, {Woodward}, \&
  {Lomheim}}]{1997SPIE.3063...77D}
{Dutton}, T.~E., {Woodward}, W.~F., \& {Lomheim}, T.~S. 1997, in Society of
  Photo-Optical Instrumentation Engineers (SPIE) Conference Series, Vol. 3063,
  Society of Photo-Optical Instrumentation Engineers (SPIE) Conference Series,
  ed. G.~C. {Holst}, 77--101

\bibitem[{{ESA}(1997)}]{1997ESASP1200.....P}
{ESA}, ed. 1997, ESA Special Publication, Vol. 1200, {The HIPPARCOS and TYCHO
  catalogues. Astrometric and photometric star catalogues derived from the ESA
  HIPPARCOS Space Astrometry Mission}

\bibitem[{{Eyer} {et~al.}(2011){Eyer}, {Suveges}, {Dubath}, {Mowlavi}, {Greco},
  {Varadi}, {Evans}, \& {Bartholdi}}]{2011EAS....45..161E}
{Eyer}, L., {Suveges}, M., {Dubath}, P., {et~al.} 2011, in EAS Publications
  Series, Vol.~45, EAS Publications Series, 161--166

\bibitem[{{Gielesen} {et~al.}(2012){Gielesen}, {de Bruijn}, {van den Dool},
  {Kamphues}, {Meijer}, {Calvel}, {Laborie}, {Monteiro}, {Coatantiec},
  {Touzeau}, {Erdmann}, \& {Gare}}]{2012SPIE.8442E..1RG}
{Gielesen}, W., {de Bruijn}, D., {van den Dool}, T., {et~al.} 2012, in Society
  of Photo-Optical Instrumentation Engineers (SPIE) Conference Series, Vol.
  8442, Society of Photo-Optical Instrumentation Engineers (SPIE) Conference
  Series

\bibitem[{{G{\'o}mez} {et~al.}(2010){G{\'o}mez}, {Helmi}, {Brown}, \&
  {Li}}]{2010MNRAS.408..935G}
{G{\'o}mez}, F.~A., {Helmi}, A., {Brown}, A.~G.~A., \& {Li}, Y.-S. 2010,
  \mnras, 408, 935

\bibitem[{{Hestroffer} {et~al.}(2010){Hestroffer}, {Dell'Oro}, {Cellino}, \&
  {Tanga}}]{hestro10_lnp}
{Hestroffer}, D., {Dell'Oro}, A., {Cellino}, A., \& {Tanga}, P. 2010, in
  Lecture Notes in Physics, Berlin Springer Verlag, Vol. 790, Lecture Notes in
  Physics, Berlin Springer Verlag, ed. {J.~Souchay \& R.~Dvorak}, 251--340

\bibitem[{{Hestroffer} \& {Tanga}(2014)}]{hestro14_cosp}
{Hestroffer}, D. \& {Tanga}, P. 2014, in COSPAR Meeting, Vol.~40, 40th COSPAR
  Scientific Assembly. Held 2-10 August 2014, in Moscow, Russia, Abstract
  B0.4-20-14., 1199

\bibitem[{{Holl} {et~al.}(2012){Holl}, {Prod'homme}, {Lindegren}, \&
  {Brown}}]{2012MNRAS.422.2786H}
{Holl}, B., {Prod'homme}, T., {Lindegren}, L., \& {Brown}, A.~G.~A. 2012,
  \mnras, 422, 2786

\bibitem[{{Jordi} {et~al.}(2010){Jordi}, {Gebran}, {Carrasco}, {de Bruijne},
  {Voss}, {Fabricius}, {Knude}, {Vallenari}, {Kohley}, \&
  {Mora}}]{2010A&A...523A..48J}
{Jordi}, C., {Gebran}, M., {Carrasco}, J.~M., {et~al.} 2010, \aap, 523, A48

\bibitem[{{Katz} {et~al.}(2011){Katz}, {Cropper}, {Meynadier}, {Jean-Antoine},
  {Allende Prieto}, {Baker}, {Benson}, {Berthier}, {Bigot}, {Blomme},
  {Boudreault}, {Chemin}, {Crifo}, {Damerdji}, {David}, {David}, {Delle Luche},
  {Dolding}, {Fr{\'e}mat}, {Gerbier}, {Gerssen}, {G{\'o}mez}, {Gosset},
  {Guerrier}, {Guy}, {Hall}, {Hestroffer}, {Huckle}, {Jasniewicz}, {Ludwig},
  {Martayan}, {Morel}, {Nguyen}, {Ocvirk}, {Parr}, {Royer}, {Sartoretti},
  {Seabroke}, {Simon}, {Smith}, {Soubiran}, {Steinmetz}, {Th{\'e}venin},
  {Turon}, {Udry}, {Veltz}, \& {Viala}}]{2011EAS....45..189K}
{Katz}, D., {Cropper}, M., {Meynadier}, F., {et~al.} 2011, in EAS Publications
  Series, Vol.~45, EAS Publications Series, 189--194

\bibitem[{{Kirkpatrick}(1979)}]{1979ITED...26.1742K}
{Kirkpatrick}, S. 1979, IEEE Transactions on Electron Devices, 26, 1742

\bibitem[{{Kohley} {et~al.}(2012){Kohley}, {Gar{\'e}}, {V{\'e}tel}, {Marchais},
  \& {Chassat}}]{2012SPIE.8442E..1PK}
{Kohley}, R., {Gar{\'e}}, P., {V{\'e}tel}, C., {Marchais}, D., \& {Chassat}, F.
  2012, in Society of Photo-Optical Instrumentation Engineers (SPIE) Conference
  Series, Vol. 8442, Society of Photo-Optical Instrumentation Engineers (SPIE)
  Conference Series

\bibitem[{{Kordopatis} {et~al.}(2011){Kordopatis}, {Recio-Blanco}, {de
  Laverny}, {Bijaoui}, {Hill}, {Gilmore}, {Wyse}, \&
  {Ordenovic}}]{2011A&A...535A.106K}
{Kordopatis}, G., {Recio-Blanco}, A., {de Laverny}, P., {et~al.} 2011, \aap,
  535, A106

\bibitem[{{Krone-Martins} {et~al.}(2013){Krone-Martins}, {Ducourant},
  {Teixeira}, {Galluccio}, {Gavras}, {dos Anjos}, {de Souza}, {Machado}, \& {le
  Campion}}]{KroneMartins2013}
{Krone-Martins}, A., {Ducourant}, C., {Teixeira}, R., {et~al.} 2013, \aap, 556,
  102

\bibitem[{Langford(1978)}]{Langford:a16721}
Langford, J.~I. 1978, Journal of Applied Crystallography, 11, 10

\bibitem[{{Lindegren} {et~al.}(2008){Lindegren}, {Babusiaux}, {Bailer-Jones},
  {Bastian}, {Brown}, {Cropper}, {H{\o}g}, {Jordi}, {Katz}, {van Leeuwen},
  {Luri}, {Mignard}, {de Bruijne}, \& {Prusti}}]{2008IAUS..248..217L}
{Lindegren}, L., {Babusiaux}, C., {Bailer-Jones}, C., {et~al.} 2008, in IAU
  Symposium, Vol. 248, IAU Symposium, ed. {W.~J.~Jin, I.~Platais, \&
  M.~A.~C.~Perryman}, 217--223

\bibitem[{{Liu} {et~al.}(2012){Liu}, {Bailer-Jones}, {Sordo}, {Vallenari},
  {Borrachero}, {Luri}, \& {Sartoretti}}]{2012MNRAS.426.2463L}
{Liu}, C., {Bailer-Jones}, C.~A.~L., {Sordo}, R., {et~al.} 2012, \mnras, 426,
  2463

\bibitem[{{Lomheim} {et~al.}(1990){Lomheim}, {Shima}, {Angione}, {Woodward}, \&
  {Asman}}]{1990ITNS...37.1876L}
{Lomheim}, T.~S., {Shima}, R.~M., {Angione}, J.~R., {Woodward}, W.~F., \&
  {Asman}, D.~J. 1990, IEEE Transactions on Nuclear Science, 37, 1876

\bibitem[{{Massart}(2012)}]{LL:GAIA.ASF.MEM.PLM.00259}
{Massart}, B. 2012, EADS~Astrium memorandum GAIA.ASF.MEM.PLM.00259
  {C}alibration {P}rocedure {P}roposal of the {R}ejection {P}arameters, issue
  1, revision 0

\bibitem[{{Mignard} {et~al.}(2007){Mignard}, {Cellino}, {Muinonen}, {Tanga},
  {Delb{\`o}}, {Dell'Oro}, {Granvik}, {Hestroffer}, {Mouret}, {Thuillot}, \&
  {Virtanen}}]{Mignard07}
{Mignard}, F., {Cellino}, A., {Muinonen}, K., {et~al.} 2007, Earth Moon and
  Planets, 101, 97

\bibitem[{{Mignard} \& {Klioner}(2010)}]{2010IAUS..261..306M}
{Mignard}, F. \& {Klioner}, S.~A. 2010, in IAU Symposium, Vol. 261, IAU
  Symposium, ed. {S.~A.~Klioner, P.~K.~Seidelmann, \& M.~H.~Soffel}, 306--314

\bibitem[{{Mora} \& {Vosteen}(2012)}]{2012SPIE.8442E..1QM}
{Mora}, A. \& {Vosteen}, A. 2012, in Society of Photo-Optical Instrumentation
  Engineers (SPIE) Conference Series, Vol. 8442, Society of Photo-Optical
  Instrumentation Engineers (SPIE) Conference Series

\bibitem[{{Mouret}(2011)}]{2011PhRvD..84l2001M}
{Mouret}, S. 2011, \prd, 84, 122001

\bibitem[{{Nelder} \& {Mead}(1965)}]{1965CJ......7..308N}
{Nelder}, J.~A. \& {Mead}, R. 1965, Computer Journal, 7, 308

\bibitem[{{Pasquier} \& {Massart}(2012)}]{LL:GAIA.ASF.TCN.PLM.00709}
{Pasquier}, J.-F. \& {Massart}, B. 2012, {EADS~Astrium technical note
  GAIA.ASF.TCN.PLM.00709 {R}adiation {C}ampaign \#5. {A}strium {S}{M}
  {D}etection {T}ests {R}eport}, issue 1, revision 0

\bibitem[{{Perryman}(2009)}]{2009aaat.book.....P}
{Perryman}, M. 2009, {Astronomical Applications of Astrometry: Ten Years of
  Exploitation of the Hipparcos Satellite Data} (Cambridge University Press)

\bibitem[{{Perryman} {et~al.}(2001){Perryman}, {de Boer}, {Gilmore}, {H{\o}g},
  {Lattanzi}, {Lindegren}, {Luri}, {Mignard}, {Pace}, \& {de
  Zeeuw}}]{2001A&A...369..339P}
{Perryman}, M.~A.~C., {de Boer}, K.~S., {Gilmore}, G., {et~al.} 2001, \aap,
  369, 339

\bibitem[{{Perryman} {et~al.}(1997){Perryman}, {Lindegren}, {Kovalevsky},
  {Ho{\o}g}, {Bastian}, {Bernacca}, {Cr{\'e}z{\'e}}, {Donati}, {Grenon},
  {Grewing}, {van Leeuwen}, {van der Marel}, {Mignard}, {Murray}, {Le Poole},
  {Schrijver}, {Turon}, {Arenou}, {Froeschl{\'e}}, \&
  {Petersen}}]{1997A&A...323L..49P}
{Perryman}, M.~A.~C., {Lindegren}, L., {Kovalevsky}, J., {et~al.} 1997, \aap,
  323, L49

\bibitem[{{Pickles}(1998)}]{1998PASP..110..863P}
{Pickles}, A.~J. 1998, \pasp, 110, 863

\bibitem[{{Pourbaix}(2008)}]{2008IAUS..248...59P}
{Pourbaix}, D. 2008, in IAU Symposium, Vol. 248, IAU Symposium, ed. {W.~J.~Jin,
  I.~Platais, \& M.~A.~C.~Perryman}, 59--65

\bibitem[{Press {et~al.}(2007)Press, Teukolsky, Vetterling, \&
  Flannery}]{Press:2007:NRE:1403886}
Press, W.~H., Teukolsky, S.~A., Vetterling, W.~T., \& Flannery, B.~P. 2007,
  Numerical Recipes 3rd Edition: The Art of Scientific Computing, 3rd edn. (New
  York, NY, USA: Cambridge University Press)

\bibitem[{{Prod'homme} {et~al.}(2012){Prod'homme}, {Holl}, {Lindegren}, \&
  {Brown}}]{2012MNRAS.419.2995P}
{Prod'homme}, T., {Holl}, B., {Lindegren}, L., \& {Brown}, A.~G.~A. 2012,
  \mnras, 419, 2995

\bibitem[{{Provost} {et~al.}(2007){Provost}, {Le Roy}, {Mamdy}, {Flandin}, \&
  {Paulsen}}]{2007ESASP.638E..39P}
{Provost}, S., {Le Roy}, M., {Mamdy}, B., {Flandin}, G., \& {Paulsen}, T. 2007,
  in ESA Special Publication, Vol. 638, DASIA 2007 - Data Systems In Aerospace

\bibitem[{{Robin} {et~al.}(2012){Robin}, {Luri}, {Reyl{\'e}}, {Isasi}, {Grux},
  {Blanco-Cuaresma}, {Arenou}, {Babusiaux}, {Belcheva}, {Drimmel}, {Jordi},
  {Krone-Martins}, {Masana}, {Mauduit}, {Mignard}, {Mowlavi},
  {Rocca-Volmerange}, {Sartoretti}, {Slezak}, \&
  {Sozzetti}}]{2012A&A...543A.100R}
{Robin}, A.~C., {Luri}, X., {Reyl{\'e}}, C., {et~al.} 2012, \aap, 543, A100

\bibitem[{{Sozzetti}(2011)}]{2011EAS....45..273S}
{Sozzetti}, A. 2011, in EAS Publications Series, Vol.~45, EAS Publications
  Series, 273--278

\bibitem[{{Spagna}(2014)}]{LL:ASP-006}
{Spagna}, A. 2014, DPAC technical note GAIA-C3-TN-OATO-ASP-006 {A}symmetric
  {A}ngular {R}esolution of the {G}aia {D}etections, issue 2, revision 0

\bibitem[{{Stancik} \& {Brauns}(2008)}]{2008VibSpec47...66S}
{Stancik}, A.~L. \& {Brauns}, E.~B. 2008, Vibrational Spectroscopy, 47, 66

\bibitem[{{Tanga} {et~al.}(2012){Tanga}, {Campins}, \&
  {Paolicchi}}]{2012P&SS...73....1T}
{Tanga}, P., {Campins}, H., \& {Paolicchi}, P. 2012, \planss, 73, 1

\bibitem[{{Thuillot} {et~al.}(2014){Thuillot}, {Carry}, {Berthier}, {David},
  {Hestroffer}, \& {Rocher}}]{thuillot14_sf2a}
{Thuillot}, W., {Carry}, B., {Berthier}, J., {et~al.} 2014, in SF2A-2014:
  Proceedings of the Annual meeting of the French Society of Astronomy and
  Astrophysics, ed. J.~{Ballet}, F.~{Martins}, F.~{Bournaud}, R.~{Monier}, \&
  C.~{Reyl{\'e}}, 445--448

\bibitem[{{Tsalmantza} {et~al.}(2009){Tsalmantza}, {Kontizas},
  {Rocca-Volmerange}, {Bailer-Jones}, {Kontizas}, {Bellas-Velidis}, {Livanou},
  {Korakitis}, {Dapergolas}, {Vallenari}, \& {Fioc}}]{Tsalmantza2009}
{Tsalmantza}, P., {Kontizas}, M., {Rocca-Volmerange}, B., {et~al.} 2009, \aap,
  504, 1071

\bibitem[{{Turon} {et~al.}(1992){Turon}, {Cr{\'e}z{\'e}}, {Egret}, {G{\'o}mez},
  {Grenon}, {Jahrei{\ss}}, {R{\'e}qui{\`e}me}, {Argue}, {Bec-Borsenberger},
  {Dommanget}, {Mennessier}, {Arenou}, {Chareton}, {Crifo}, {Mermilliod},
  {Morin}, {Nicolet}, {Nys}, {Pr{\'e}vot}, {Rousseau}, {Perryman}, \& {et
  al.}}]{1992ESASP1136.....T}
{Turon}, C., {Cr{\'e}z{\'e}}, M., {Egret}, D., {et~al.}, eds. 1992, ESA Special
  Publication, Vol. 1136, {The HIPPARCOS input catalogue}

\bibitem[{{Tylka} {et~al.}(1997){Tylka}, {Adams}, {Boberg}, {Brownstein},
  {Dietrich}, {Flueckiger}, {Petersen}, {Shea}, {Smart}, \&
  {Smith}}]{1997ITNS...44.2150T}
{Tylka}, A.~J., {Adams}, J.~H., {Boberg}, P.~R., {et~al.} 1997, IEEE
  Transactions on Nuclear Science, 44, 2150

\bibitem[{{van de Hulst} \& {Reesinck}(1947)}]{1947ApJ...106..121V}
{van de Hulst}, H.~C. \& {Reesinck}, J.~J.~M. 1947, \apj, 106, 121

\bibitem[{{Wertheim} {et~al.}(1974){Wertheim}, {Butler}, {West}, \&
  {Buchanan}}]{Wertheim:1974}
{Wertheim}, G.~K., {Butler}, M.~A., {West}, K.~W., \& {Buchanan}, D. N.~E.
  1974, Review of Scientific Instruments, 45, 1369

\bibitem[{{Wilkinson} {et~al.}(2005){Wilkinson}, {Vallenari}, {Turon},
  {Munari}, {Katz}, {Bono}, {Cropper}, {Helmi}, {Robichon}, {Th{\'e}venin},
  {Vidrih}, {Zwitter}, {Arenou}, {Baylac}, {Bertelli}, {Bijaoui}, {Boschi},
  {Castelli}, {Crifo}, {David}, {Gomboc}, {G{\'o}mez}, {Haywood}, {Jauregi},
  {de Laverny}, {Lebreton}, {Marrese}, {Marsh}, {Mignot}, {Morin}, {Pasetto},
  {Perryman}, {Pr{\v s}a}, {Recio-Blanco}, {Royer}, {Sellier}, {Siviero},
  {Sordo}, {Soubiran}, {Tomasella}, \& {Viala}}]{2005MNRAS.359.1306W}
{Wilkinson}, M.~I., {Vallenari}, A., {Turon}, C., {et~al.} 2005, \mnras, 359,
  1306

\bibitem[{{Xapsos} {et~al.}(1999){Xapsos}, {Summers}, {Barth},
  {Stassinopoulos}, \& {Burke}}]{1999ITNS...46.1481X}
{Xapsos}, M.~A., {Summers}, G.~P., {Barth}, J.~L., {Stassinopoulos}, E.~G., \&
  {Burke}, E.~A. 1999, IEEE Transactions on Nuclear Science, 46, 1481

\bibitem[{{Xapsos} {et~al.}(2000){Xapsos}, {Summers}, {Barth},
  {Stassinopoulos}, \& {Burke}}]{2000ITNS...47..486X}
{Xapsos}, M.~A., {Summers}, G.~P., {Barth}, J.~L., {Stassinopoulos}, E.~G., \&
  {Burke}, E.~A. 2000, IEEE Transactions on Nuclear Science, 47, 486

\end{thebibliography}

\Online

\end{document}